\documentclass[authoryear,3p,times,preprint,review,fleqn]{elsarticle}
\usepackage{amsmath,natbib,mathrsfs,stmaryrd,titletoc,subfigure}
\usepackage[ruled,vlined]{algorithm2e}
\usepackage[scaled=0.92]{helvet}	% set Helvetica as the sans-serif font
\usepackage[pdftex,
CJKbookmarks=false, bookmarksnumbered=true, bookmarksopen=true,
colorlinks=true, % Ã¯Â¿Â½Ã¯Â¿Â½Ã¯Â¿Â½Ã¯Â¿Â½Ã¯Â¿Â½Ã¯Â¿Â½ÃÂ¡Ã¯Â¿Â½Ã¯Â¿Â½ÃÂ±Ã¯Â¿Â½Ã¯Â¿Â½Ã¯Â¿Â½Ã¯Â¿Â½Ã¯Â¿Â½ÃÂ¸ÃÂ³Ã¯Â¿Â½false Ã¯Â¿Â½Ã¯Â¿Â½Ã¯Â¿Â½Ã¯Â¿Â½Ã¯Â¿Â½Ã¯Â¿Â½Ã¯Â¿Â½Ã¯Â¿Â½Ã¯Â¿Â½Ã¥Â¶Â¼Ã¯Â¿Â½ÃÂºÃ¯Â¿Â½ÃÂ«
citecolor=blue, linkcolor=red, anchorcolor=green,
urlcolor=blue]{hyperref}
\usepackage[capitalise]{cleveref}
\usepackage{tabularx}
\usepackage{booktabs}
\usepackage[usenames]{color}
\usepackage[textsize=tiny]{todonotes}

\usepackage[section]{placeins}

\newcommand{\bfsigma}{\boldsymbol{\sigma}}
\newcommand{\bfepsilon}{\boldsymbol{\epsilon}}

\newcommand{\ts}{\text{sym}}

\newcommand{\td}{\text{d}}

\newcommand{\itw}{\mathrm{\mathit{w}}}

\numberwithin{equation}{section}

\newcounter{remark}[section]
\numberwithin{remark}{section}
\newcommand{\remark}{\noindent \textbf{Remark} \refstepcounter{remark} \textbf{\theremark} \;}

\SetAlCapSkip{1em} %=> vertical space between algorithms and text
\crefname{figure}{Figure}{Figure}
\crefname{equation}{Eq.}{Eqs.}
\definecolor{darkgray}{rgb}{0.95,0.95,0.95}

\begin{document}

\begin{frontmatter}

% Title, authors and addresses

% use the tnoteref command within \title for footnotes;
% use the tnotetext command for theassociated footnote;
% use the fnref command within \author or \address for footnotes;
% use the fntext command for theassociated footnote;
% use the corref command within \author for corresponding author footnotes;
% use the cortext command for theassociated footnote;
% use the ead command for the email address,
% and the form \ead[url] for the home page:
% \title{Title\tnoteref{label1}}
% \tnotetext[label1]{}
% \author{Name\corref{cor1}\fnref{label2}}
% \ead{email address}
% \ead[url]{home page}
% \fntext[label2]{}
% \cortext[cor1]{}
% \address{Address\fnref{label3}}
% \fntext[label3]{}

\title{\textbf{A generalized phase-field cohesive zone model (\textit{$\mu$PF-CZM}) for fracture}}

%\tnotetext[label1]{Dedicate to }

% use optional labels to link authors explicitly to addresses:
% \author[label1,label2]{}
% \address[label1]{}
% \address[label2]{}

\author{Jian-Ying Wu\corref{cor1}}
\cortext[cor1]{Tel.: (+86) 20-87112787} \ead{jywu@scut.edu.cn}
\address{State Key Laboratory of Subtropical Building and Urban Science, South China University of Technology, 510641 Guangzhou, China.}

%\author[label1,label2]{Jian-Ying Wu\corref{cor1}}
%\cortext[cor1]{Tel.: (+86) 20-87112787} \ead{jywu@scut.edu.cn}
%\address[label1]{State Key Laboratory of Subtropical Building Science, South China University of Technology, 510641 Guangzhou, China.}
%\address[label2]{Department of Civil Engineering, South China University of Technology, 510641 Guangzhou, China.}
%
%\author[label2]{Luo-Jia Zhuang}
%
%\author[label3]{Vinh Phu Nguyen}
%\address[label3]{Department of Civil Engineering, Monash University, Clayton, Victoria 3800, Australia.}

\begin{abstract}

In this work a generalized phase-field cohesive zone model (\texttt{$\mu$PF-CZM}) is proposed within the framework of the unified phase-field theory for brittle and cohesive fracture. With the introduction of an extra dissipation function for the crack driving force, in addition to the geometric function for the phase-field regularization and the degradation function for the constitutive relation, theoretical and application scopes of the original \texttt{PF-CZM} are broadened greatly. These characteristic functions are analytically determined from the conditions for the length scale insensitivity and a non-shrinking crack band in a universal, optimal and rationalized manner, for almost any specific traction--separation law. In particular, with an optimal geometric function, the crack irreversibility can be considered without affecting the target traction--separation softening law. Not only concave softening behavior but also high-order cohesive traction, both being limitations of the previous works, can be properly dealt with. The global fracture responses are insensitive not only to the phase-field length scale but also to the traction order parameter, though the crack bandwidth might be affected by both. Despite the loss of variational consistency in general cases, the resulting \texttt{$\mu$PF-CZM} is still thermodynamically consistent. Moreover, the existing numerical implementation can be adopted straightforwardly with minor modifications. Representative numerical examples are presented to validate the proposed \texttt{$\mu$PF-CZM} and to demonstrate its capabilities in capturing brittle and cohesive fracture with general softening behavior. The insensitivity to both the phase-field length scale and the traction order parameter is also sufficiently verified.

\end{abstract}

\begin{keyword}
% keywords here, in the form: keyword \sep keyword

% PACS codes here, in the form: \PACS code \sep code

% MSC codes here, in the form: \MSC code \sep code
% or \MSC[2008] code \sep code (2000 is the default)

Phase-field model; cohesive fracture; cohesive zone model; energy dissipation; quasi-brittle failure; softening.

\end{keyword}

\end{frontmatter}

%\begin{center} \textbf{
%Discussion: Energy dissipation in phase-field modeling of cohesive fracture \\
%Comments on the article ``Phase-field modelling of cohesive fracture'' \\
%by Lin Chen and Ren\'{e} de Borst [European Journal of Mechanics / A Solids 90 (2021) 104343]} \\ \vspace{5mm}
%
%Jian-Ying Wu \\
%State Key Laboratory of Subtropical Building Science, South China University of Technology, \\ 510641 Guangzhou, China. Email: jywu@scut.edu.cn
%\end{center}
\section{Introduction}

Fracture is one of the most frequently encountered failure modes in solids and structures. Ever since the birth of fracture mechanics in 1921 \citep{Griffith1921}, it has evolved into an important branch of solid mechanics, and still remains an open research area even after one century. 

Among many historic contributions \citep{Irwin1957,Eshelby1957,Barenblatt1959,Rice1968,HMP1976,XN1994,BP1997,MDB1999,Silling2000} during the last century, the variational phase-field approach to fracture, set forth by \cite{FM1998,BFM2000} in the mathematics community, attracts extensive attentions in the mechanics and engineering communities due to the promotion of the late Professor C. Miehe \citep{MWH2010a,MHW2010c}. In the phase-field model (PFM) for fracture, the sharp crack is regularized into a localized crack band with the bandwidth characterized by a small length scale parameter. A spatially continuous field variable, say, the crack phase-field, and its spatial gradient, are introduced to characterize the cracking state. Usually, the crack phase-field takes a value of zero for the intact state and of unity for the completely broken one, or vice versa, with the intermediate value representing the partially fractured one. The sharp crack surface area---a surface integral---is then represented by a volume integral in terms of the crack phase-field and its spatial gradient, such that the energy dissipation during fracture can be quantified. Once the strain energy density of the cracking solid is also properly defined, either the variational principle \citep{BFM2008} or the irreversible thermodynamics \citep{MWH2010a} is adopted to build a set of coupled partial differential equations (PDEs) governing the displacement field and the crack phase-field. Complex fracture phenomena, e.g., crack nucleation, propagation, branching and merging, etc., can be captured in a standalone framework \citep{AGL2015,WNNSBS2018}. Note that the gradient damage model (GDM) \citep{FN1996,LA1999} shares some similarities to the PFM and the in-between difference becomes diminishing provided the phase-field regularization of sharp cracks is also incorporated \citep{PAMM2011,Wu2018}. 

%The gradient damage model (GDM) \citep{FN1996,LA1999} shares some similarities to the phase-field model (PFM). Aiming to suppress the mesh size and alignment dependence inherent to a local damage model \citep{MPC1989}, the GDM also introduces the spatial gradient of the damage variable in the context of nonlocal continuum mechanics \citep{Polizzotto2003}. Either the principle of virtual work or the thermodynamics for generalized standard continua \citep{GFA2010} can be employed to derive a similar PDE based evolution law for the damage variable. However, as the GDM intrinsically lacks the material property of fracture energy and the regularization of sharp cracks, the energy functional of damaging solids is generally postulated in an \textit{ad hoc} manner. By learning from the phase-field regularization of sharp cracks \citep{PAMM2011,Wu2018}, the difference between GDM and PFM becomes subtle and is almost diminishing. %In this work they are not discriminated.

Based on the \cite{AT1990} elliptic regularization of the \cite{MS1989} functional in image segmentation, almost all the existing PFMs incorporate two characteristic functions \textit{exclusively} in terms of the crack phase-field into the formulation. One is the \textit{geometric function} in phase-field regularization of the sharp crack, and the other is the \textit{degradation function} in definition of the strain energy density and constitutive relation within the regularized crack band. The earlier variational phase-field models, e.g., the \texttt{AT2} \citep{BFM2000,MWH2010a} and \texttt{AT1} \citep{PAMM2011,MBK2015}, apply only to brittle fracture and converge to the \cite{Griffith1921} linear elastic fracture mechanics (LEFM) for a vanishing length scale. Despite their wide acceptance, such PFMs are inapplicable to cohesive fracture with non-negligible softening behavior due to mild release of the strain energy and progressive energy dissipation during failure. Moreover, in order to model crack nucleation the phase-field length scale has to be treated as a material property \citep{TLBMM2018}. As far as cohesive fracture in quasi-brittle solids is concerned, the length scale parameter so determined is usually not small enough, leading to severe over-estimation of the load capacity and energy dissipation \citep{MBK2015,MNH2019}. 

Being a ground-breaking work over the \cite{Griffith1921} LEFM, the \cite{Barenblatt1959} cohesive zone model (CZM) introduces a finite failure strength to deal with crack nucleation and a fracture energy dependent traction--separation softening law to describe crack propagation. Therefore, the PFM that converges upon a vanishing length scale to the CZM is undoubtedly more favored. Along this line, \citet{VdeB2013} proposed a phase-field model for interface fracture; see also \citet{CdeB2022}. An auxiliary field, together with the displacement field and the crack phase-field, is introduced to approximate the crack opening needed in the CZM. However, the crack path has to be known \textit{a priori}, making this model restrictive in the modeling of arbitrary crack propagation. Based on the \texttt{AT2} model for brittle fracture, \cite{CFI2016,FI2017,LCM2023} proposed a PFM for cohesive fracture, assuming that the degradation function not only depends on the crack phase-field but also are proportional to the length scale parameter. The $\varGamma$-convergence \citep{Braides1998} to the \cite{Barenblatt1959} CZM was proved in the 1D case. Similarly, \cite{LG2011,Lorentz2017} modified the \texttt{AT1} model with a length scale dependent degradation function and developed a GDM/PFM for cohesive fracture; see also \cite{GLHTD2019,FC2020}. Though these two PFMs both converges to the CZM upon a vanishing length scale, only rather limited traction--separation laws can be captured and general softening behavior cannot be considered.

%An ideal PFM for cohesive fracture should also incorporate the failure strength and fracture energy as two independent, physically well-defined quantities into the formulation. 

%Combination of the principle of energy conservation, the stability criterion and the crack irreversibility gives the governing equations. 

Almost at the same time, \cite{Wu2017} proposed a unified phase-field theory for damage and fracture. Two parameterized characteristic functions, i.e., a quadratic polynomial for the geometric function and a rational fraction for the degradation function, respectively, are postulated. The \texttt{AT1/2} models for brittle fracture can be recovered as its particular instances. More importantly, a phase-field cohesive zone model (\texttt{PF-CZM}) emerges naturally from this framework. Although the \texttt{PF-CZM} is originally devised for the modeling of cohesive fracture in quasi-brittle solids \citep{Wu2018}, the author soon realized that it applies also to brittle fracture \citep{WN2018}. With all the parameters involved in the characteristic functions calibrated from the standard material properties, those commonly adopted linear and convex (exponential, hyperbolic and \cite{CHR1986}) traction--separation softening laws for cohesive fracture can be reproduced or approximated with high precision. Basically, the \texttt{PF-CZM} has the following features: \textit{(i)} it converges to the CZM for a vanishing phase-field length scale,  \textit{(ii)} it incorporates the finite failure strength and fracture energy as two independent material properties, and \textit{(iii)} it is insensitive to both the phase-field length scale and the mesh discretization. Accordingly, the phase-field length scale is usually treated as a numerical parameter that can be taken as small as possible such that the $\varGamma$-convergence to the \cite{Barenblatt1959} CZM is preserved. Moreover, the strength-based criterion for crack nucleation, the energy-based criterion for crack propagation and the (quasi-)variational principle-based crack path chooser are all incorporated into the formulation \citep{WHZN2021}. Extensive numerical simulations demonstrate that the \texttt{PF-CZM} is able to capture crack nucleation, propagation and the resulting size effect in quasi-brittle solids \citep{Wu2018b,FW2018} and in brittle ones \citep{WHNM2022}, in both 2D and 3D cases \citep{WHZN2021}. Remarkably, except the crack bandwidth, all the other numerical predictions (crack patterns, load--displacement curves, etc.) are insensitive to the length scale parameter, as long as the latter is small enough and well resolved by the spatial discretization; see \cite{WNNSBS2018} for a review.

%These merits make it capable of capturing simultaneously crack nucleation in the absence of stress singularity and propagation of large cracks in both brittle and quasi-brittle solids, almost regardless of the incorporated length scale. Due to these merits, the \texttt{PF-CZM} has been successfully applied to the modeling of mechanical fracture in solids \citep{FW2018,WQNZT2018,MNW2019,MBBPB2020,LPB2020,WHZN2021,WHNM2022,WY2022}.Moreover, owing to its intrinsically inter-field coupled nature, the extension to fracture in more complex multi-physical problems \citep{WMN2020,MNW2021,MNWNdeV2021,WC2021a,WC2022,CW2022,WCZ2022} is straightforward. 

The \texttt{PF-CZM} still needs to be improved. Firstly, despite the generic and arguably simplest forms, the involved characteristic functions are postulated in a heuristic manner with some mathematical and physical conditions fulfilled \textit{a priori}. Secondly, even with an optimal geometric function, the \texttt{PF-CZM} cannot deal with concave traction--separation laws, like the \cite{PPR2009} softening for adhesives, since otherwise the resulting crack band would shrink \citep{GLML2023}. Such a shrinking crack band would lead to either violation of the crack irreversibility or discrepancies between the target and predicted softening laws \citep{Wu2022}. Recently, within the framework of the unified phase-field theory, \cite{FFL2021} proposed a particular version of the \texttt{PF-CZM}. A similar rational fraction relationship between the characteristic functions and a linear parameterized cohesive traction were assumed. Though such assumptions are still non-physically motivated, the geometric function can now be analytically solved from Abel's integral equation. However, such a geometric function is usually not optimal such that a non-shrinking crack band cannot be guaranteed even for those convex softening curves, e.g., the exponent one, the \cite{CHR1986} one, etc. Upon the same assumptions, \cite{XR2024} proposed an extended GDM with the introduction of an extra characteristic function. Compared to the \texttt{PF-CZM}, the governing equation for the crack phase-field (damage variable therein) was postulated with the aim to decoupling the crack profile and the traction--separation softening law. The relation between the degradation function and the geometric function, rather than the latter itself, was analytically determined. Nevertheless, the optimal geometric function that intrinsically guarantees a non-shrinking crack band is still missing. Moreover, the physical interpretation of the extra characteristic function and the relationship between this extended GDM and the unified phase-field theory are not known.

%Ever since the great success of the variational phase-field models for brittle fracture, research efforts were exerted to account for cohesive one. Based on the energy functional \eqref{eq:energy-functional-cohesive}, 

In order to address the above issues, in this work a generalized phase-field cohesive zone model (\texttt{$\mu$PF-CZM}) is proposed within the framework of the unified phase-field theory for brittle and cohesive fracture. With the merits of the original \texttt{PF-CZM} and relevant works preserved as much as possible, it greatly extends the theoretical and application scopes in several crucial aspects: \textit{(i)} an extra dissipation function, not necessary coincident with the degradation function, is introduced to characterize crack driving force; \textit{(ii)} the involved characteristic functions are analytically determined from arguably the simplest conditions for a non-shrinking crack band and length scale insensitivity; \textit{(iii)} the cohesive traction of any order can be considered without affecting the target traction--separation softening law; \textit{(iv)} almost any softening curve, whatever it is convex, linear and concave, can be dealt with in a optimal, universal and rationalized manner. Despite the loss of variational consistency in general cases, the resulting \texttt{$\mu$PF-CZM} is still thermodynamically consistent, and is insensitive to the incorporated length scale and traction order parameters. Moreover, the existing numerical implementation of the original \texttt{PF-CZM} can be adopted straightforwardly with minor modifications. Representative numerical examples are presented to validate the proposed \texttt{$\mu$PF-CZM} and to demonstrate its capabilities in capturing brittle and cohesive fracture with general softening behavior.

%\begin{itemize}
%\item The unified phase-field theory for fracture is extended to the variationally inconsistent but thermodynamically consistent scenarios with an extra dissipation function distinct from the degradation function.
%
%\item In the variationally inconsistent formulation the crack bandwidth is independent of the degradation function, leading to more flexibility in construction of the involved geometric and dissipation functions.
%
%\item In the \texttt{$\mu$PF-CZM} the geometric and dissipation functions are constructed \textit{a priori} upon the crack irreversibility and length scale insensitive conditions whereas the degradation function is solved analytically.
%
%
%\item The \texttt{$\mu$PF-CZM} preserves the length scale insensitivity of the original \texttt{PF-CZM} and applies to cohesive fracture with concave softening laws without violating the crack irreversibility condition.
%
%\item The lowest-order \texttt{$\mu$PF-CZM} is proposed with the arguably simplest dissipation function, an optimal geometric function and the degradation function analytically solved from a specific traction--separation law.
%\end{itemize}

The remainder of this work is structured as follows. \Cref{sec:unified-pfm} presents extension of the original unified phase-field theory for fracture to variationally inconsistent but still thermodynamically consistent scenarios. In addition to the crack geometric function and the stress degradation function, an extra energy dissipation function is introduced to characterize evolution of the crack phase-field. In  \cref{sec:analytical-solution} the extended unified phase-field theory is applied to cohesive fracture in a softening bar, with the traction--separation law and crack bandwidth semi-analytically derived. The necessity for a non-shrinking crack band and the conditions for length scale insensitive traction--separation laws are discussed. \Cref{sec:generalized-pfczm} addresses the generalized phase-field cohesive zone model (\texttt{$\mu$PF-CZM}) for fracture, with the geometric function, degradation function and dissipation function determined analytically from the conditions for a non-shrinking crack band and length scale insensitivity. \Cref{sec:numerical-implementation} is devoted to the numerical implementation of the proposed \texttt{$\mu$PF-CZM}. The existing multi-field finite element implementation and alternate minimization algorithm are adopted with minor modifications. \cref{sec:numerical-examples} presents representative numerical examples to validate the proposed \texttt{$\mu$PF-CZM} and to demonstrate its capabilities in capture brittle and cohesive fracture in solids with convex, linear and concave softening behavior. The insensitivity to the length scale and traction order of the \texttt{$\mu$PF-CZM} is verified in all examples. The most relevant conclusions are drawn in \cref{sec:conclusions}. For the sake of completeness, four appendices are attached for the reference to the solution to Abel's equation, the softening curves commonly adopted for cohesive fracture, the polynomial fitting of the \cite{CHR1986} softening and \cite{PPR2009} softening, and the characteristic functions involved in the proposed \texttt{$\mu$PF-CZM}, respectively.

\section{The unified phase-field theory for fracture}
\label{sec:unified-pfm}

In this section, the unified phase-field theory for brittle and cohesive fracture proposed in \cite{Wu2017} is extended to more general scenarios such that the variationally inconsistent formulation can be incorporated as well.

Let $\varOmega \subset \mathbb{R}^{n_{\dim}}$ ($n_{\dim} = 1, 2, 3$) be the reference configuration of a cracking solid. The external boundary is denoted by $\partial \varOmega \subset \mathbb{R}^{n_{\dim} - 1}$, with $\boldsymbol{n}^{\ast}$ being the outward normal vector. Deformations of the solid, resulting from the body forces (per unit volume) $\boldsymbol{b}^{\ast}$ distributed in the domain $\varOmega$ and the tractions $\boldsymbol{t}^{\ast}$ applied to the boundary $\partial \varOmega_{t} \subset \partial \varOmega$, are described by the displacement field $\boldsymbol{u} (\boldsymbol{x})$ and infinitesimal strain field $\bfepsilon (\boldsymbol{x}) := \nabla^{\ts} \boldsymbol{u} (\boldsymbol{x})$, for the symmetric gradient operator $\nabla^{\ts} (\cdot)$ with respect to the spatial coordinate $\boldsymbol{x}$. The complementary boundary $\partial \varOmega_{u} = \partial \varOmega \backslash \partial \varOmega_{t}$ is subjected to prescribed displacements $\boldsymbol{u}^{\ast}$. The set of sharp cracks, one dimension less than the domain $\varOmega$, is denoted by $\mathcal{S}$, with the normal vector signified by $\boldsymbol{n}_{_\mathcal{S}}$.

\subsection{Phase-field regularization of sharp cracks}

In phase-field models for fracture, the sharp crack $\mathcal{S}$ is regularized over a crack band $\mathcal{B} \subseteq \varOmega$ in which the crack phase-field $d (\boldsymbol{x})$ localizes, with the exterior domain $\varOmega \backslash \mathcal{B}$ being intact. In accordance with \cite{Braides1998}, the sharp crack surface area $A_{_\mathcal{S}}$ is approximated by the following volume integral
\begin{subequations}\label{eq:functional-phase-field}
\begin{align}
  A_{_\mathcal{S}}
    = \int_{\mathcal{S}} \; \td A
      \approx \int_{\mathcal{B}} \gamma (d; \nabla d) \; \td V
	  = A_{_d}
\end{align}
where the crack surface density function $\gamma (d; \nabla d)$ is expressed in terms of the crack phase-field $d$ and its spatial gradient $\nabla d$. 

The following generic crack surface density function is frequently considered \citep{Wu2017}
\begin{align}\label{eq:crack-density-function}
  \gamma (d; \nabla d)
    = \dfrac{1}{c_{\alpha}} \bigg[ \dfrac{1}{b} \alpha (d) 
    + b \big| \nabla d \big|^{2} \bigg] 
  \qquad \text{with} \qquad
  c_{\alpha}
    = 4 \int_{0}^{1} \sqrt{\alpha (\beta)} \; \td \beta
\end{align}
\end{subequations}
where the \textit{geometric function} $\alpha (d)$ characterizes homogeneous crack evolution; the phase-field length scale $b$ is a regularization parameter that controls the localized crack bandwidth; $c_{\alpha} > 0$ is a scaling parameter such that the sharp crack surface $A_{_\mathcal{S}}$ is recovered for a fully softened crack. %In \cref{eq:crack-density-function}, the first term penalizes deviations from the state $d = 0$, and the second one regularizes sharp variations of the crack phase-field through the gradient $\nabla d$. For such regularized models, the measure $A_{_d}$ converge weakly to the surface area $A$.

\subsection{Constitutive relation and degradation function}

As usual, the strain energy density function $\psi (\bfepsilon, d)$ in the cracking solid is defined as
\begin{align}\label{eq:strain-energy}
	\psi (\bfepsilon, d)
		= \omega(d) \psi_{0} (\bfepsilon), \qquad
%		= \omega(d) \dfrac{1}{2} \bfepsilon : \mathbb{E}_{0} : \bfepsilon
%		= \dfrac{1}{\omega (d)} \dfrac{1}{2} \bfsigma 
%		: \mathbb{E}_{0}^{-1} : \bfsigma
	\psi_{0} (\bfepsilon)
		= \dfrac{1}{2} \bfepsilon : \mathbb{E}_{0} : \bfepsilon
%		= \dfrac{1}{2} \bar{\bfsigma} : \mathbb{E}_{0}^{-1} : \bar{\bfsigma}
%		= \dfrac{1}{\omega^{2} (d)} \dfrac{1}{2} \bfsigma 
%		: \mathbb{E}_{0}^{-1} : \bfsigma
\end{align}
where $\mathbb{E}_{0}$ represents the fourth-order elasticity tensor of the material; the initial strain energy density function $\psi_{0} (\bfepsilon)$ is degenerated by the \textit{degradation function} $\omega(d)$ with the derivative $\omega' (d) := \partial \omega / \partial d$. The resulting Cauchy stress $\bfsigma$ is given by
\begin{align}\label{eq:stress-pfm-nosplit}
	\bfsigma
		= \dfrac{\partial \psi}{\partial \bfepsilon}
		= \omega (d) \mathbb{E}_{0} : \bfepsilon
		= \omega (d) \bar{\bfsigma}
%		= \mathbb{E}_{0} : \bfepsilon^{\text{e}}
\end{align}
for the (undamaged) effective stress $\bar{\bfsigma} := \mathbb{E}_{0} : \bfepsilon$. 

\begin{figure}[ht] \centering
  \includegraphics[width=0.55\textwidth]{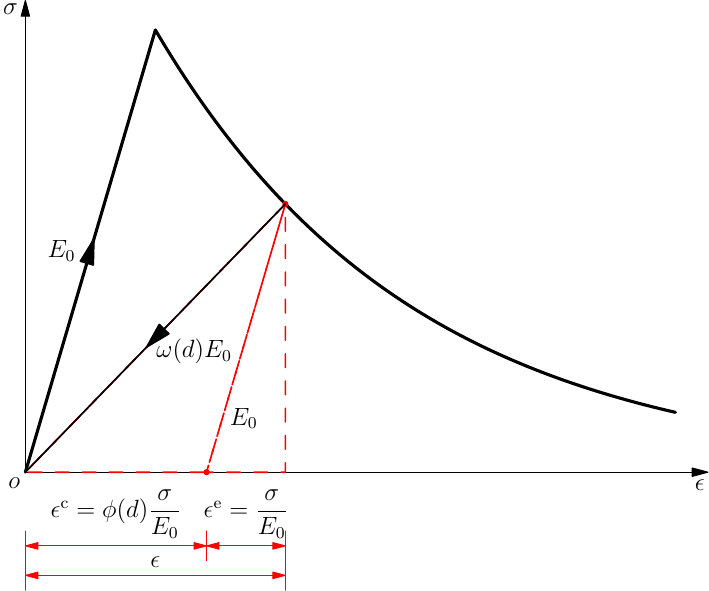}
  \caption{Decomposition of the total strain into the elastic and inelastic cracking components}
  \label{fig:strain-decompositions}  
\end{figure}

As shown in \cref{fig:strain-decompositions}, the strain field $\bfepsilon$ in the cracking solid allows an additive split, i.e.,
\begin{align}
	\bfepsilon
		= \bfepsilon^{\text{e}} + \bfepsilon^{\text{c}}, \qquad
	\bfepsilon^{\text{c}}
		= \bfepsilon - \bfepsilon^{\text{e}}
		= \bigg[ \dfrac{1}{\omega (d)} - 1 \bigg] 
			\mathbb{E}_{0}^{-1} : \bfsigma
		= \phi (d) \mathbb{E}_{0}^{-1} : \bfsigma
\end{align}
where $\bfepsilon^{\text{e}} := \mathbb{E}_{0}^{-1} : \bfsigma$ represents the elastic strain; the inelastic strain $\bfepsilon^{\text{c}}$ is characterized by the \textit{cracking function} $\phi (d)$
\begin{align}\label{eq:cracking-characteristic-function}
	\phi (d)
		= \frac{1}{\omega (d)} - 1
	\qquad \Longrightarrow \qquad
	\omega (d)
		= \dfrac{1}{1 + \phi (d)}, \qquad
	\omega' (d)
		=-\omega^{2} (d) \phi' (d)
\end{align}
with the derivative $\phi' (d) := \partial \phi / \partial d$. Note that in the original \texttt{PF-CZM} \citep{Wu2017,Wu2018,WN2018} and the relevant works \citep{FFL2021,XR2024}, the rational fraction function \eqref{eq:cracking-characteristic-function} were postulated \textit{a priori} in an \textit{ad hoc} manner. As will be shown later, the cracking function $\phi (d)$ plays the central role in determination of the crack opening (i.e., separation) needed for the CZM.

%\remark \label{rmk:AT-elliptic-regularization} The energy functional \eqref{eq:total-energy-functional-pfm} is  the \cite{AT1990} elliptic regularization of the \cite{Griffith1921} LEFM for sharp cracks
%\begin{align}\label{eq:energy-functional-LEFM}
%	\mathscr{E} (\boldsymbol{u}, \mathcal{S})
%		= \int_{\varOmega} \psi_{0} (\bfepsilon^{\text{e}}) \; \td V
%		+ G_{\text{f}} A_{_\mathcal{S}}
%		- \mathscr{F} (\boldsymbol{u}) 
%\end{align}
%The \texttt{AT2} \citep{BFM2000,BFM2008} and \texttt{AT1} \citep{PAMM2011,MBK2015} models are two popular instances of this family; . $\Box$

\remark \label{rmk:decomposition-strain-energy} In order to capture the tension/compression asymmetry of fracture, sometimes the elastic strain energy density $\psi_{0} (\bfepsilon)$ is decomposed into the so-called positive and negative parts $(\psi_{0}^+(\bfepsilon), \psi_{0}^-(\bfepsilon))$, and only the tensile part $\psi_{0}^+(\bfepsilon)$ is affected by the crack phase-field $d (\boldsymbol{x})$ through the degradation function $\omega (d)$, i.e., 
\begin{align}\label{eq:strain-energy-density-split}
	\psi (\bfepsilon, d)
		= \omega(d) \psi_{0}^{+} (\bfepsilon) 
		+ \psi_{0}^{-} (\bfepsilon)
	\qquad \Longrightarrow \qquad
	\bfsigma
		= \dfrac{\partial \psi}{\partial \bfepsilon}
		= \omega(d) \bar{\bfsigma}^{+}
    + \bar{\bfsigma}^{-}
\end{align}
for the positive and negative effective stress tensors $\bar{\bfsigma}^{\pm} := \partial \psi_{0}^{\pm} / \partial \bfepsilon$, respectively. In the literature various decomposition schemes have been considered; see \cite{WNNSBS2018} for the discussion. $\Box$

\subsection{Governing equations}

For quasi-static fracture of solids under the infinitesimal strain regime, the mechanical equilibrium equation reads
\begin{align}\label{eq:equilibrium-governing-equations-pfm}
\begin{cases}
	\nabla \cdot \bfsigma
		+ \boldsymbol{b}^{\ast}
		= \boldsymbol{\mathit{0}} & \qquad \text{in} \; \varOmega \\
%	\bfsigma \cdot \boldsymbol{n}_{_\mathcal{S}}
%		= \boldsymbol{\mathit{0}} & \qquad \text{on} \; \mathcal{S} \\
	\bfsigma \cdot \boldsymbol{n}^{\ast}
		= \boldsymbol{t}^{\ast} & \qquad \text{on} \; \partial \varOmega_{t}
\end{cases}
\end{align}
for the Cauchy stress $\bfsigma$ given in \cref{eq:stress-pfm-nosplit} or \cref{eq:strain-energy-density-split}.

Regarding evolution of the crack phase-field, the Kuhn-Tucker loading/unloading conditions apply \citep{Wu2017}
\begin{subequations}\label{eq:damage-criterion}
\begin{align}
  \dot{d} \ge 0, \qquad
  g (Y, d) \le 0, \qquad
  \dot{d} g (Y, d) \equiv 0 \qquad \text{in } \mathcal{B}
\end{align}
together with the Neumann boundary condition
\begin{align}
	\nabla d \cdot \boldsymbol{n}_{_\mathcal{B}} \ge 0 
			\qquad \text{on } \partial \mathcal{B}
\end{align}
\end{subequations}
where $\boldsymbol{n}_{_\mathcal{B}}$ is the outward unit normal vector of the boundary $\partial \mathcal{B}$. Note that due to the crack irreversibility $\dot{d} (\boldsymbol{x}) \ge 0$, the phase-field evolution law \eqref{eq:damage-criterion} is described by inequalities rather than equalities. The failure function $g (Y, d)$ is expressed as
\begin{align}
  g (Y, d) :
	 = Y - G_{\text{f}} \; \delta_{d} \gamma 
%		 \begin{cases}
%			 = 0 & \qquad \dot{d} > 0 \\
%			 < 0 & \qquad \dot{d} = 0 \\
%		 \end{cases}
	\qquad \text{with} \qquad
  \delta_{d} \gamma :
    = \partial_{d} \gamma - \nabla \cdot \big( \partial_{\nabla d} 
      \gamma \big)
    = \dfrac{1}{c_{\alpha}} \bigg[ \dfrac{1}{b} \alpha' (d) 
    - 2 b \Delta d \bigg]	
\end{align}
where the \cite{Griffith1921} fracture energy $G_{\text{f}}$ represents the energetic cost of complete fracture; $\delta_{d} \gamma$ is the variational derivative of the crack surface density \eqref{eq:crack-density-function} for the Laplacian $\Delta d = \nabla \cdot \nabla d$ and the derivative $\alpha' (d) := \partial \alpha / \partial d$; the energy release rate $Y$  is conjugate to the crack phase-field $d$, i.e.,
\begin{align}\label{eq:crack-driving-force}
	Y :
    =-\dfrac{\partial \psi}{\partial d}
    =-\omega' (d) \bar{Y} \ge 0, \qquad
  \bar{Y} :
  	= \dfrac{\partial \psi}{\partial \omega}
  	= \psi_{0} (\bfepsilon) \; \text{or} \; \psi_{0}^{+} (\bfepsilon)
\end{align}
for the effective counterpart $\bar{Y}$, respectively; see  \cref{rmk:effective-crack-driving-force-hybrid}.

The crack driving force \eqref{eq:crack-driving-force} is both variationally and thermodynamically consistent. Usually it is more flexible to use only the thermodynamically consistent formulation as in \cite{Wu2017,Wu2018,Wu2018b}
\begin{align}\label{eq:crack-driving-force-generalized}
	Y 
		=-\varpi' (d) \bar{Y} \ge 0 
	\qquad \Longrightarrow \qquad
	\dot{\mathscr{D}}
		= \int_{\mathcal{B}} Y \dot{d} \ge 0
\end{align}
where the \textit{dissipation function} $\varpi (d)$ that characterizes the energy dissipation rate $\dot{\mathscr{D}}$ is not necessarily coincident with the \textit{degradation function} $\omega (d)$ for the constitutive relation $\bfsigma (\bfepsilon, d)$. Note that it is the derivative $\varpi' (d)$, rather than the function $\varpi (d)$ itself, that is necessary in evaluation of the crack driving force $Y$. For the sake of consistency with the counterparts \eqref{eq:cracking-characteristic-function} and \eqref{eq:crack-driving-force}, the following expression is considered
\begin{align}\label{eq:energetic-degradation-function-derivative}
	\varpi' (d)
		=-\omega^{2} (d) \mu' (d) %\le 0 \qquad \text{with} \qquad
%	\mu' (d) \ge 0
\end{align}
in terms of an auxiliary function $\mu (d)$ with the derivative $\mu' (d) := \partial \mu / \partial d$. The variationally consistent formulation is recovered for the particular case with $\varpi' (d) = \omega' (d)$, or, equivalently, $\mu (d) = \phi (d)$.

\remark In the variational phase-field approach to fracture, the displacement field $\boldsymbol{u}$ and the crack phase-field $d$ are minimizers \citep{BFM2008}
\begin{align}\label{eq:minimization-problem}
	(\boldsymbol{u}; d)
	  = \text{Arg} \Big\{ \min_{ \hat{\boldsymbol{u}}, \; \hat{d} }
	  	\mathscr{E} (\hat{\boldsymbol{u}}; \hat{d}) \Big\}
	\qquad \text{with} \qquad
  \mathscr{E} (\boldsymbol{u}, d) 
    = \int_{\varOmega} \psi (\bfepsilon (\boldsymbol{u}), d) \; \td V
    + \int_{\mathcal{B}} G_\text{f} \; \gamma (d; \nabla d) \; \td V 
    - \mathscr{F} (\boldsymbol{u})
\end{align}
for the potential energy $\mathscr{F} (\boldsymbol{u})$ of external forces. For the variationally inconsistent formulation, the minimization problem \eqref{eq:minimization-problem} is replaced by the following modified one \citep{Wu2018b}
\begin{align}
\begin{cases}
	\boldsymbol{u} \displaystyle 
		= \text{Arg} \Big\{ \min_{\hat{\boldsymbol{u}}}
			\mathscr{E} (\hat{\boldsymbol{u}}, \hat{d}) \Big\} \vspace{2mm} \\
	d \displaystyle 
		= \text{Arg} \Big\{ \min_{\hat{d}}
			\bar{\mathscr{E}} (\hat{\boldsymbol{u}}, \hat{d}) \Big\}
\end{cases}
\end{align}
where the modified strain energy density functional $\bar{\mathscr{E}} (\boldsymbol{u}, d)$ is defined by
\begin{align}
  \bar{\mathscr{E}} (\boldsymbol{u}, d) 
    = \int_{\varOmega} \bar{\psi} (\bfepsilon (\boldsymbol{u}), d) \; \td V
    + \int_{\mathcal{B}} G_\text{f} \; \gamma (d; \nabla d) \; \td V 
    - \mathscr{F} (\boldsymbol{u})
\end{align}
with
\begin{align}
  \bar{\psi} (\bfepsilon, d)
		= \varpi(d) \psi_{0} (\bfepsilon)
  \qquad \text{or} \qquad
  \bar{\psi} (\bfepsilon, d)
		= \varpi(d) \psi_{0}^{+} (\bfepsilon) + \psi_{0}^{-} (\bfepsilon)
\end{align}
That is, in the variationally inconsistent formulation the Cauchy stress $\bfsigma$ and the crack driving force $Y$ are associated with distinct energy potentials. Note that the proposed formulation resembles the non-associated plasticity with distinct yield and potential functions. Though the variational principle of energy minimization \citep{FM1998,BFM2000}, is abandoned, it is still thermodynamically consistent. Moreover, it is also justified by the ``local variational principle for fracture'' recently advocated by Larsen and coworkers \citep{Larsen2024,LDLP2024}. $\Box$

\remark \label{rmk:effective-crack-driving-force-hybrid} Besides the decomposition of the strain energy density mentioned in \cref{rmk:decomposition-strain-energy}, another strategy capturing the tension/compression asymmetry of fracture is to use a variationally inconsistent effective crack driving force $\bar{Y}$, resulting in the so-called hybrid formulation \citep{AGL2015}. For tension-dominant brittle and quasi-brittle materials, the effective crack driving force $\bar{Y}$ can be modified as \citep{Wu2017,Wu2018,WN2018}
\begin{align}\label{eq:equivalent-effective-stress}
	\bar{Y} :
		= \dfrac{1}{2 E_{0}} \bar{\sigma}_{\text{eq}}^{2}, \qquad
	\bar{\sigma}_{\text{eq}} (\bar{\bfsigma})
		= \langle \bar{\sigma}_{1} \rangle
\end{align}
where $E_{0}$ is Young's modulus of the material. Here, the equivalent effective stress $\bar{\sigma}_{\text{eq}}$ is defined as the major principal value $\bar{\sigma}_{1}$ of the effective stress tensor $\bar{\bfsigma}$, resembling the Rankine criterion. Other expressions for the equivalent effective stress $\bar{\sigma}_{\text{eq}}$, e.g., the modified von Mises criterion \citep{WHN2020}, can be adopted for mixed-mode failure. Though in the 2D and 3D cases the variational consistency of the resulting phase-field model is altered by this redefinition, it is maintained in the 1D case so long as $\varpi' (d) = \omega' (d)$ is adopted. $\Box$

\subsection{Characteristic functions}

The above unified phase-field theory for fracture is particularized by specifying the involved characteristic functions, i.e., the \textit{geometric function} $\alpha (d)$, the \textit{degradation function} $\omega (d)$ (or the cracking function $\phi (d)$), and the \textit{dissipation function} $\varpi (d)$ (or, equivalently, $\mu (d)$). For brittle fracture there are many possibilities in choosing them \citep{WNNSBS2018} and only cohesive fracture is considered in this work. 

\subsubsection{Geometric function}

As in the crack surface density function \eqref{eq:crack-density-function} the length scale parameter $b$ has already been incorporated, it is sufficient that the geometric function $\alpha (d) : [0, 1] \to [0, +\infty)$ depends \textit{exclusively} on the crack phase-field $d (\boldsymbol{x})$. 

\begin{figure}[h!] \centering
  \subfigure[$\alpha (d) = 2d - d^{2}$]{
  \includegraphics[width=0.31\textwidth]{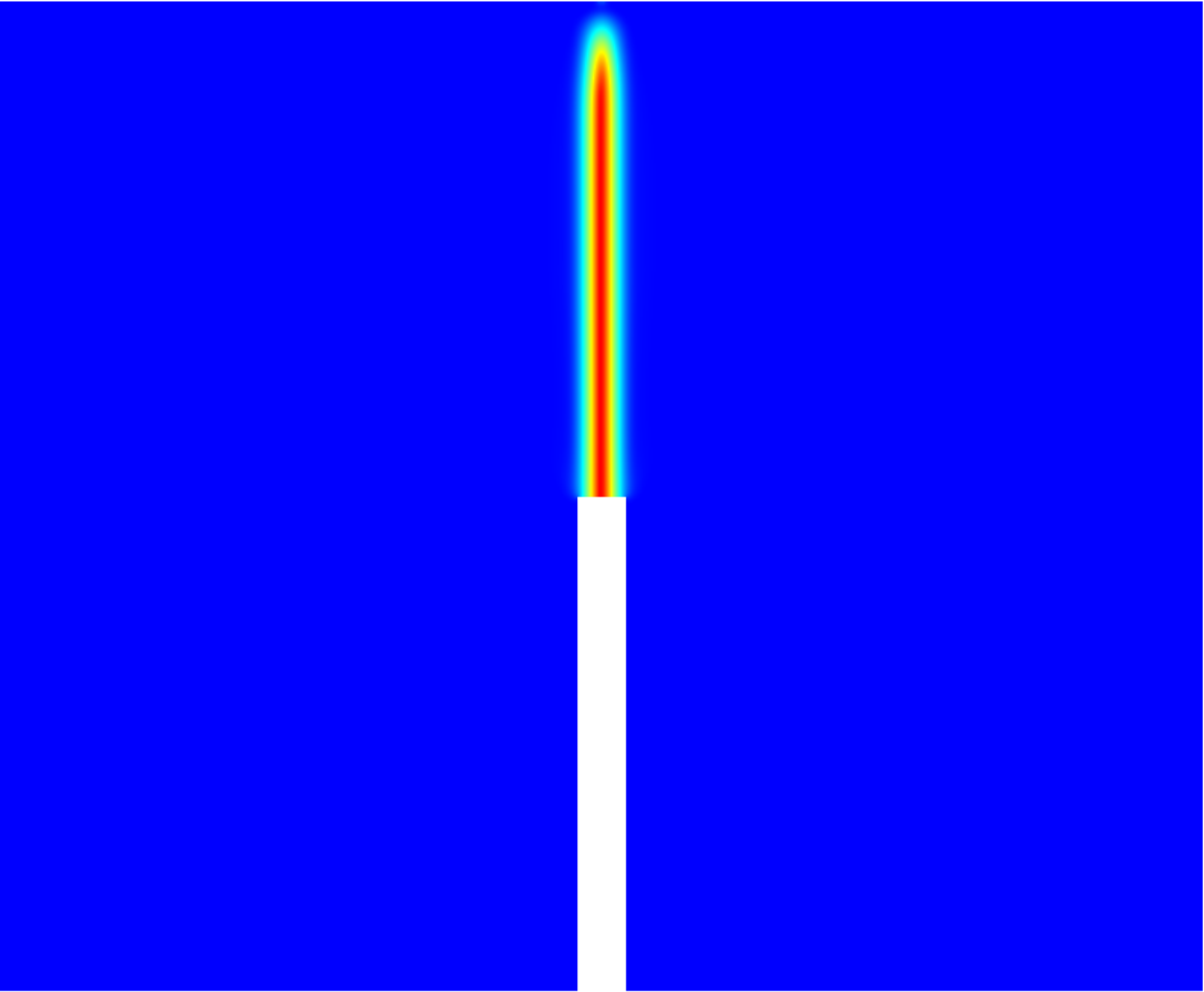}
  \label{fig:crack-bullet}} \hfill
  \subfigure[$\alpha (d) = d \big(1 - d \big)$]{
  \includegraphics[width=0.31\textwidth]{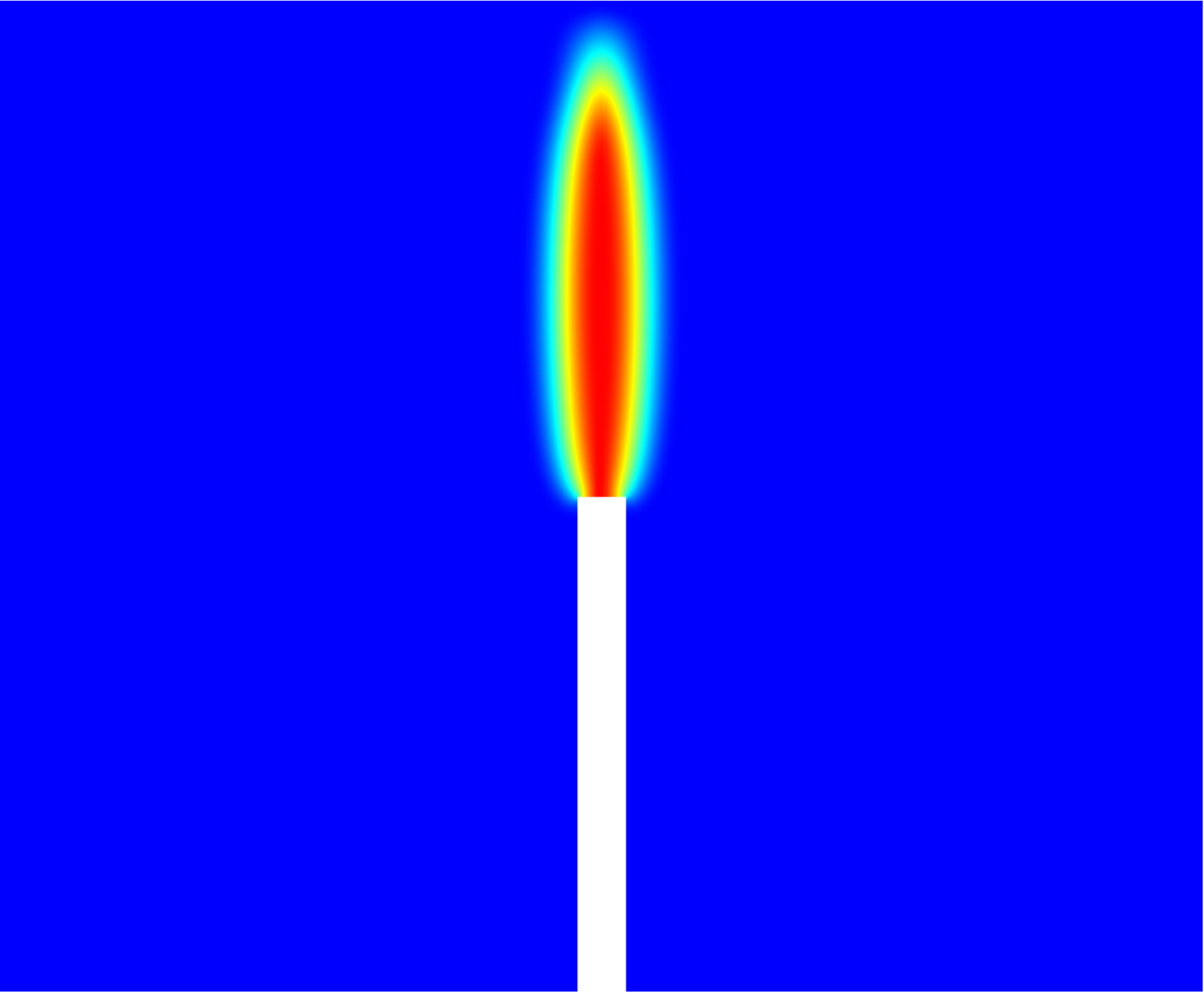}
  \label{fig:crack-spindle2}} \hfill
  \subfigure[$\alpha (d) = d^{2} \big(1 - d \big)^{2}$]{
  \includegraphics[width=0.31\textwidth]{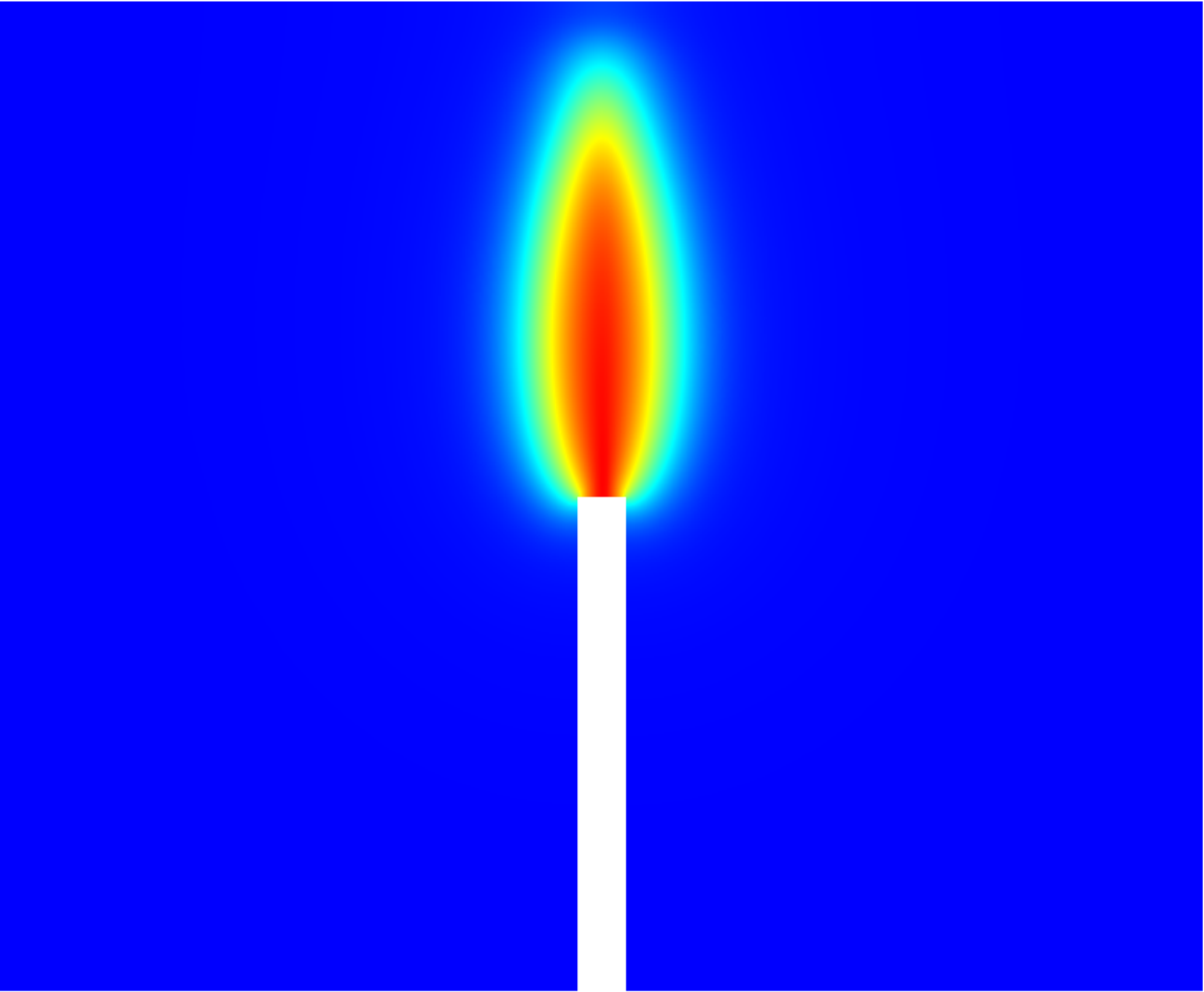}
  \label{fig:crack-spindle}}
  \caption{Crack profiles for various geometric functions.}
  \label{fig:crack-profiles}  
\end{figure}

In phase-field models for fracture, it is generally assumed that the geometric function $\alpha (d)$ fulfills the conditions
\begin{align}\label{eq:geometric-function-conditions}
	\alpha (d) \ge 0, \qquad
	\alpha (0) 
		= 0, \qquad
	\alpha (1)
		= 1, \qquad 
	\alpha' (d) :
		= \partial \alpha / \partial d \ge 0
\end{align}
The first two conditions are indispensable for $\varGamma$-convergence of the phase-field regularization \eqref{eq:functional-phase-field}, while the last two guarantee that $\alpha (d) \in [0, 1]$ holds as well, resulting in the expected \textit{bullet}-shaped crack profile (i.e., cylindrical body with a sharp tip) shown in \cref{fig:crack-bullet}.

%In order to achieve the bullet-shaped crack profile, an extra condition $\alpha (1) > 0$ needs to be imposed. In particular, the natural condition $\alpha (1) = 1$ is generally considered, i.e.,
%\begin{align}
%	\alpha (d) \ge 0, \qquad
%	\alpha (0) 
%		= 0, \qquad
%	\alpha (1)
%		= 1
%\end{align}
%However, the boundedness of the crack phase-field $d (\boldsymbol{x}) \in [0, 1]$ can only be automatically guaranteed in the \texttt{AT2} model, whereas special strategies have to be introduced for other choices.

\remark \label{rmk:geometric-function} Note that $\alpha (1) = 1$ is not an indispensable condition for $\varGamma$-convergence of the phase-field regularization of sharp cracks \citep{Braides1998}. For instance, in the physics community the well-known double-well function
\begin{align}\label{eq:geometric-crack-function-KKL}
	\alpha (d)
		= d^{2} \big(1 - d \big)^{2} \qquad \text{with} \qquad
	\alpha (1)
		= 0
\end{align}
has been adopted for brittle fracture \citep{KKL2001}. As shown in \cref{fig:crack-spindle2} and \cref{fig:crack-spindle}, for cohesive fracture with non-negligible softening behavior such a geometric function yields \textit{spindle}-shaped crack profiles. $\Box$ 

\subsubsection{Degradation and dissipation functions}

The degradation and dissipation functions play important roles in phase-field models for cohesive fracture --- they determine how the stress and crack driving force respond to evolution of the crack phase-field. 

The degradation function $\omega (d) : [0, 1] \rightarrow [1, 0]$ has to satisfy the following conditions \citep{MWH2010a,Wu2017}
\begin{align}
	\omega (0) 
		= 1, \qquad
	\omega (1) 
		= 0; \qquad
	\omega' (d) \le 0
\end{align}
or, equivalently,
\begin{align}\label{eq:conditions-cracking-function}
	\phi (0)
		= 0, \qquad
	\phi (1)
		=+\infty, \qquad
%	\phi (d)
%		= \dfrac{1}{b} \dfrac{Q (d)}{\big(1 - d \big)^{p}}, \qquad
	\phi' (d) \ge 0
\end{align}
for the cracking function $\phi (d)$ introduced in \cref{eq:cracking-characteristic-function}. 

Similarly, the auxiliary function $\mu (d)$ and the resulting dissipation function $\varpi (d)$ need to fulfill the following conditions
\begin{align}\label{eq:conditions-degradation-functions}
	\mu (0)
		= 0, \qquad
	\mu (1)
		=+\infty, \qquad
	\mu' (d) 
		= O (\omega^{2} (d)) \ge 0 \qquad \Longrightarrow \qquad
	\varpi' (1)
		= 0
\end{align}
Owing to the last condition $\varpi' (1) = 0$, the crack driving force \eqref{eq:crack-driving-force-generalized} vanishes for a fully softened crack (i.e., $d = 1$) and the resulting crack band does not become widened as usually observed in the classical nonlocal/gradient-enhanced damage models \citep{deBV2016}.

\remark For brittle fracture with negligible softening behavior, it is sufficient that the degradation function $\omega (d)$ and the dissipation function $\varpi (d)$ depend only on the crack phase-field $d (\boldsymbol{x})$. For instance, the quadratic polynomial $\omega (d) = \varpi (d) = \big( 1 - d \big)^{2}$ is adopted in the well-known \texttt{AT1/2} models for brittle fracture. Comparatively, for cohesive fracture it is crucial to characterize the softening behavior due to mild release of the strain energy and the progressive energy dissipation during failure. It has been shown that, in order for a phase-field model based on the energy functional \eqref{eq:minimization-problem} to be applicable to cohesive fracture, proper incorporation of the length scale $b$ into the degradation functions $\omega (d)$ and the dissipation function $\varpi (d)$ might be the indispensable. For instance, the Conti-Freddi-Iurlano-Lammen (\texttt{CFIL}) model \citep{CFI2016,FI2017,LCM2023} and the original phase-field cohesive zone model (\texttt{PF-CZM}) \citep{Wu2017,Wu2018,WN2018} both belong to this family, and so does the generalized \texttt{PF-CZM} proposed later in this work. $\Box$

\section{Application to a softening bar: 1-D analytical solution}
\label{sec:analytical-solution}

In this section the extended unified phase-field theory is applied to a softening bar under uniaxial stretching. The analytical results are used to determine the involved characteristic functions appropriate for cohesive fracture. 

\begin{figure}[h!] \centering
  \includegraphics[width=0.85\textwidth]{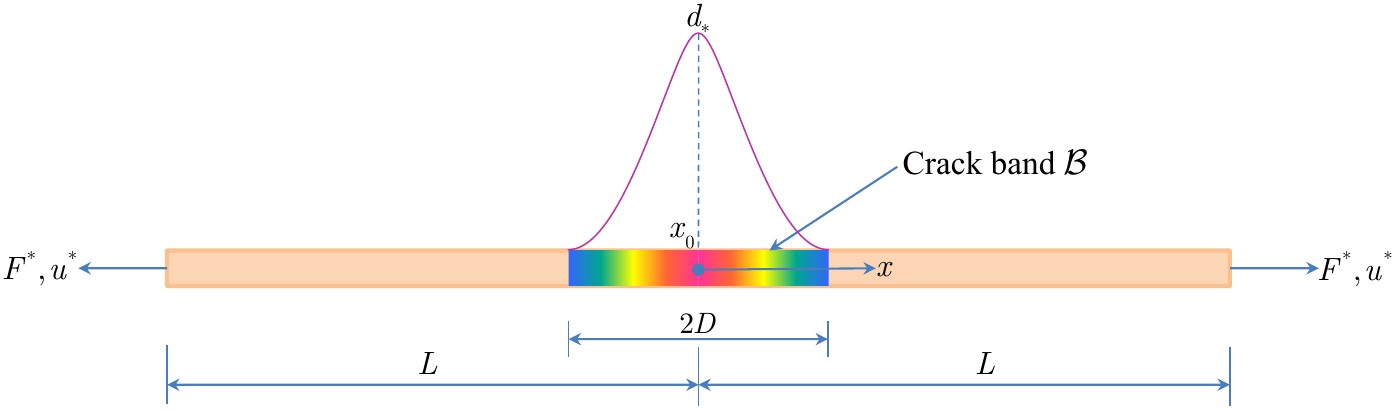}
  \caption{A softening bar under uniaxial stretching}
  \label{fig:uniaxial-stretching-bar}  
\end{figure}

As shown in \cref{fig:uniaxial-stretching-bar}, consider a bar $x \in [-L, L]$ sufficiently long such that crack evolution is not affected by the boundary effects. The bar is loaded at both ends by two increasing displacements along opposite directions. The localized solution is interested, in which the inelastic strain field and the crack phase-field localize into a narrow crack band. Assume that the crack is initiated at the centroid \footnote[1]{The precise location of the crack has no effect in this case so long as crack evolution is not polluted by the boundary effects.} $x_{0} = 0$ and the crack band is localized within the domain $\mathcal{B} := \Big\{ x \big| x \in [-D, D] \Big\}$, with $D \ll L$ being the half bandwidth not necessarily constant during failure.

%Comparatively, the strain field $\epsilon (x)$ is no longer uniform along the bar, due to the presence of the cracking strain $\epsilon^{\text{c}}$
%\begin{align}\label{eq:displacement-jump-apparent}
%	\epsilon^{\text{c}}
%%		= \epsilon - \epsilon^{\text{e}}
%%		= \dfrac{\sigma}{E_{0}} \bigg[ \dfrac{1}{\omega (d)} - 1 \bigg]
%		= \phi (d) \dfrac{\sigma}{E_{0}} 
%	\qquad \Longrightarrow \qquad
%  \itw
%  	= \int_{-L}^{L} \epsilon^{\text{c}} \; \td x
%    = \dfrac{2 \sigma}{E_{0}} \int_{0}^{D} \phi (d) \; \td x
%\end{align}
%for the \textit{apparent} displacement jump $\itw$ across the crack band. 

The distributed body forces are neglected. It then follows from the equilibrium equation $\td \sigma / \td x = 0$ that the stress field $\sigma (x)$ remains to be homogeneous along the bar even after crack initiation occurs. Provided the crack maintains in the loading state, i.e., $\dot{d} (x) > 0$ \textit{anywhere} within the crack band $\forall x \in \mathcal{B}$, the phase-field evolution law \eqref{eq:damage-criterion} becomes an identity \citep{Wu2017,WNNSBS2018}
\begin{align}\label{eq:governing-equations-1D}
	\dfrac{\sigma^{2}}{2 E_{0}} \mu' (d) 
		- \frac{G_{\text{f}}}{c_{\alpha} b} \Big[ \alpha' (d) 
		- 2 b^{2} \Delta d \Big]
		= 0
	\qquad \Longrightarrow \qquad
	\dfrac{\sigma^{2}}{2 E_{0}} \mu (d) 
		- \frac{G_{\text{f}}}{c_{\alpha} b} \Big[ \alpha (d) 
		- \big( b \nabla d \big)^{2} \Big]
	 	= 0
\end{align}
where uniformity of the stress field $\sigma (x)$ along the bar has been considered for arriving at \cref{eq:governing-equations-1D}$_{2}$. Note that the above governing equation is almost identical to that in the original \texttt{PF-CZM} --- the only difference is that the cracking function $\phi (d)$ is replaced by the auxiliary one $\mu (d)$ defining the dissipation function $\varpi (d)$.

\subsection{Traction--separation law and crack bandwidth}

The traction (stress in 1D) $\sigma$ and the \textit{apparent} displacement jump $\itw$ across the crack band $\mathcal{B}$ are evaluated as \citep{Wu2017,WNNSBS2018}
\begin{subequations}\label{eq:stress-opening-1D}
\begin{align}\label{eq:localised-stress-1D}
 	\sigma (d_{\ast})
 	 &= \sqrt{\frac{2 E_{0} G_{\text{f}}}{c_{\alpha}} \eta (d_{\ast})}
 	  = f_{\text{t}} \sqrt{\frac{1}{\eta_{0}} \eta (d_{\ast})}
 	\qquad \quad\;\;\; \qquad \text{with} \qquad    
	\eta (d)
		= \dfrac{1}{b} \dfrac{\alpha (d)}{\mu (d)} \\
 	\label{eq:displacement-jump-apparent}
  \itw (d_{\ast}) 
%	 &= \int_{-D}^{D} \epsilon^{\text{c}} \; \td x
%    = \dfrac{2 \itw_{\text{cL}}}{c_{\alpha}} \sqrt{\eta_{0} 
%    	\eta (d_{\ast})} \int_{0}^{d_{\ast}} b \phi (\vartheta)
%			\sqrt{\dfrac{\eta (\vartheta)}{\alpha (\vartheta)} }    	 
%    	\cdot \dfrac{1}{\sqrt{\eta (\vartheta) - \eta (d_{\ast})}} 
%    	\; \td \vartheta
	 &= \dfrac{2 \itw_{\text{cL}}}{c_{\alpha}} \sqrt{\eta_{0} 
		 	\eta (d_{\ast})} \int_{0}^{d_{\ast}} b \phi (\vartheta)
		 	\mathscr{H} (\vartheta; d_{\ast}) \; \td \vartheta
	\qquad \text{with} \qquad
  \mathscr{H} (d; d_{\ast})
%    = \sqrt{\alpha (d) - \dfrac{1}{A_{0}} \sigma^{2} d (d)} 
%    = \sqrt{ \alpha (d) - \dfrac{\alpha (d_{\ast})}{\phi (d_{\ast})} 
%    	\phi (d)}
		= \sqrt{\dfrac{\eta (d) / \alpha (d)}{\eta (d) - \eta (d_{\ast})}}	
\end{align}
\end{subequations}
where the maximum value $d_{\ast} := d (x = x_{0})$ of the crack phase-field $d (x)$ measures the loading level; upon crack initiation the critical stress (i.e., failure strength) $f_{\text{t}}$ is given by
\begin{align}\label{eq:critical-stress-general}
	f_{\text{t}}
		= \lim_{d_{\ast} \to 0} \sigma (d_{\ast})
% 		= \sqrt{A_{0} \dfrac{\alpha (d_{\ast})}{
% 			\phi (d_{\ast})}}
 		= \sqrt{\frac{2 E_{0} G_{\text{f}}}{c_{\alpha}} \eta_{0}}
% 	\qquad	
%	\eta (d)
%		= \dfrac{1}{b} \dfrac{\alpha (d)}{\phi (d)}
 	\qquad \text{with} \qquad
 	\eta_{0}
		= \lim_{d_{\ast} \to 0} \eta (d_{\ast})
\end{align}
and $\itw_{\text{cL}} := 2 G_{\text{f}} / f_{\text{t}}$ denotes the ultimate crack opening of the linear softening law \eqref{eq:linear-softening-cohesive}.

The (inverse) profile of the crack phase-field $x (d; d_{\ast})$ and the half width $D (d_{\ast})$ of the crack band $\mathcal{B}$ are determined as \citep{Wu2017,WNNSBS2018}
\begin{align}\label{eq:half-bandwidth-1D}
  x (d; d_{\ast})
    = b \int_{d}^{d_{\ast}} \mathscr{H} (\vartheta; d_{\ast}) 
    	\; \td \vartheta, \qquad %\text{with} \qquad
  D (d_{\ast}) :
  	= x (0; d_{\ast})
    = b \int_{0}^{d_{\ast}} \mathscr{H} (\vartheta; d_{\ast}) 
    	\; \td \vartheta
\end{align}
In particular, the initial and ultimate half bandwidths are given by
\begin{align}\label{eq:crack-bandwidths-general}
	D_{0} :
	  = D (d_{\ast} = 0)
		= \pi b \sqrt{\dfrac{2}{\eta_{0} b \mu'' (0) - \alpha'' (0)}}, 
	\qquad
	D_{\text{u}} :
	  = D (d_{\ast} = 1)
		= b \int_{0}^{1} \dfrac{1}{\sqrt{\alpha (\vartheta)}} \; \td \vartheta
\end{align}
As expected, the phase-field length scale parameter $b$ measures the crack band width.

As can be seen, all the global responses of the softening bar are determined uniquely by the characteristic functions $\alpha (d)$, $\mu (d)$ and $\phi (d)$. Specifically, the cracking function $\phi (d)$ only affects the crack opening $\itw (d_{\ast})$, while the crack bandwidth $D (d_{\ast})$ depends on the geometric function $\alpha (d)$ and the auxiliary function $\mu (d)$. Similarly to \cite{XR2024}, this decoupling leads to more flexibility in determination of the involved characteristic functions .

\remark \label{rmk:initial-slope-general} For the traction--separation law \eqref{eq:stress-opening-1D}, the initial slope $k_{0}$ of the softening curve $\sigma (\itw)$ is evaluated as
\begin{align}\label{eq:initial-slope-general}
	k_{0} :
		= \lim_{d{\ast} \to 0} \dfrac{\partial \sigma}{\partial \itw}
%		= \dfrac{\td \sigma}{\td d_{0}^{\ast}} \cdot 
%			\dfrac{E_{0}}{D} \bigg( 
%			\dfrac{\td \sigma^{\ast}}{\td d_{0}^{\ast}} \phi^{\ast} 
%		+ \dfrac{\td \phi^{\ast}}{\td d_{0}^{\ast}} \sigma^{\ast} 
%			\bigg)^{-1} \bigg|_{d_{0}^{\ast} = 0}		
%		= \dfrac{E_{0}}{D} \big( f_{\text{t}} \phi' (0) 
%			\big)^{-1} \dfrac{\td \sigma}{\td d_{0}^{\ast}} 
%			\bigg|_{d_{0}^{\ast} = 0}		
		= \dfrac{E_{0}}{D_{0} f_{\text{t}} \phi' (0)} 
			\lim_{d{\ast} \to 0} \dfrac{\partial \sigma}{\partial d_{\ast}} 
	  = \dfrac{E_{0}}{2 D_{0} \phi' (0)} \cdot \dfrac{1}{\eta_{0}}
	 		\lim_{d{\ast} \to 0} \eta' (d_{\ast})
%	  =-\dfrac{c_{\alpha} \pi^{2}}{4 \xi^{2}} 
%	  	\dfrac{f_{\text{t}}^{2}}{G_{\text{f}}} 
%	  	\bigg( \dfrac{b}{D_{0}} \bigg)^{3} 
\end{align}
Note that the initial slope is one of the most crucial characteristics for a specific softening curve and has non-negligible effects on the global responses (e.g., the peak load); see \cite{BP1997} for more discussion. $\Box$

\subsection{Length scale insensitive conditions}

\Cref{eq:stress-opening-1D} constitutes a parameterized traction--separation law of the \cite{Barenblatt1959} CZM, provided that the traction \eqref{eq:localised-stress-1D}, crack opening \eqref{eq:displacement-jump-apparent} and failure strength \eqref{eq:critical-stress-general} are all independent of the phase-field length scale $b$. 

As the geometric function $\alpha (d)$ depends exclusively on the crack phase-field $d (x)$, the following length scale \textit{insensitive} conditions have to be fulfilled
\begin{align}\label{eq:length-scale-insensitivity-condition}
\begin{cases}
	f_{\text{t}} \in (0, +\infty) & \qquad \Longleftrightarrow \qquad
	\eta_{0} 
		= \dfrac{c_{\alpha}}{2 l_{\text{ch}}} 
			\in (0, +\infty) \vspace{2mm} \\
	\mu (d) \propto \dfrac{1}{b} & \qquad \Longleftrightarrow \qquad
	\exists \; \bar{\mu} (d) 
		= b \mu (d) \quad \text{and} \quad
	\eta (d) :
		= \dfrac{1}{b} \dfrac{\alpha (d)}{\mu (d)}
		= \dfrac{\alpha (d)}{\bar{\mu} (d)} \vspace{2mm} \\
	\phi (d) \propto \dfrac{1}{b} & \qquad \Longleftrightarrow \qquad
	\exists \; \bar{\phi} (d) 
		= b \phi (d)
\end{cases}
\end{align}
where Irwin's internal length $l_{\text{ch}} := E_{0} G_{\text{f}} / f_{\text{t}}^{2}$ measures the brittleness of the material --- the smaller it is, the more brittle the material behaves. In other words, the parameter $\eta_{0}$ needs to be a constant of finite value, and the dependence of the functions $\bar{\mu} (d) = b \mu (d)$ and $\bar{\phi} (d) = b \phi (d)$ on the phase-field length scale $b$ has to be excluded. %That is, the explicit dependence of all the functions $\bar{\mu} (d) = b \mu (d)$, $\bar{\phi} (d) = b \phi (d)$ and $\eta (d)$ on the phase-field length scale $b$ has to be excluded. %Equivalently, the cracking function $\phi (d)$ introduced in \cref{eq:cracking-characteristic-function} needs to be inversely proportional to the length scale $b$, i.e., $\phi (d) \propto \frac{1}{b}$.

Upon the length scale insensitive conditions \eqref{eq:length-scale-insensitivity-condition}, the half crack bandwidth $D (d_{\ast})$ in \cref{eq:half-bandwidth-1D} and in particular, the initial and ultimate half bandwidths $D_{0}$ and $D_{\text{u}}$ in \cref{eq:crack-bandwidths-general}, are all proportional to the length scale $b$, while the traction--separation law $\sigma (\itw)$ is length scale insensitive as expected.

\subsection{Crack irreversibility and non-shrinking crack band}

The above analytical results hold upon the identity \eqref{eq:governing-equations-1D}. That is, the crack has to be maintained in the loading state during the whole failure process of the bar, i.e., $\dot{d} (x) > 0 \; \forall x \in \mathcal{B}$. However, even for this simplest 1D softening bar subjected to a monotonically increasing displacement, only the maximum value $d_{\ast}$ at the centroid of the crack band, which measures the loading level, always increases during failure, i.e., $\dot{d}_{\ast} > 0$. Accordingly, the crack irreversibility condition $\dot{d} (x) \ge 0$ still has to be dealt with elsewhere carefully. 

\begin{figure}[h!] \centering
  \subfigure[Shrinking crack band]{
  \includegraphics[width=0.485\textwidth]{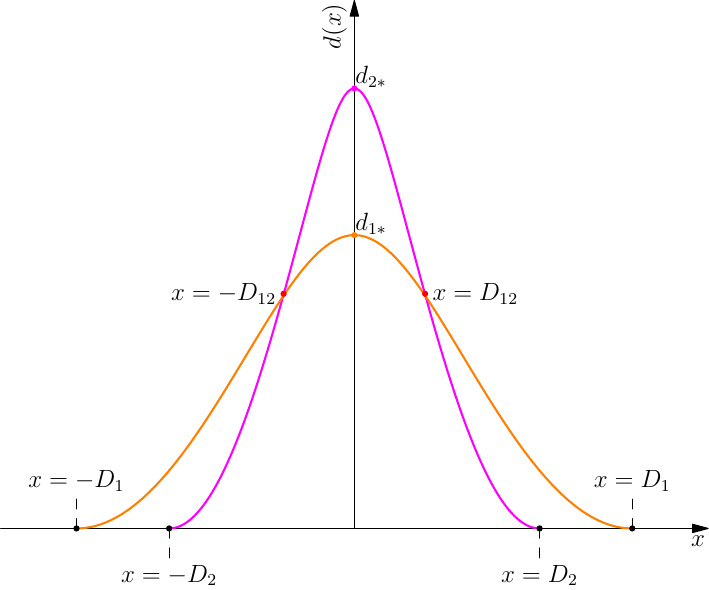}
  \label{fig:damage-profile-shrinking}}
  \subfigure[Non-shrinking crack band]{
  \includegraphics[width=0.485\textwidth]{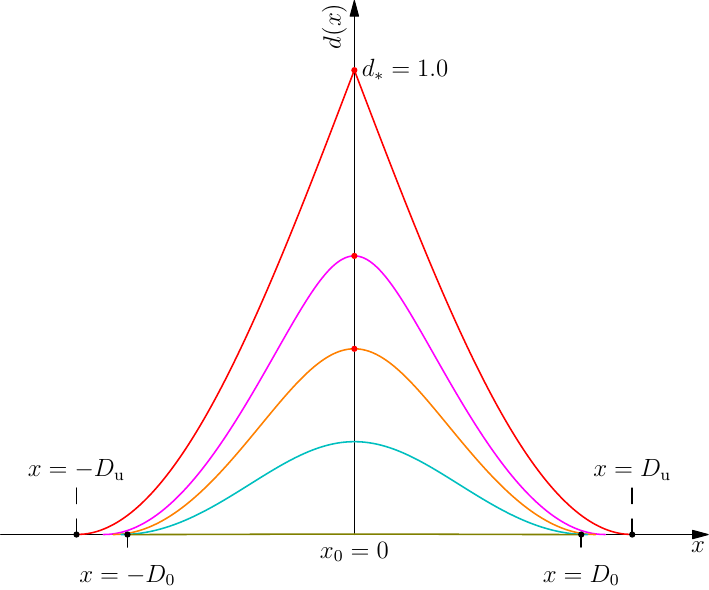}
  \label{fig:damage-profile-3th}} \hfill
  \caption{Evolution of the crack phase-field within the crack band.}
  \label{fig:gpfczm-damage-profiles}  
\end{figure}

Without loss of generality, the following two cases can be discriminated:
\begin{itemize}
\item A shrinking crack band shown in \cref{fig:damage-profile-shrinking}. In this case, the crack band $\mathcal{B}$ shrinks during failure, i.e.,
\begin{align}
	\dot{D} (d_{\ast})
		= \dfrac{\partial D}{\partial d_{\ast}} \dot{d}_{\ast} < 0
	\qquad \Longrightarrow \qquad
	\dfrac{\partial D}{\partial d_{\ast}} < 0
\end{align}
Let the crack bandwidths at two time instants $t_{1} < t_{2}$ be denoted by $D_{1} > D_{2}$. The corresponding crack profiles $d (x, d_{1 \ast})$ and $d (x, d_{2 \ast})$ intersects at the points with coordinates $x = \pm D_{12}$). Even though the maximum values $d_{2 \ast} > d_{1 \ast}$, some regions within the crack band, i.e., $x \in [D_{12}, D_{1}] \cup [-D_{1}, -D_{12}]$, exhibit healing of the crack phase-field. This spurious crack healing cannot be treated by enforcing the crack irreversibility condition $\dot{d} (x) \ge 0$ --- if so, the resulting crack profile $d (x, d_{2 \ast})$ is no longer represented by the analytical result \eqref{eq:half-bandwidth-1D}, i.e., the magenta curve in \cref{fig:damage-profile-shrinking}, and the traction--separation law \eqref{eq:stress-opening-1D} is distinct from the target one. Therefore, this case should be excluded. 

\item A non-shrinking crack band depicted in \cref{fig:damage-profile-3th}. That is, the crack band $\mathcal{B}$ does not shrink, or, equivalently, the half bandwidth $D (d_{\ast})$ is non-decreasing, i.e.,
\begin{align}\label{eq:non-shrinking-crack-band}
	\dot{D} (d_{\ast})
		= \dfrac{\partial D}{\partial d_{\ast}} \dot{d}_{\ast} \ge 0
	\qquad \Longrightarrow \qquad
	\dfrac{\partial D}{\partial d_{\ast}} \ge 0
\end{align}
Regarding this 1D problem with $\dot{d}_{\ast} > 0$ guaranteed \textit{a priori}, the crack irreversibility $\dot{d} (x) \ge 0$ is automatically fulfilled, so long as the crack band is non-shrinking as stated by the condition \eqref{eq:non-shrinking-crack-band}. In such a case, enforcement of the crack irreversibility $\dot{d} (x) \ge 0$ does not affect the traction--separation softening law \eqref{eq:stress-opening-1D}. 

For general 2D and 3D problems, it is not easy to determine uniquely the center line/surface of the crack band and to ensure the increasing monotonicity $\dot{d}_{\ast} > 0$. As a consequence, the crack irreversibility $\dot{d} (\boldsymbol{x}) \ge 0$ cannot be guaranteed by the condition \eqref{eq:non-shrinking-crack-band} alone. Rather, special treatments, e.g., the bound constraints \citep{AMM2009}, the history variable \citep{MHW2010c}, etc., have to be introduced explicitly or implicitly. In such cases, the condition \eqref{eq:non-shrinking-crack-band} for a non-shrinking crack band is indispensable such that the traction--separation softening law \eqref{eq:stress-opening-1D} holds unaffected by enforcement of the crack irreversibility $\dot{d} (\boldsymbol{x}) \ge 0$. 

\end{itemize}
In summary, the condition \eqref{eq:non-shrinking-crack-band} for a non-shrinking crack band has to be fulfilled in order for the target traction--separation law \eqref{eq:stress-opening-1D} to hold upon satisfaction of the crack irreversibility $\dot{d} (\boldsymbol{x}) \ge 0$. It is this reason that the original \texttt{PF-CZM} \citep{Wu2017} applies only to linear and convex softening curves in which the condition \eqref{eq:non-shrinking-crack-band} is satisfied \textit{a priori}, though it has been recently used for the \cite{PPR2009} softening law \citep{GLML2023} which is concave. 

Usually, it is difficult, if not impossible, to determine the monotonicity \eqref{eq:non-shrinking-crack-band} of the half bandwidth $D (d_{\ast})$ analytically for general characteristic functions and only numerical results can be considered. However, as will be shown later, the analytical result is still available for some particular and optimal choices.

Alternatively, provided the monotonicity of the crack bandwidth does not vary during failure, the condition \eqref{eq:non-shrinking-crack-band} implies that the initial half bandwidth $D_{0}$ has to be not larger than the ultimate one $D_{\text{u}}$, i.e.,
\begin{align}\label{eq:irreversibility-condition}
	\dfrac{\partial D}{\partial d_{\ast}} \ge 0 
	\qquad \Longrightarrow \qquad
  D_{0} \le D_{\text{u}}
\end{align}
As the cracking function $\phi (d)$ does not affect the crack bandwidth $D (d_{\ast})$, one needs only to consider the optimal functions $\alpha (d)$ and $\mu (d)$ that guarantee a non-shrinking crack band.

%As shown in \cref{fig:gpfczm-damage-profiles}, in this 1D case the \textit{crack irreversibility condition} $\dot{d} (x) \ge 0$ implies that the crack band $\mathcal{B}$ cannot shrink, or, equivalently, the half bandwidth $D (d_{\ast})$ needs to be non-decreasing (the \textit{crack band non-shrinking condition})
%\begin{align}\label{eq:irreversibility-condition-general}
%	\dot{d} (x) \ge 0 \qquad \Longleftrightarrow \qquad
%	\dot{D} (d_{\ast})
%		= \dfrac{\partial D}{\partial d_{\ast}} \dot{d}_{\ast} \ge 0
%	\qquad \Longrightarrow \qquad
%	\dfrac{\partial D}{\partial d_{\ast}} \ge 0
%\end{align}
%On one hand, the condition for a non-shrinking crack band  $\dot{D} (\boldsymbol{x}) \ge 0$ or $D_{0} \le D_{\text{u}}$ cannot be simply enforced by numerical schemes --- otherwise the resulting traction--separation law would no longer be described by the expected one \eqref{eq:stress-opening-1D} since in this case \cref{eq:governing-equations-1D} does not apply any more. On the other hand, f

\section{Generalized phase-field cohesive zone model}
\label{sec:generalized-pfczm}

In this section, a generalized \texttt{PF-CZM} (\texttt{$\mu$PF-CZM}) is presented with the involved characteristic functions properly determined, extending the original \texttt{PF-CZM} \citep{Wu2017,Wu2018,WN2018} to more general scenarios. As the variational principle of energy minimization is not persisted even in the 1D case, it follows that
\begin{align}
	\mu (d) \ne \phi (d), \qquad
	\varpi' (d) \ne \omega' (d)
\end{align}
in the general cases, though in some particular cases the identities may hold.

\subsection{General formulation}

Note that the functions $\mu (d)$ and $\alpha (d)$ both satisfy the conditions $\mu (0) = \alpha (0) = 0$. Moreover, the condition \eqref{eq:conditions-degradation-functions}$_{2}$, i.e., $\mu (1) = +\infty$, needs to hold upon complete failure. Arguably the simplest expression for the function $\mu (d)$ is of the following form
\begin{align}\label{eq:gpfczm-dissipation-function}
	\mu (d)
		= a_{0} \dfrac{\alpha (d)}{\big(1 - d \big)^{2p}}
%		= a_{0} \dfrac{\alpha (d)}{h (d)}
%	\bar{\mu} (d)
%		= b a_{0} \dfrac{Q (d)}{\big(1 - d \big)^{p}}
	\qquad \Longrightarrow \qquad
%	\bar{\mu} (d)
%		= b a_{0} \dfrac{\alpha (d)}{\big(1 - d \big)^{2p}}
%		= b a_{0} \dfrac{\alpha (d)}{h (d)}
	\eta (d)
		= \dfrac{\alpha (d)}{b \mu (d)}
		= \dfrac{1}{b a_{0}} \big(1 - d \big)^{2 p},
	\qquad
	\eta_{0}
		= \lim_{d \to 0} \eta (d)
%		= \dfrac{1}{b a_{0}} \lim_{d \to 0} h (d)
		= \dfrac{1}{b a_{0}}
%		= \dfrac{1}{b a_{0}} \lim_{d \to 0} \dfrac{\alpha' (d)}{Q' (d)}
%		= \dfrac{1}{b a_{0}} \lim_{d \to 0} \dfrac{\alpha'' (d)}{Q'' (d)}
%		= \cdots
\end{align}
%with the function $h (d)$ given by
%\begin{align}
%	h (d)
%		= \big(1 - d \big)^{2 p}
%\end{align}
for the parameter $a_{0} > 0$ and the exponent $p \ge 1$ to be determined \footnote[2]{The traction order parameter $p < 1$ would give a zero ultimate crack opening, generally implying a softening curve with snap-back. Contrariwise, an exponent $p \ge 1$ would lead to a positive failure crack opening $\itw_{\text{c}} > 0$; see \cite{Wu2017} for more details. Note that the exponent $p \ge 1$ is not necessarily an integer and can be a real number.}. Note that for a polynomial geometric function $\alpha (d)$, the above function $\mu (d)$ is a particular case of that suggested in the original \texttt{PF-CZM} \citep{Wu2017,Wu2018,WN2018}.

In order to fulfill the length scale insensitivity conditions \eqref{eq:length-scale-insensitivity-condition}, both functions $\bar{\mu} (d) = b \mu (d)$ and $\eta (d)$ have to be independent of the length scale parameter $b$, and $\eta_{0}$ has to be a constant of finite value. These facts transform into
\begin{align}
	\eta_{0}
		= \dfrac{1}{b a_{0}}
		= \dfrac{c_{\alpha}}{2 l_{\text{ch}}}
	\qquad \Longrightarrow \qquad
	a_{0}
		= \dfrac{2}{c_{\alpha}} \dfrac{l_{\text{ch}}}{b}
\end{align}
That is, the parameter $a_{0}$ is inversely proportional to the length scale $b$ and so is the function $\mu (d)$, i.e., $a_{0} \propto b^{-1}$ and $\mu (d) \propto b^{-1}$. This conclusion is also consistent with the original \texttt{PF-CZM}.

Accordingly, the traction--separation law \eqref{eq:stress-opening-1D} becomes
\begin{subequations}\label{eq:pfczm-tsl-general}
\begin{align}\label{eq:traction-gpfczm-general}
	\sigma (d_{\ast})
	 &= f_{\text{t}} \big(1 - d_{\ast} \big)^{p} \\
	\label{eq:crack-opening-gpfczm-general}
	\itw (d_{\ast})
%	 &= \dfrac{4 G_{\text{f}}}{c_{\alpha} f_{\text{t}}} 
%   		\sqrt{\dfrac{\big(1 - d_{\ast} \big)^{p}}{P (d_{\ast})}}
%      \int_{0}^{d_{\ast}} \sqrt{ \dfrac{1}{ \big(1 - \vartheta \big)^{p} 
%    / P (\vartheta) - \big(1 - d_{\ast} \big)^{p} / P (d_{\ast})} 
%      \cdot \dfrac{\alpha (\vartheta) P (\vartheta)}{\big(1 - \vartheta \big)^{p}} } \; \td \vartheta \notag \\
%	 &= \dfrac{4 G_{\text{f}}}{c_{\alpha} f_{\text{t}}}    		
%      \int_{0}^{d_{\ast}} \bigg[ \dfrac{1}{h (d_{\ast})} 
%    - \dfrac{1}{h (\vartheta)} \bigg]^{-\frac{1}{2}} 
%    	\cdot \dfrac{\sqrt{\alpha (\vartheta)}}{h (\vartheta)} 
%    	\; \td \vartheta
	 &= \dfrac{2}{c_{\alpha} a_{0}} \itw_{\text{cL}} 
	 		\big(1 - d_{\ast} \big)^{p} \int_{0}^{d_{\ast}} 
	 		\dfrac{\phi (\vartheta) (1 - \vartheta)^{p}}{
	 		\sqrt{\alpha (\vartheta)}} \cdot
	 		\dfrac{1}{\sqrt{(1 - \vartheta)^{2p} 
	 	- (1 - d_{\ast})^{2p}}} \; \td \vartheta
\end{align}
\end{subequations}
As can be seen, the exponent $p \ge 1$ represents the order of the parameterized traction. In \cite{FFL2021,XR2024}, the exponent $p = 1$ is assumed, implying that the traction $\sigma (d_{\ast})$ is assumed \textit{a priori} to be linearly dependent on the maximum value $d_{\ast}$ of the crack phase-field. The (inverse) profile of the crack phase-field $x (d; d_{\ast})$ and the half crack bandwidth $D (d_{\ast})$ are given by
\begin{align}\label{eq:pfczm-bandwidth-general}
	x (d; d_{\ast})
	  = \int_{0}^{d_{\ast}} \dfrac{(1 - \vartheta)^{p}}{
	 		\sqrt{\alpha (\vartheta)}} \cdot \dfrac{1}{
	 		\sqrt{(1 - \vartheta)^{2p} - (1 - d_{\ast})^{2p}}} 
	 		\; \td \vartheta, \qquad
	D (d_{\ast})
		= x (d = 0, d_{\ast})
\end{align}
Once the geometric function $\alpha (d)$ and cracking function $\phi (d)$ are available, the global responses of the resulting \texttt{$\mu$PF-CZM} can be obtained. However, except for some particular cases, it is not easy to derive the analytical results for the traction--separation law \eqref{eq:pfczm-tsl-general} and the half crack bandwidth \eqref{eq:pfczm-bandwidth-general}. In this case, the characteristics of the softening curve $\sigma (\itw)$ can be obtained analytically as in \cref{rmk:characteristics-softening-curve}. Alternatively, numerical quadrature can be considered. 

Vice versa, if the target traction--separation law $\sigma (\itw)$ is given, \cref{eq:crack-opening-gpfczm-general} is Abel's equation \citep{PM2008} from which the kernel function is solved as (see \ref{sec:solution-Abel-equation})
\begin{align}\label{eq:solution-Abel-equation}
	\dfrac{\phi (d) (1 - d)^{p}}{\sqrt{\alpha (d)}}
	  = a_{0} p \dfrac{c_{\alpha}}{\pi} \dfrac{\partial}{\partial d} 
	  	\Bigg[ \int_{0}^{d} \dfrac{\bar{\itw} (\vartheta) 
	  	\big(1 - \vartheta \big)^{p - 1}}{\sqrt{(1 - \vartheta)^{2p} 
	  - (1 - d)^{2p}}} \; \td \vartheta \Bigg]
\end{align}
where $\bar{\itw} (d_{\ast}) := \itw (d_{\ast}) / \itw_{\text{cL}}$ represents the crack opening normalized with respect to the ultimate value $\itw_{\text{cL}}$ of the linear softening curve \eqref{eq:linear-softening-cohesive}. Accordingly, once either the geometric function $\alpha (d)$ or the cracking function $\phi (d)$ is known, the other one can be determined uniquely. 

In this work, we choose to determine the expression for the much simpler function $\alpha (d)$ and solve the more complex one $\phi (d)$ analytically, i.e.,
\begin{align}\label{eq:cracking-function-Abel}
	\phi (d)
		= a_{0} p \dfrac{c_{\alpha}}{\pi} 
			\dfrac{\sqrt{\alpha (d)}}{\big(1 - d \big)^{p + 1}} \varXi (d)
	\qquad \text{with} \qquad \;
	\varXi (d)
		= \big(1 - d \big) \dfrac{\partial}{\partial d} 
	  	\Bigg[ \int_{0}^{d} \dfrac{\bar{\itw} (\vartheta) 
	  	\big(1 - \vartheta \big)^{p - 1}}{\sqrt{(1 
	  - \vartheta)^{2p} - (1 - d)^{2p}}} \; \td \vartheta	\Bigg]
\end{align}
such that the crack opening \eqref{eq:crack-opening-gpfczm-general} simplifies as
\begin{align}\label{eq:crack-opening-gpfczm-final}
	\itw (d_{\ast})
%	 &= \dfrac{2}{c_{\alpha} a_{0}} \itw_{\text{cL}} 
%	 		\big(1 - d_{\ast} \big)^{p} \int_{0}^{d_{\ast}} 
%	 		\dfrac{\varXi (\vartheta)}{\sqrt{(1 - \vartheta)^{2p} 
%	 	- (1 - d_{\ast})^{2p}}} \; \td \vartheta \notag \\
	 	= \itw_{\text{cL}} \bar{\itw} (d_{\ast}) 
	\qquad \text{with} \qquad
	\bar{\itw} (d_{\ast})
	  = \big(1 - d_{\ast} \big)^{p} \dfrac{1}{\pi} \int_{0}^{d_{\ast}} 
	 		\dfrac{2p \varXi (\vartheta) / (1 - \vartheta)}{
	 		\sqrt{(1 - \vartheta)^{2p} - (1 - d_{\ast})^{2p}}} 
	 		\; \td \vartheta
\end{align}
Note that the above cracking functions $\phi (d)$ fulfill the conditions \eqref{eq:conditions-cracking-function} and are also inversely proportional to the length scale $b$ as expected. 

The function $\varXi (d)$ in \cref{eq:cracking-function-Abel}$_{2}$ is nothing else but the solution to Abel's equation \eqref{eq:crack-opening-gpfczm-final}$_{2}$ for a given function $\bar{\itw} (d_{\ast})$. Consequently, though the function $\phi (d)$ and the crack opening $\itw (d_{\ast})$ both depend on the exponent $p \ge 1$, the resulting traction--separation law $\sigma (\itw)$ is independent of it. In other words, the specific value of the traction order parameter $p \ge 1$ does not affect the global responses (at least in the 1D case) predicted from the \texttt{$\mu$PF-CZM}. Moreover, the crack opening $\itw (d_{\ast})$ and the resulting traction--separation law $\sigma (\itw)$ are independent of the geometric function $\alpha (d)$, bringing great flexibility in determination of the involved characteristics functions. 

\remark \label{rmk:characteristics-softening-curve} Though the traction--separation softening law \eqref{eq:pfczm-tsl-general} cannot be expressed in closed-form for general characteristic functions, it is still possible to derive some crucial characteristics of the softening curve analytically. For instance, the initial slope $k_{0}$ is evaluated from \cref{eq:initial-slope-general} as
\begin{align}
	k_{0} 
%		= \lim_{d{\ast} \to 0} \dfrac{\partial \sigma}{\partial \itw}
%	  =-\dfrac{E_{0}}{D_{0} \phi' (0)} p
%		=-\dfrac{E_{0}}{\pi b \phi' (0)} p
%			\sqrt{2 p \alpha' (0)} %\notag \\
%	  = \bigg[ \dfrac{c_{\alpha} a_{0}}{\pi \phi' (0)} 
%	 		p \sqrt{2 p \alpha' (0)} \bigg] k_{0\text{L}}
	  = \bar{k}_{0} k_{0\text{L}}
	\qquad \text{with} \qquad
	\bar{k}_{0} :
		= \sqrt{2 p \alpha' (0)} \; \Bigg[ \lim_{d \to 0} 
			\sqrt{\alpha (d)} \varXi' (d) +	\frac{1}{2} \alpha' (0) 
			\dfrac{\varXi (d)}{\sqrt{\alpha (d)}} \Bigg]^{-1}
\end{align}
for the initial slope $k_{0\text{L}} =-\frac{1}{2} f_{\text{t}}^{2} / G_{\text{f}}$ of the linear softening curve \eqref{eq:linear-softening-cohesive}. The ultimate crack opening $\itw_{\text{c}}$ is given by
\begin{align}
	\itw_{\text{c}} :
	  = \lim_{d_{\ast} \to 1} \itw (d_{\ast})
	  = \itw_{\text{cL}} \lim_{d_{\ast} \to 1} \varXi (d_{\ast})
\end{align}
%Whether the ultimate crack opening $\itw_{\text{c}}$ is of finite or infinite value depends on the function $\varXi (d)$. 
Note that in the original \texttt{PF-CZM} \citep{Wu2017,Wu2018,WN2018}, the characteristic functions $\alpha (d)$ and $\phi (d) = \mu (d)$ were postulated \textit{a priori} in the parameterized form. The initial slope and ultimate crack opening of the resulting softening curve can be similarly determined from which the involved parameters are calibrated. $\Box$

%\begin{align}
%	\phi' (d)
%		= a_{0} p \dfrac{c_{\alpha}}{\pi} \dfrac{\big(1 - d \big) 
%			\alpha (d) \varXi' (d) + \varXi (d) \Big[ \frac{1}{2} 
%			\big(1 - d \big) \alpha' (d) 
%		+ \big( p + 1 \big) \alpha (d) \Big]}{\big(1 - d \big)^{p + 2} 
%			\sqrt{\alpha (d)}}
%\end{align}
%\begin{align}
%	\phi' (0)
%		= a_{0} p \dfrac{c_{\alpha}}{\pi} 
%			\lim_{d \to 0} \dfrac{ \alpha (d) \varXi' (d) 
%		+ \frac{1}{2} \varXi (d) \alpha' (d)}{\sqrt{\alpha (d)}}
%\end{align}
%\begin{align}
%	k_{0} 
%	  = \Bigg[ 2 \sqrt{\dfrac{2 p}{\alpha' (0)}} \lim_{d \to 0}
%	  	\dfrac{\sqrt{\alpha (d)}}{\varXi (d)} \Bigg] k_{0\text{L}}
%%	  = \dfrac{\pi^{2} c_{\alpha} a_{0}}{2 \phi' (0) \alpha' (0)} 
%%	  	\bigg( \dfrac{b}{D_{0}} \bigg)^{3} k_{\text{0L}} 
%\end{align}

\remark \label{rmk:variationally-consistent-formulation} For the variationally consistent formulation, i.e., $\phi (d)	= \mu (d)$, it follows from \cref{eq:gpfczm-dissipation-function,eq:cracking-function-Abel} that
\begin{align}\label{eq:geometric-function-pfczm-feng}
	a_{0} \dfrac{\alpha (d)}{\big(1 - d \big)^{2 p}}
		= a_{0} p \dfrac{c_{\alpha}}{\pi} 
			\dfrac{\sqrt{\alpha (d)}}{\big(1 - d \big)^{p + 1}} \; 
			\varXi (d) 
	\qquad \Longrightarrow \qquad
%	\sqrt{\alpha (d)}
%		= p \dfrac{c_{\alpha}}{\pi} \big(1 - d \big)^{p - 1} \varXi (d)
%\end{align}
%or, equivalently,
%\begin{align}
	\sqrt{\alpha (d)}
		= \big(1 - d \big)^{p - 1} \varXi (d), \qquad
	c_{\alpha}
%		= 4 \int_{0}^{1} \sqrt{\alpha (\vartheta)} \; \td \vartheta
		= \dfrac{1}{p} \pi
\end{align}
Accordingly, the cracking function $\phi (d)$ is determined as
\begin{align}\label{eq:cracking-function-gpfczm-associated}
	\phi (d)
		= \mu (d)
%		= a_{0} \dfrac{\alpha (d)}{\big(1 - d \big)^{2p}}
		= a_{0} \dfrac{\varXi^{2} (d)}{\big(1 - d \big)^{2}}
	\qquad \text{with} \qquad
	a_{0}
		= \dfrac{2p}{\pi} \dfrac{l_{\text{ch}}}{b}
\end{align}
together with the following crack bandwidth $D (d_{\ast})$
\begin{align}\label{eq:crack-bandwidth-pfczm-feng}
  D (d_{\ast})
    = b \int_{0}^{d_{\ast}} \dfrac{1 - \vartheta}{\varXi (\vartheta)} 
    	\cdot \dfrac{1}{\sqrt{(1 - \vartheta)^{2p} - (1 - d_{\ast})^{2p}}} 
    	\; \td \vartheta
\end{align}
Note that only for the lowest-order case with linear traction (i.e., $p = 1$) does the geometric function \eqref{eq:geometric-function-pfczm-feng} fulfill the increasing monotonicity \eqref{eq:geometric-function-conditions}$_{2}$ and can the expected crack profile of the bullet shape be achieved. This choice recovers the particular version of the \texttt{PF-CZM} developed in \cite{FFL2021}. As will be shown later, the solved geometric function \eqref{eq:geometric-function-pfczm-feng} is generally not optimal in guaranteeing the condition \eqref{eq:non-shrinking-crack-band} or \eqref{eq:irreversibility-condition} for a non-shrinking crack band; see \cref{rmk:shrinking-consistent-formulation} for more details. $\Box$

\subsection{Specific traction--separation softening laws}

Specifically, let us consider some softening laws commonly adopted for cohesive fracture; see also \ref{sec:softening-curves}.

\begin{itemize}
\item Linear softening. With the parameterized traction \eqref{eq:traction-gpfczm-general}, the linear traction--separation law \eqref{eq:linear-softening-cohesive} gives the following crack opening
\begin{align}
	\itw (d_{\ast})
		= \itw_{\text{cL}} \Big[ 1 - \big(1 - d_{\ast} \big)^{p} \Big] 
	\qquad \Longrightarrow \qquad
	\bar{\itw} (d_{\ast}) 
		= 1 - \big(1 - d_{\ast} \big)^{p}
\end{align}
The solution \eqref{eq:cracking-function-Abel}$_{2}$ to Abel's equation gives
\begin{align}\label{eq:cracking-function-Abel-linear}
	\varXi (d)
		= s (d)
		= \sqrt{1 - (1 - d)^{2p}} \qquad \text{with} \qquad
	s (d)
		= \sqrt{1 - (1 - d)^{2p}}
\end{align}
It then follows that
\begin{align}
	\bar{k}_{0}
		= 1, \qquad
	\displaystyle \lim_{d_{\ast} \to 1} \varXi (d_{\ast})
		= 1		
\end{align}
such that the initial slope $k_{0}$ and the ultimate crack opening $\itw_{\text{c}}$ of the linear softening curve \eqref{eq:linear-softening-cohesive} are reproduced.

\item Exponential softening. For the parameterized traction \eqref{eq:traction-gpfczm-general}, the crack opening associated with the traction--separation law \eqref{eq:softening-curve-exponential} is expressed as
\begin{align}
	\itw (d_{\ast})
		=-\dfrac{1}{2} \itw_{\text{cL}} \ln \big(1 - d_{\ast} \big)^{p}
	\qquad \Longrightarrow \qquad
	\bar{\itw} (d_{\ast}) 
		=-\dfrac{1}{2} \ln \big(1 - d_{\ast} \big)^{p}
\end{align}
The solution \eqref{eq:cracking-function-Abel}$_{2}$ to Abel's equation gives 
\begin{align}\label{eq:cracking-function-Abel-exponential}
	\varXi (d)
		= \dfrac{1}{2} \text{arctanh} \big( s(d) \big)
		= \dfrac{1}{2} \text{arctanh} \sqrt{1 - (1 - d)^{2p}}
\end{align}
such that
\begin{align}
	\bar{k}_{0}
		= 2, \qquad
	\displaystyle \lim_{d_{\ast} \to 1} \varXi (d_{\ast})
		= +\infty				
\end{align}
Again, the initial slope $k_{0}$ and the ultimate crack opening $\itw_{\text{c}}$ of the exponential softening curve \eqref{eq:softening-curve-exponential} are also reproduced.

\item Polynomial softening. Some traction--separation laws, such as the \cite{CHR1986} softening for concrete \eqref{eq:softening-curve-Cornelissen}, the \cite{PPR2009} softening for adhesives \eqref{eq:PPR-softening}, etc., involve complicate functions, making them difficult to derive the analytical solution \eqref{eq:cracking-function-Abel}. In order to address this issue, high-order parameterized polynomials corresponding to the traction \eqref{eq:traction-gpfczm-general} can be used to fit the crack opening
\begin{align}\label{eq:polynomial-softening}
	\itw (d_{\ast})
		=-\displaystyle \itw_{\text{cL}} \sum_{n = 0}^{N} c_{n} 
			\big(1 - d_{\ast} \big)^{n \cdot p}
	\qquad \Longrightarrow \qquad
	\bar{\itw} (d_{\ast}) 
		=-\sum_{n = 0}^{N} c_{n} \big(1 - d_{\ast} \big)^{n \cdot p}
\end{align}
such that the solution \eqref{eq:cracking-function-Abel}$_{2}$ to Abel's equation yields
\begin{align}\label{eq:cracking-function-polynomial}
	\varXi (d)
		= \sum_{n = 0}^{N} c_{n} \beta_{n} (d)
	\qquad \text{with} \qquad
	\beta_{n} (d)
		=-\big(1 - d \big) \dfrac{\partial}{\partial d} 
			\Bigg( \int_{0}^{d} \dfrac{(1 - \vartheta)^{n \cdot p}}{
			\sqrt{(1 - \vartheta)^{2p} - (1 - d)^{2p}}} 
			\; \td \vartheta \Bigg)
\end{align}
where the fitting coefficients $c_{n}$ and the explicit expressions of the functions $\beta_{n} (d)$ are given in \ref{sec:coefficients-polynomial-softening-curves}.

For instance, the sixth-order polynomial, i.e., $N = 6$, is usually sufficient for fitting, leading to
\begin{align}\label{eq:cracking-function-polynomial-6terms}
	\varXi (d)
		= \bar{c}_{1} s + \bar{c}_{3} s^{3}
		+ \bar{c}_{5} s^{5} + \big( \bar{c}_{2} s_{1}^{2}	
		+	\bar{c}_{4} s_{1}^{4} + \bar{c}_{6} s_{1}^{6} \big)
			\; \text{arctanh} \big( s(d) \big)
%	\phi (d)
%		= a_{0} \dfrac{\sqrt{2 d - d^{2}}}{\big( 1 - d \big)^{2}} \cdot
%			\varXi (d)
\end{align}
where the coefficients $\bar{c}_{n} \; (n = 1, 2, \cdots 6)$ are given by
\begin{subequations}\label{eq:cracking-function-polynomial-6terms-coefficients}
\begin{align}
	\bar{c}_{1}
	 &= c_{1} + c_{2} + 3 c_{3} + \dfrac{5}{2} c_{4} + 5 c_{5} 
		+ \dfrac{33}{8} c_{6}, \qquad
	\bar{c}_{2}
		= c_{2} \\
	\bar{c}_{3}
	 &=-\bigg( 2 c_{3} + \dfrac{3}{2} c_{4} + \dfrac{20}{3} c_{5} + 5 c_{6}
	 		\bigg), \qquad
	\bar{c}_{4}
	  = \dfrac{3}{2} c_{4}, \qquad
	\bar{c}_{5}
		= \dfrac{8}{3} c_{5} + \dfrac{15}{8} c_{6}, \qquad
	\bar{c}_{6}
		= \dfrac{15}{8} c_{6}
\end{align}
\end{subequations}
for the function $s_{1} (d) = (1 - d)^{p}$. It can be verified that the following identities hold
\begin{align}
	\bar{k}_{0}
		= \big( \bar{c}_{1} + \bar{c}_{2} + \bar{c}_{4} 
		+ \bar{c}_{6} \big)^{-1}
		= \bigg( \sum_{n = 0}^{6} n \cdot c_{n} \bigg)^{-1}, \qquad
	\displaystyle \lim_{d_{\ast} \to 1} \varXi (d_{\ast})
%		= \bar{c}_{1} - \bar{c}_{3} + \bar{c}_{5}
		= \sum_{n = 1}^{6} c_{n}
		=-c_{0}
\end{align}
Accordingly, the initial slope $k_{0}$ and the ultimate crack opening $\itw_{\text{c}}$ of the polynomial softening curve \eqref{eq:polynomial-fitting-conditions} are also reproduced.

\end{itemize}

\begin{figure}[b!] \centering
  \subfigure[Linear softening curve]{
  \includegraphics[width=0.485\textwidth]{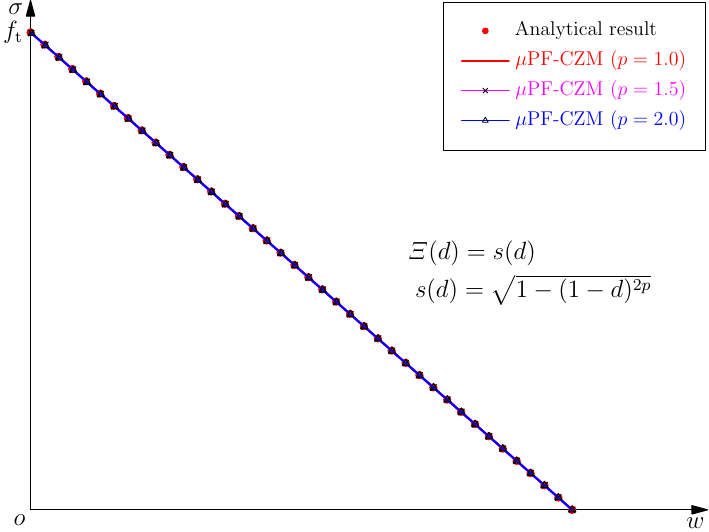}
  \label{fig:gpfczm-softening-curves-linear}} \hfill
  \subfigure[Exponential softening curve]{
  \includegraphics[width=0.485\textwidth]{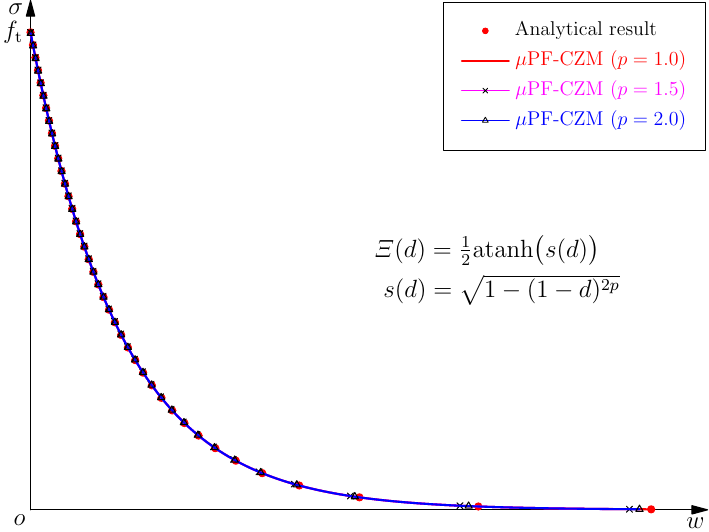}
  \label{fig:gpfczm-softening-curves-exponential}}
\end{figure}
\begin{figure}[t!] \centering
  \subfigure[\cite{CHR1986} softening curve]{
  \includegraphics[width=0.485\textwidth]{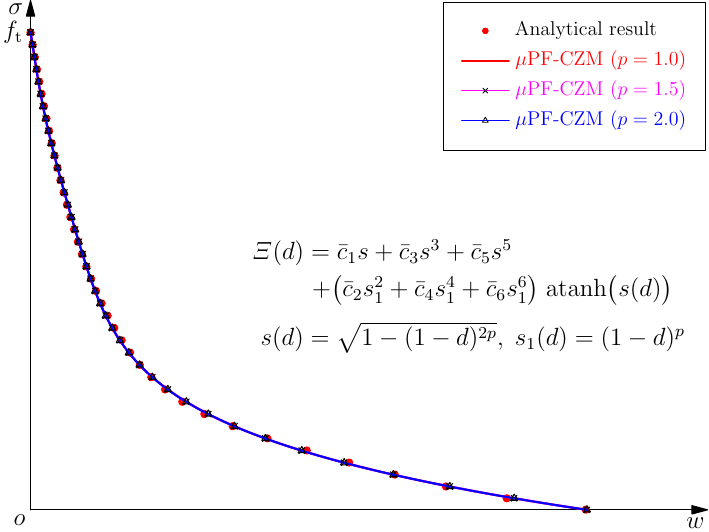}
  \label{fig:gpfczm-softening-curves-cornelissen}} \hfill
  \subfigure[\cite{PPR2009} softening curve]{
  \includegraphics[width=0.485\textwidth]{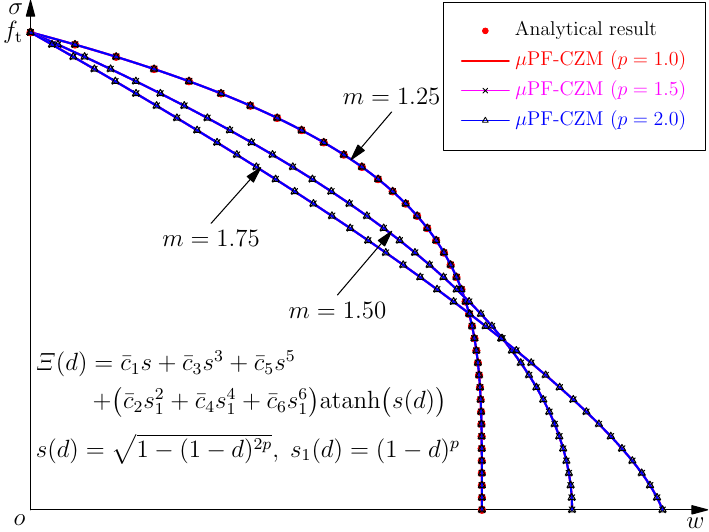}
  \label{fig:gpfczm-softening-curves-PPR}}
  \subfigure[Comparison of various softening curves ($p = 1$)]{
  \includegraphics[width=0.485\textwidth]{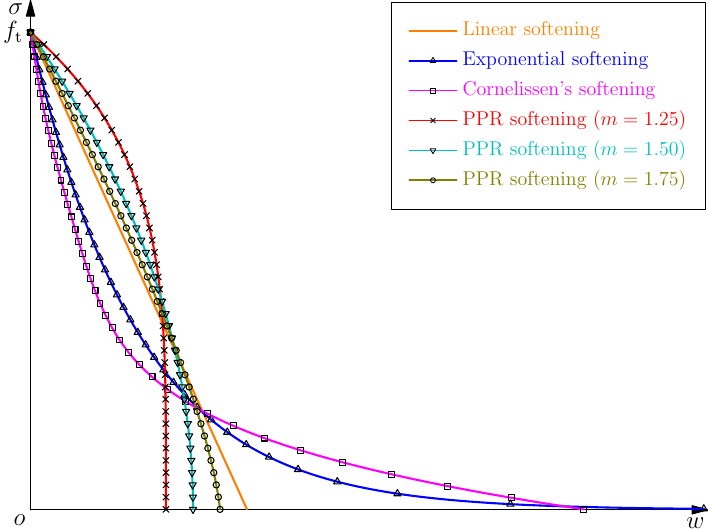}
  \label{fig:gpfczm-softening-curves-comparison}}  
  \caption{Softening curves predicted by the \texttt{$\mu$PF-CZM}.}
  \label{fig:gpfczm-softening-curves}  
\end{figure}

The softening curves given by Eqs. \eqref{eq:traction-gpfczm-general} and \eqref{eq:crack-opening-gpfczm-final} are compared in \cref{fig:gpfczm-softening-curves} against the target ones. As can be seen, the linear and exponential softening curves are exactly reproduced, while for the \cite{CHR1986} and \cite{PPR2009} softening curves the invisible discrepancies are caused by the polynomial fitting. Remarkably, the softening curves given by the \texttt{$\mu$PF-CZM} are indeed independent of the exponent $p \ge 1$.

\subsection{Optimal geometric function}
\label{sec:optimal-geometric-function}

As mentioned before, the traction--separation softening law $\sigma (\itw)$ given by the \texttt{$\mu$PF-CZM} is independent of the geometric function $\alpha (d)$. However, this conclusion holds only when the condition \eqref{eq:non-shrinking-crack-band} or \eqref{eq:irreversibility-condition} for a non-shrinking crack band is fulfilled. It follows from \cref{eq:crack-bandwidths-general} that
\begin{align}\label{eq:irreversibility-condition-pfczm}
	D_{0}
		= \dfrac{\pi b}{\sqrt{2 p \alpha' (0)}}\le 
	D_{\text{u}}
		= b \int_{0}^{1} \dfrac{1}{\sqrt{\alpha (\vartheta)}} \; \td \vartheta
	\qquad \Longrightarrow \qquad
	2 p \alpha' (0) \ge \Big( \dfrac{\pi b}{D_{\text{u}}} 
	\Big)^{2}
\end{align}
That is, the larger values $p$ and $\alpha' (0)$ are, the smaller the initial half bandwidth $D_{0}$ is and the more easily the condition for a non-shrinking crack band can be fulfilled. 

\subsubsection{Second-order polynomial geometric function}

\begin{figure}[t!] \centering
  \subfigure[2nd-order polynomial geometric function \eqref{eq:crack-geometric-function}]{
  \includegraphics[width=0.485\textwidth]{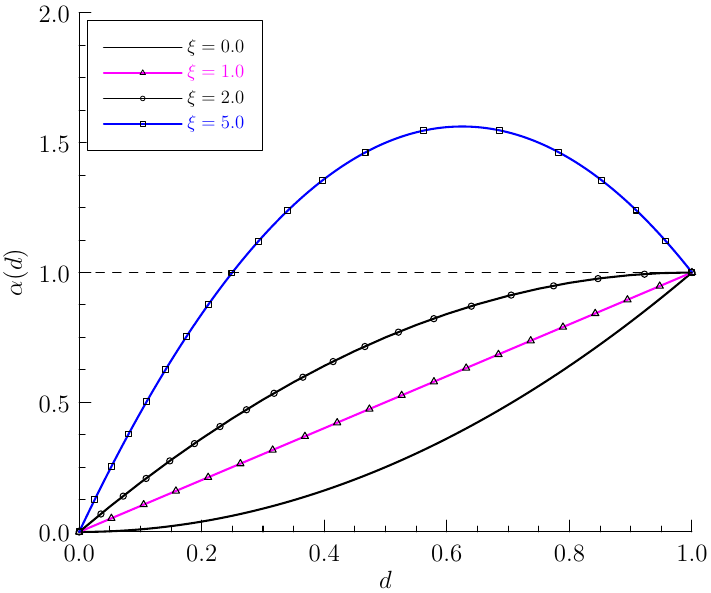}
  \label{fig:crack-geometric-function-parameterized}} \hfill
  \subfigure[Initial and ultimate half bandwidths]{
  \includegraphics[width=0.485\textwidth]{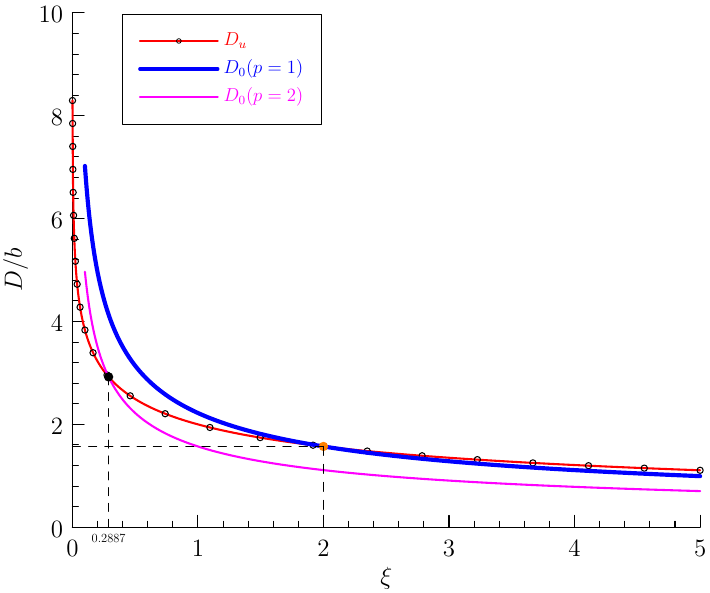}
  \label{fig:half-bandwidths-p2}}
  \caption{The parameterized 2nd-order polynomial geometric function $\alpha (d)$ suggested in \cite{Wu2017}. The non-negativity and increasing monotonicity are guaranteed only for the parameter $\xi \in [0, 2]$ and only $\xi = 2$ is optimal in guaranteeing the condition \eqref{eq:irreversibility-condition-pfczm} for any traction order parameter $p \ge 1$.}
  \label{fig:optimal-geometric-function}  
\end{figure}

Though higher-order polynomials or other types of functions can be considered as well, the parameterized polynomial suggested in \cite{Wu2017} might be the simplest geometric function satisfying the conditions \eqref{eq:geometric-function-conditions} and is adopted in this work
\begin{align}\label{eq:crack-geometric-function}
	\alpha (d)
		= \xi d + \big( 1 - \xi \big) d^{2}, \qquad
	\alpha' (d)
		= \xi + 2 \big( 1 - \xi \big) d
\end{align}
such that the ultimate half crack band width $D_{\text{u}}$ is evaluated as
\begin{align}\label{eq:half-bandwidth-ultimate}
	D_{\text{u}} (\xi)
   &= \begin{cases}
        \dfrac{b}{\sqrt{1 - \xi}} \ln \bigg[ \dfrac{2 
            \sqrt{1 - \xi} + 2 - \xi}{\xi} \bigg]  
          & \qquad \xi \in (0, 1] \vspace{2mm} \\
        \dfrac{b}{\sqrt{\xi - 1}} \bigg( \dfrac{\pi}{2} 
	        - \arcsin \dfrac{2 - \xi}{\xi} \bigg)
          & \qquad \xi \in [1, +\infty)
      \end{cases}
\end{align}
%such that the scaling constant $c_{\alpha}$ is evaluated as
%\begin{align}\label{eq:scaling-constant-explicit}
%  c_{\alpha}
%    = \begin{cases}
%        \dfrac{1}{\big(1 - \xi \big)^{3/2}} \bigg[ \dfrac{1}{2} \xi^{2} 
%            \ln \bigg( \dfrac{\xi}{2 \sqrt{1- \xi} + 2 - \xi} \bigg)
%          + \big(2 - \xi \big) \sqrt{1 - \xi} \bigg]
%          & \qquad \xi \in [0, 1] \vspace{2mm} \\       
%        \dfrac{1}{\big( \xi - 1 \big)^{3/2}} 
%		        \bigg[ \dfrac{1}{2} \xi^{2} 
%            \bigg( \dfrac{\pi}{2} - \arcsin \dfrac{2 - \xi}{\xi} 
%            \bigg) - \big(2 - \xi \big) \sqrt{\xi - 1} \bigg] 
%          & \qquad \xi \in [1, +\infty)      
%      \end{cases}	
%\end{align}
for the parameter $\xi \in [0, 2]$. As commented in \cref{rmk:geometric-function}, though the condition $\xi \le 2$ might be unnecessary for $\varGamma$-convergence of the approximation \eqref{eq:functional-phase-field}, it is still assumed here for the monotonicity; see \cref{fig:crack-geometric-function-parameterized}.

%\begin{figure}[h!] \centering
%  \includegraphics[width=0.485\textwidth]{crack-geometric-function-parametrized.pdf}
%  \caption{The parameterized 2nd-order polynomial geometric function $\alpha (d)$ suggested in \cite{Wu2017}. The non-negativity and increasing monotonicity are guaranteed only for the parameter $\xi \in [0, 2]$.}
%	\label{fig:crack-geometric-function-parameterized}
%\end{figure}

%In the physics community, the well-known double-well function \citep{KKL2001}
%\begin{align}
%	\alpha (d)
%		= d^{2} \big(1 - d \big)^{2}
%	\qquad \Longrightarrow \qquad
%	c_{\alpha}
%		= \dfrac{2}{3}
%\end{align}
%has been dominantly adopted; see \cite{AGL2015} for the review. Similarly, the following polynomial 
%\begin{align}
%	\alpha (d)
%		= d \big(1 - d \big)
%	\qquad \Longrightarrow \qquad
%	c_{\alpha}
%		= \dfrac{\pi}{2}
%\end{align}
%can also be considered. Both geometric functions also vanish at the end point $d = 1$, i.e., $\alpha (1) = 0$; see \cref{fig:crack-geometric-function-particular}. As will be shown later, this kind of geometric functions usually yields \textit{spindle}-shaped profiles for the crack phase-field.

%The analytical results presented in \cref{sec:analytical-solution} and in particular, Eqs. \eqref{eq:pfczm-tsl-general} $\sim$ \eqref{eq:crack-opening-gpfczm-final}, apply only if the crack irreversibility \eqref{eq:irreversibility-condition} is fulfilled, i.e.,
%\begin{align}\label{eq:irreversibility-condition-pfczm}
%	\dot{D} (d_{\ast}) \ge 0
%	\qquad \Longrightarrow \qquad
%	D_{0}
%		= \dfrac{\pi b}{\sqrt{2 p \xi}} \le D_{\text{u}}
%\end{align}

As shown in Figure \ref{fig:half-bandwidths-p2}, for the geometric function \eqref{eq:crack-geometric-function} the initial and ultimate half bandwidths are both monotonically decreasing functions of the parameter $\xi \in [0, +\infty)$, but the former decreases more rapidly than the latter. Accordingly, a larger parameter $\xi \in [0, 2]$ or $p \ge 1$ is more favorable in the \texttt{$\mu$PF-CZM}.

%\begin{figure}[h!] \centering
%  \includegraphics[width=0.485\textwidth]{half-bandwidths-gpfczm.pdf}
%  \caption{Initial and ultimate half bandwidths for $p = 1$ and $p = 2$}
%  \label{fig:half-bandwidths-p2}
%\end{figure}

For the worst case with the smallest exponent $p = 1$, the condition \eqref{eq:irreversibility-condition-pfczm} becomes
\begin{align}\label{eq:irreversibility-condition-pfczm-wu}
	p
		= 1, \quad
	D_{0}
		= \dfrac{\pi b}{\sqrt{2 \xi}} \le D_{\text{u}}
	\qquad \Longrightarrow \qquad
	\xi \ge 2
\end{align}
Provided that the conditions \eqref{eq:geometric-function-conditions} are mandatory, only the geometric function \eqref{eq:crack-geometric-function} with the parameter $\xi = 2$, i.e.,
\begin{align}\label{eq:optimal-geometric-function1}
	\xi
		= 2, \quad
	p
		= 1 \qquad \Longrightarrow \qquad
	\alpha (d)
		= 2 d - d^{2}, \qquad
%		= 1 - \big(1 - d \big)^{2}
%	\alpha' (0)
%		= 2, \qquad
	c_{\alpha}
		= \pi, \qquad
%	d_{\text{u}} (x)
%		= 1 - \sin \Big( \dfrac{|x|}{b} \Big), \qquad		
	D_{0}
		= D_{\text{u}}
		= \frac{\pi}{2} b
\end{align}
applies for the lowest-order \texttt{$\mu$PF-CZM} with $p = 1$. Moreover, for the higher-order \texttt{$\mu$PF-CZM} with $p > 1$, it follows that
\begin{align}
	\xi 
		= 2, \quad p > 1 \qquad \Longrightarrow \qquad
	D_{0}
		= \dfrac{\pi b}{2 \sqrt{p}} < \dfrac{1}{2} \pi b
		= D_{\text{u}}
\end{align}
Therefore, the condition \eqref{eq:irreversibility-condition-pfczm} is automatically fulfilled for the exponent $p \ge 1$. In other words, the geometric function \eqref{eq:optimal-geometric-function1} is optimal for the \texttt{$\mu$PF-CZM} with traction of any order $p \ge 1$ and will be adopted throughout this work. Note that the same geometric function is also optimal for the original \texttt{PF-CZM} which applies only to non-concave (i.e., convex and linear) softening curves. 

Upon the optimal geometric function \eqref{eq:optimal-geometric-function1}, the cracking function $\phi (d)$ in \cref{eq:cracking-function-Abel} becomes
\begin{align}\label{eq:optimal-geometric-function}
	\alpha (d)
		= 2 d - d^{2} \qquad \Longrightarrow \qquad
	\phi (d)
		= a_{0} p \dfrac{\sqrt{2 d - d^{2}}}{\big(1 - d \big)^{p + 1}}
			\cdot \varXi (d)
\end{align}
where the function $\varXi (d)$ is given in \cref{eq:cracking-function-Abel-linear} for linear softening, in \cref{eq:cracking-function-Abel-exponential} for exponential one and in \cref{eq:cracking-function-polynomial-6terms} for the sixth-order polynomial one, respectively.

\remark For the linear traction--separation law \eqref{eq:linear-softening-cohesive} and the lowest traction order $p = 1$, the variationally inconsistent formulation with the optimal geometric function \eqref{eq:optimal-geometric-function} gives
\begin{align}
	\varXi (d)
		= \sqrt{2 d - d^{2}} \qquad \Longrightarrow \qquad
	\alpha (d)
		= 2 d - d^{2}, \qquad
	\phi (d)
		= \mu (d)
		= a_{0} \dfrac{2 d - d^{2}}{\big(1 - d \big)^{2}}
\end{align}
which are the same as those given by the variationally consistent formulation presented in \cref{rmk:variationally-consistent-formulation} and by the original \texttt{PF-CZM} \citep{Wu2017,Wu2018,WN2018}. In other words, for linear softening the lowest-order \texttt{$\mu$PF-CZM} with $p = 1$ is variationally consistent and coincident with the original \texttt{PF-CZM}. $\Box$

\subsubsection{Crack profile and crack bandwidth}
\label{sec:crack-profiles-bandwidth}

As already shown, the traction order parameter $p \ge 1$ does not affect the traction--separation softening law $\sigma (\itw)$. However, it does affect the crack bandwidth $D (d_{\ast})$ and the distribution of the crack phase-field $d (x; d_{\ast})$ within the crack band $\mathcal{B}$. 

\begin{figure}[h!] \centering
  \subfigure[Profiles of the crack phase-field]{
  \includegraphics[width=0.475\textwidth]{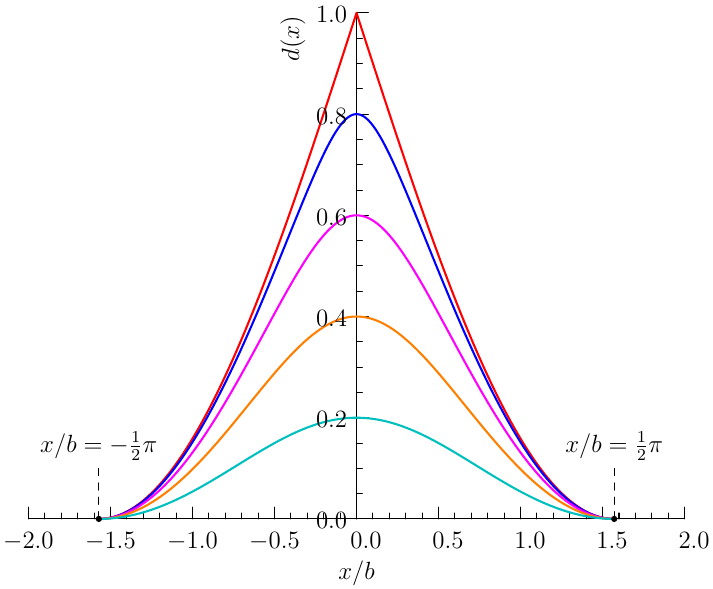}
  \label{fig:damage-profile}} \hfill
  \subfigure[Evolution of the half crack bandwidth]{
  \includegraphics[width=0.475\textwidth]{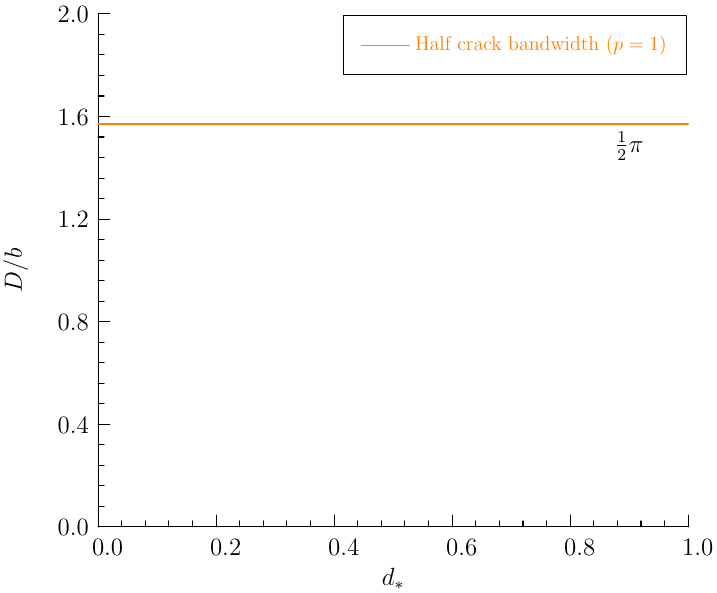}
  \label{fig:gpfczm-bandwidth}}
  \caption{Profiles of the crack phase-field and evolution of the crack bandwidth given by the lowest-order \texttt{$\mu$PF-CZM} with $p = 1$.}
  \label{fig:gpfczm-profiles-bandwdith} 
\end{figure}

For the lowest-order \texttt{$\mu$PF-CZM} with $p = 1$, upon the optimal geometric function \eqref{eq:optimal-geometric-function} it allows deriving from \cref{eq:pfczm-bandwidth-general} the following analytical results
\begin{align}\label{eq:damage-profile-optimal}
	d (x)
		= 1 - \sqrt{1 - \big( 2 d_{\ast} - d_{\ast}^{2} \big)
			\cos^{2} \Big( \dfrac{|x|}{b} \Big)}, \qquad
  D (d_{\ast})
%   &= b \int_{0}^{d_{\ast}} \dfrac{1 - \vartheta}{\sqrt{
%    	\big[ 1 - (1 - \vartheta)^{2} \big] \big[ (1 - \vartheta)^{2} 
%    - (1 - d_{\ast})^{2} \big]}} \; \td \vartheta \notag \\
%   &= b \int_{a}^{1} \dfrac{t}{\sqrt{ \big( 1 - t^{2} \big) 
%   		\big( t^{2} - a^{2} \big)}} \; \td t \notag \\
%   &= b \int_{\arcsin(a)}^{\frac{\pi}{2}} 
%   		\dfrac{\sin (x) }{\sqrt{\sin^{2} (x) - a^{2}}} \; \td x
%    = b \arcsin \dfrac{\cos (x)}{\sqrt{1 - a^{2}}} 
%    	\bigg|_{\frac{\pi}{2}}^{\arcsin (a)}
    = \dfrac{\pi}{2} b
\end{align}
As shown in \cref{fig:gpfczm-profiles-bandwdith}, the crack bandwidth $D (d_{\ast})$ remains constant during the whole failure process such that the non-shrinking condition $\dot{D} (d_{\ast}) \ge 0$ is always fulfilled. 

\begin{figure}[h!] \centering
  \subfigure[Profiles of the crack phase-field ($p = 2$)]{
  \includegraphics[width=0.475\textwidth]{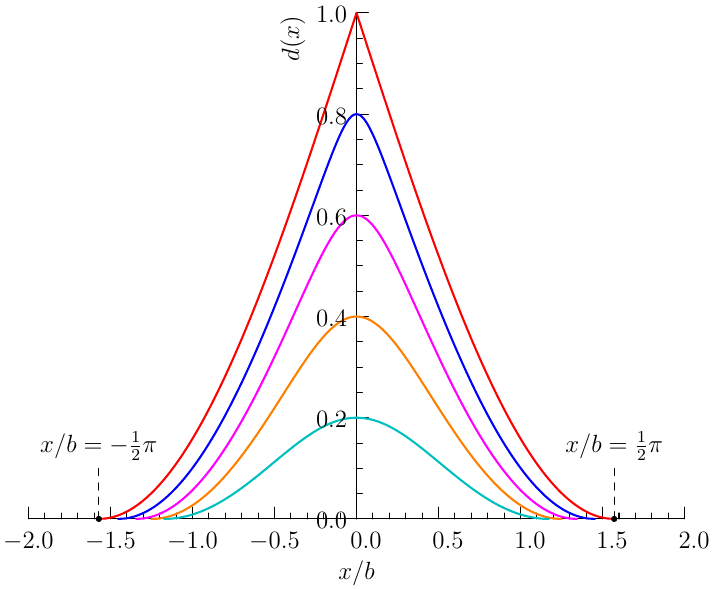}
  \label{fig:damage-profile-4th}} \hfill
  \subfigure[Evolution of the half crack bandwidth]{
  \includegraphics[width=0.475\textwidth]{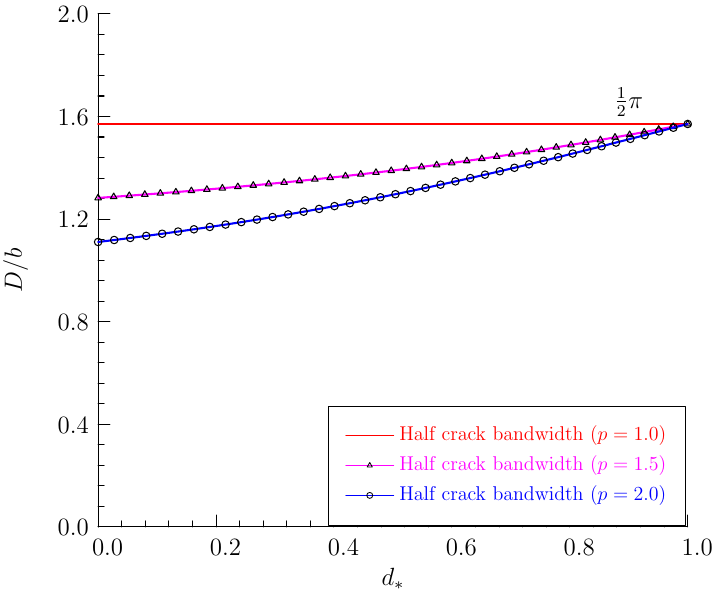}
  \label{fig:gpfczm-bandwidth-orders}}
  \caption{Profiles of the crack phase-field and evolution of the crack bandwidth given by the higher-order \texttt{$\mu$PF-CZM}.}
  \label{fig:gpfczm-profiles-bandwdith-4th}  
\end{figure}

For the higher-order \texttt{$\mu$PF-CZM} with $p > 1$, it is, however, very difficult, if not impossible, to derive the analytical results. In such cases, the numerical quadrature is adopted to give the crack profiles depicted in, e.g., \cref{fig:damage-profile-3th} for $p = 1.5$ and \cref{fig:damage-profile-4th} for $p = 2$. As can be seen, though the traction order parameter $p \ge 1$ does not affect the traction--separation softening law (at least in the 1D case) and the resulting global responses, it does have some effects on the crack profile and evolution of the crack bandwidth during failure. In particular, the crack bandwidth $D (d_{\ast})$ becomes larger with the increasing loading level $d_{\ast}$ and eventually approaches the ultimate value $D_{\text{u}} = \frac{1}{2} \pi b$. 

Evolution of the half crack bandwidth for various exponents $p \ge 1$ is presented in \cref{fig:gpfczm-bandwidth-orders}. As expected, the non-shrinking condition $\dot{D} (d_{\ast}) \ge 0$ is always fulfilled for the optimal geometric function \eqref{eq:optimal-geometric-function}. Note that the ultimate profile of the crack phase-field $d_{\text{u}} (x)$ and the ultimate half crack bandwidth $D_{\text{u}}$, corresponding to $d_{\ast} = 1$, are identical for various exponents $p \ge 1$ since they depends only on the geometric function $\alpha (d)$. Comparatively, during the intermediate stage, i.e., $d_{\ast} \in [0, 1)$, the crack profile $d (x)$ does not coincide and the half crack bandwidth $D (d_{\ast})$ is smaller in the higher-order \texttt{$\mu$PF-CZM} with $p > 1$ than that in the lowest-order one with $p = 1$.

\remark For the lowest-order \texttt{$\mu$PF-CZM} with $p = 1$, it follows from the crack profile \eqref{eq:damage-profile-optimal} that
\begin{align}
	\dot{d}_{\ast} > 0
	\qquad \Longrightarrow \qquad
	\dot{d} (x)
		= \big(1 - d_{\ast} \big) \bigg[ 1 - \big( 2 d_{\ast} 
		- d_{\ast}^{2} \big) \cos^{2} \Big( \dfrac{|x|}{b} \Big) 
			\bigg]^{-\frac{1}{2}} \dot{d}_{\ast} \ge 0
\end{align}
Comparatively, for the higher-order \texttt{$\mu$PF-CZM} with $p > 1$ the crack bandwidth is increasing
\begin{align}
	\dot{d}_{\ast} > 0, \; \dfrac{\partial D}{\partial d_{\ast}} > 0
	\qquad \Longrightarrow \qquad
	\dot{D} (d_{\ast}) > 0, \; \dot{d} (x) \ge 0 
\end{align}
upon the optimal geometric function \eqref{eq:optimal-geometric-function}.
Therefore, in the 1D case the proposed \texttt{$\mu$PF-CZM} automatically guarantees the crack irreversibility $\dot{d} \ge 0$ so long as the monotonically increasing condition $\dot{d}_{\ast} > 0$ is fulfilled at the centroid of the crack band. $\Box$

\begin{figure}[h!] \centering
%  \subfigure[Linear softening]{
%  \includegraphics[width=0.485\textwidth]{cohesive_pfczm_linear_feng_bandwidth.pdf}
%  \label{fig:cohesive_pfczm_linear_feng_bandwidth}}
  \includegraphics[width=0.5\textwidth]{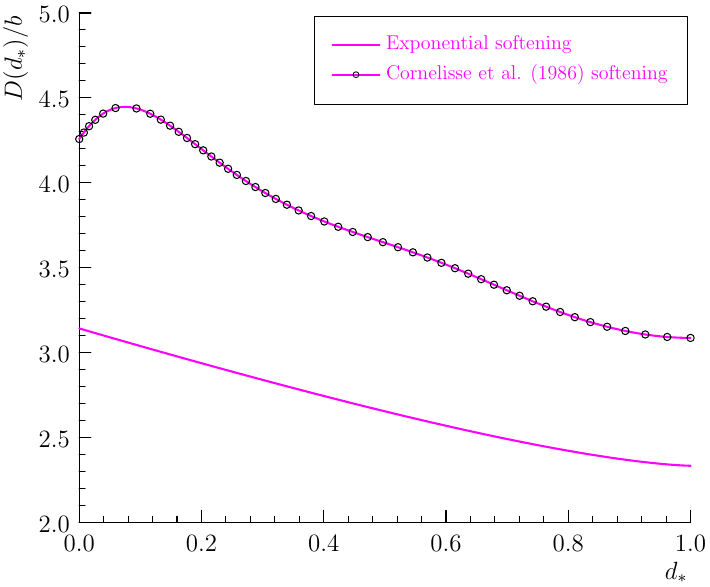}
  \caption{Evolution of the crack band width given by the lowest-order ($p = 1$) variationally consistent $\mu$\texttt{PF-CZM} for the exponential softening. The material properties $f_{\text{t}} = 3.0$ MPa and $G_{\text{f}} = 0.12$ N/mm are adopted.}
  \label{fig:cohesive_pfczm_exponential_feng_bandwidth} 
\end{figure}

\remark \label{rmk:shrinking-consistent-formulation} Regarding the variationally consistent formulation presented in \cref{rmk:variationally-consistent-formulation}, the solved geometric function \eqref{eq:geometric-function-pfczm-feng} is generally not optimal in guaranteeing the condition \eqref{eq:irreversibility-condition-pfczm} for a non-shrinking crack band, with the only exception of linear softening. In order to be more specific, let us consider evolution of the crack bandwidth \eqref{eq:crack-bandwidth-pfczm-feng}. As can be seen from \cref{fig:cohesive_pfczm_exponential_feng_bandwidth}, for the exponential softening the crack band shrinks as cracking proceeds, leading to two mutually exclusive scenarios: either \textit{(i)} some cracked regions experience self-healing if no constraint is enforced on the crack irreversibility, or \textit{(ii)} the target traction--separation softening law is not reproduced and more importantly, the predicted fracture parameters, e.g., the fracture energy, the ultimate crack opening, etc., are distinct from the given ones, if the crack irreversibility is enforced. The same conclusion also applies to the \cite{CHR1986} softening, not even saying other general ones \citep{Wu2022,XR2024}. Therefore, the variationally consistent formulation (including the \cite{FFL2021} model corresponding to $p = 1$) should be used cautiously in the modeling of cohesive fracture with general softening behavior. Besides the proposed variational inconsistent $\mu$\texttt{PF-CZM}, one feasible remedy is to use the original \text{PF-CZM} \citep{Wu2017,Wu2018,WN2018} with the same optimal geometric function \eqref{eq:optimal-geometric-function} --- however, it applies only to non-concave softening laws. $\Box$

%Moreover, the half width of the crack band is given by
%\begin{align}
%  D (d_{\ast})
%    = \dfrac{\pi}{2} b, \qquad p = 1
%\end{align}
%which remains constant during the failure process.

%\begin{align}
%	d (x)
%		= 1 - \sqrt{1 - \big( 2 d_{\ast} - d_{\ast}^{2} \big)
%			\cos^{2} \Big( \dfrac{\pi}{2} \dfrac{|x|}{D} \Big)}
%\end{align}
%\begin{align}
%	\dot{d} (x)
%		= \bigg[ 1 - \big( 2 d_{\ast} - d_{\ast}^{2} \big) \cos^{2} 
%			\Big( \dfrac{|x|}{b} \Big) \bigg]^{-\frac{1}{2}} 
%			\bigg[ \big(1 - d_{\ast} \big) \dot{d}_{\ast}
%		+ \big( 2 d_{\ast} - d_{\ast}^{2} \big) \sin 
%			\Big( \pi \dfrac{|x|}{D} \Big) \dfrac{\pi |x|}{2 D^{2}} 
%			\dot{D} \bigg]
%\end{align}
%

%\begin{align}
%	A_{_d}
%		= \int_{\varOmega} \gamma (d; \nabla d) \; \td V, \qquad
%	\gamma (d; \nabla d)
%	  = \dfrac{1}{\pi} \bigg[ \dfrac{1}{b} \alpha (d) 
%	  + b \big| \nabla d \big|^{2} \bigg] 
%\end{align}
%\begin{align}
%	d (x)
%		= 1 - \sqrt{1 - \big( 2 d_{\ast} - d_{\ast}^{2} \big)
%			\cos^{2} \Big( \dfrac{|x|}{b} \Big)}
%\end{align}
%\begin{align}
%	A_{_d}
%		= A_{_\mathcal{S}} \int_{-\frac{1}{2} \pi b}^{\frac{1}{2} \pi b} 
%			\gamma (d; \nabla d) \; \td x
%\end{align}

\section{Numerical implementation}
\label{sec:numerical-implementation}

In this section, the proposed \texttt{$\mu$PF-CZM} is numerically implemented in the multi-field FEM. As it also falls into the unified phase-field theory, the numerical implementation presented in \cite{Wu2018b} is adopted here with minor modifications. 

\subsection{Weak form}

In accordance with the weighted residual method, the governing equations \eqref{eq:equilibrium-governing-equations-pfm} and \eqref{eq:damage-criterion} are rewritten as the following weak form
\begin{align}\label{eq:weak-form-phase-field}
\begin{cases} 
	\displaystyle 
  \int_{\varOmega} \bfsigma : \nabla^{\ts} \delta \boldsymbol{u} 
			\; \td V
		= \delta \mathscr{P}^{\ast} & \displaystyle 
	\qquad \text{with} \qquad
	\delta \mathscr{P}^{\ast}
		= \int_{\varOmega} \boldsymbol{b}^{\ast} 
			\cdot \delta \boldsymbol{u} \; \td V
		+ \int_{\partial \varOmega_{t}} \boldsymbol{t}^{\ast} 
			\cdot \delta \boldsymbol{u} \; \td A \vspace{2mm} \\
	\displaystyle 	    
  \int_{\mathcal{B}} \boldsymbol{q} \cdot \nabla \delta d \; \td V 
  		\ge \int_{\mathcal{B}} q \cdot \delta d \; \td V &
  \qquad \text{with} \qquad
  \boldsymbol{q}
  	= \dfrac{2 b}{c_{\alpha}} G_{\text{f}} \nabla d
\end{cases}
\end{align}
for the stress tensor $\bfsigma$ and the source term $q$
\begin{align}
  \bfsigma
  	= \omega (d) \bar{\bfsigma} \;\; \text{or} \;\;
    	\omega (d) \bar{\bfsigma}^{+} + \bar{\bfsigma}^{-}, \qquad
	q
		=-\varpi' (d) \bar{Y} 
		- \alpha' (d) \dfrac{1}{c_{\alpha} b} G_{\text{f}}
\end{align}
where $(\delta \boldsymbol{u}, \delta d)$ represent the variations of the displacement field and the crack phase-field. The involved characteristic functions $\alpha (d)$, $\omega (d)$ and $\varpi (d)$ and the derivatives are summarized in \ref{sec:characteristic-functions}.

The governing equations \eqref{eq:weak-form-phase-field} constitute the weak form for a multi-field coupled problem, with the crack phase-field subjected to the bound constraints $d \in [0, 1]$ and $\dot{d} \ge 0$. 

\subsection{Finite element discretization}

In accordance with \cite{WHZN2021}, the whole computational domain $\varOmega^{h}$ is divided into two sub-domains (the super-index $h$ indicates the typical mesh size): the cracking sub-domain $\mathcal{B}^{h}$ where cracks are expected to present and the remaining region $\varOmega^{h} \backslash \mathcal{B}^{h}$ that is elastic. Usually a much smaller cracking sub-domain $\mathcal{B}^{h} \subseteq \varOmega^{h}$ is considered so long as it encompasses potential crack propagation paths.  The element size $h$ within the cracking sub-domain $\mathcal{B}^{h}$ needs to be small enough, e.g., $h \le \frac{1}{5}b$, in order to resolve the incorporated length scale parameter $b$ \citep{Wu2017}. Those nodes belonging to the cracking sub-domain $\mathcal{B}^{h}$ have both displacement and crack phase-field degrees of freedoms (dofs). Comparatively, the nodes within the elastic sub-domain $\varOmega^{h} \backslash \mathcal{B}^{h}$ have only displacement dofs and the standard linear elastic constitutive model applies. %The displacement dofs of all element nodes $I$ in the computational domain $\varOmega$ are grouped into the vector $\mathbf{a} := \big\{ \mathbf{a}_{_I} \big\}^{\text{T}}$, and those crack phase-field dofs of all element nodes $J$ within the crack band $\mathcal{B}^{h}$ are collected into $\bar{\mathbf{a}} := \big\{ \bar{a}_{_J} \big\}^{\text{T}}$. %In this way, one can decrease the total number of dofs and thus the cost of the phase-field sub-problem can be reduced.

%For numerical implementation of 2-D problems, either the unstructured triangular piecewise linear element \citep{PAMM2011,Wu2017,Wu2018,Wu2018b,WN2018} or the structured quadrilateral bilinear one \citep{LG2011,LMGP2016} is usually considered in the discretization. In this case, each element node in the sub-domain $\mathcal{B}^{h}$ has three nodal degrees of freedom (dofs): two for the displacement and one for the damage, while those belonging to $\varOmega^{h} \backslash \mathcal{B}^{h}$ have only two displacement dofs. Similar arguments can be straightforwardly extended to general 3-D cases.

Upon spatial discretization, the weak form \eqref{eq:weak-form-phase-field} gives the following equilibrium equations in residual form \citep{Wu2018b}
\begin{subequations}\label{eq:discrete-governing-equations}
\begin{align}
  \label{eq:discrete-balance-macro}
 &\mathbf{r} :
    = \mathbf{f}^{\text{ ext}} 
    - \int_{\varOmega^{h}} \mathbf{B}^{\text{T}} \bfsigma \; \td V
    = \mathbf{0} \\    
  \label{eq:discrete-balance-micro}
 &\bar{\mathbf{r}} :
    = \int_{\mathcal{B}^{h}} \bar{\mathbf{N}}^{\text{T}} q \; \td V \; 
    - \int_{\mathcal{B}^{h}} \bar{\mathbf{B}}^{\text{T}} \mathbf{q}
    	\; \td V \; \le \mathbf{0}
\end{align}
\end{subequations}
where $\mathbf{B}$ is the displacement--strain matrix and $\mathbf{f}^{\text{ ext}}$ represents the standard external force vector \citep{Hughes2000}; the interpolation matrix $\bar{\mathbf{N}}$ and the gradient operator $\bar{\mathbf{B}}$ approximate the crack phase-field and its gradient, respectively.

%Note that the discretized governing equation \cref{eq:discrete-balance-micro} hold upon the damage boundedness $d \in [0, 1]$ and the irreversibility conditions $\dot{d} \ge 0$. %The above bound constraints will be dealt with by an appropriate solver.

\subsection{Numerical algorithm}\label{eq:AM-solver}

The system of nonlinear equations \eqref{eq:discrete-governing-equations} is solved in an incremental procedure. That is, during the time interval $[0, T]$ of interest, all the state variables are considered at the discrete interval $[T_{n}, T_{n+1}]$ for $n = 0, 1, 2, \cdots$. For a typical time interval $[T_{n}, T_{n+1}]$ of increment length $\varDelta T := T_{n+1} - T_{n}$, all the state variables are known at the instant $T_{n}$. The bound constraints $d \in [0, 1]$ and $\dot{d} \ge 0$, are replaced by \citep{AMM2009}
\begin{align}\label{eq:irreversibility-condition-numerical}
	0 \le \bar{a}_{_{J, n}} \le \bar{a}_{_{J, n + 1}} \le 1
	\qquad \forall J
\end{align}
for all nodal crack phase-field dofs $\bar{a}_{_J}$ at two consequent time instants $T_{n}$ and $T_{n+1}$, respectively. Accordingly, the governing equation \eqref{eq:discrete-balance-micro} can be solved by a proper bound constrained optimization solver \citep{FM2017}. %Such a problem is solved in this work by the reduced-space active set Newton method \citep{BM2006}. 

Though the BFGS quasi-newton monolithic algorithm \citep{WHN2020} can be used to solve the discrete equilibrium equations \eqref{eq:discrete-governing-equations} with the bound constraint \eqref{eq:irreversibility-condition-numerical}, in this work the alternate minimization (AM) solver \citep{BFM2000,BFM2008} is adopted as in the original \texttt{PF-CZM} \citep{Wu2018b}. More specifically,  
\begin{itemize}
\item Fix the crack phase-field dofs $\bar{\mathbf{a}}$ and solve the nodal displacements $\mathbf{a}$ from the mechanical sub-problem \eqref{eq:discrete-balance-macro} using the Newton-Raphson method. Linearization of the governing equation \eqref{eq:discrete-balance-macro} yields
\begin{align}
	\mathbf{K}_{uu} \delta \mathbf{a}
		= \mathbf{r} \qquad \text{with} \qquad
	\mathbf{K}_{uu}
		= \int_{\varOmega^{h}} \mathbf{B}^{\text{T}} \left( 
			\dfrac{\partial \boldsymbol{\sigma}}{\partial \boldsymbol{\epsilon}} 
			\right) \mathbf{B} \; \td V
\end{align}
where the material tangent $\partial \bfsigma / \partial \bfepsilon$ depends on the specific constitutive relation \eqref{eq:stress-pfm-nosplit} or \eqref{eq:strain-energy-density-split}. %Note that upon fixed , the material tangent and the resulting system matrix are both of major symmetry but not necessarily positive-definite. Moreover, path-following strategies, e.g., the indirect displacement control \citep{deB1987,WS1990}, can be incorporated to deal with the material softening induced snap-backs such that the post-peak unstable behavior can be tracked; see \cite{Wu2018b} for the details. 

\item Solve the crack phase-field dofs $\bar{\mathbf{a}}$ with the updated nodal displacements $\mathbf{a}$ from the cracking sub-problem \eqref{eq:discrete-balance-micro}. The Newton-Raphson linearized equation reads
\begin{align}
	\mathbf{K}_{dd} \delta \bar{\mathbf{a}}
		= \bar{\mathbf{r}} \qquad \text{with} \qquad
	\mathbf{K}_{dd}
	 	= \int_{\mathcal{B}^{h}} \left[ \bar{\mathbf{N}}^{\text{T}} 
	 		\Big( \varpi'' \bar{Y} + \alpha'' \dfrac{1}{c_{\alpha} b} 
	 		G_{\text{f}} \Big) \bar{\mathbf{N}}
		+ \dfrac{2 b}{c_{\alpha}} G_{\text{f}}
			\bar{\mathbf{B}}^{\text{T}}
			\bar{\mathbf{B}} \right] \td V
\end{align} 
subjected to the bound constraints \eqref{eq:irreversibility-condition-numerical}.

%That is, in each iteration
%\begin{itemize}
%\item[(1)] The set of active nodes is first determined from the conditions ($\bar{a}_{_{I, n + 1}} = \bar{a}_{_{I, n}}, R_{d, I} < 0$) or ($\bar{a}_{_{I, n + 1}} = 1, R_{d, I} > 0$), and the solution associated with the active nodes is set to zero.
%
%\item[(2)] The sub-system of Eq. \eqref{eq:discrete-balance-micro} associated with the inactive nodes (the remaining ones) is solved by the Newton-Raphson scheme, with the following linearized equation
%\begin{align}
%	\mathbf{K}_{dd} \delta \bar{\mathbf{a}}
%		= \mathbf{R}_{d}
%\end{align}
%for the symmetric tangent stiffness 
%\begin{align}
%	\mathbf{K}_{dd}
%	 	= \int_{\mathcal{B}} \Big( -\bar{\mathbf{N}}^{\text{T}} 
%	 		\dfrac{\partial r}{\partial d} \bar{\mathbf{N}}
%		+ \dfrac{2 b}{c_{0}} G_{\text{f}} \bar{\mathbf{B}}^{\text{T}}
%			\bar{\mathbf{B}} \Big) \td V, \qquad
%	\dfrac{\partial r}{\partial d}
%		=-\omega'' (d) \bar{Y} 
%		- \alpha'' (d) \dfrac{1}{c_{\alpha} b} G_{\text{f}}			
%\end{align}
%The solution is then updated and projected onto the bounds (i.e., if $\bar{a}_{I} > 1$ we set $\bar{a}_{I} = 1$).
%
%\item[(3)] The truncated residuals (excluding those corresponding to the inactive nodes) is computed, and the above procedure is repeated until the solution converges.
%\end{itemize}
\end{itemize} 
The sub-problems \eqref{eq:discrete-balance-macro} and \eqref{eq:discrete-balance-micro} are solved by fixing the crack phase-field dofs $\bar{\mathbf{a}}$ or the displacement dofs $\mathbf{a}$ alternately, until the final solution converges. The solution based stop criterion \citep{BFM2000}, i.e., $\big| \bar{a}_{n+1}^{(k + 1)} - \bar{a}_{n+1}^{(k)} \big| < \varepsilon$ for a small positive number, say, $\varepsilon = 1.0 \times 10^{-5}$, between two consecutive iterations, is adopted in the numerical implementation. Despite its slow convergence rate, the resulting algorithm is extremely robust, particularly when it is enhanced with the indirect displacement control as in \cite{Wu2018b}. 

%\remark The proposed \texttt{$\mu$PF-CZM} applies to 3D problems with no extra theoretical or numerical difficulty, it suffers from the burdensome computational cost as other phase-field models for fracture. This is due, on the one hand, to the very fine mesh and high number of degrees of freedom (dofs) necessary to resolve the highly localized damage distribution, and on the other  hand, to the alternative minimization or staggered solver which is robust but extremely inefficient. The first weakness can be dealt with mesh adaptation, e.g., the anisotropic version proposed in \citet{AFMP2015}, while the second one relies on more efficient solvers \citep{FM2017}. These topics will be addressed elsewhere.

\section{Representative numerical examples}
\label{sec:numerical-examples}

In this section the proposed \texttt{$\mu$PF-CZM} is validated against several benchmark problems. The objectives of the presented numerical examples are twofold: \textit{(i)} to verify that the $\mu$PF-CZM is insensitive to both the phase-field length scale and the traction order parameter, and \textit{(ii)} to demonstrate the necessity of modeling concave softening curves and the applicability of the $\mu$PF-CZM in such cases. The first objective was tested by all the presented 1D and 2D examples, while the second one was illustrated by the last example. %In particular, the insensitivity to the incorporated length scale and traction order parameters $(b, p)$ is studied.  

Our in-house FE code \textsc{feFrac} based on the open-source library \textsc{jive} \citep{NNVMW2020} was adopted in the numerical simulations. Finite element
meshes were generated using \textsc{Gmsh} \citep{gmsh2009} and
visualization was performed in \textsc{Paraview} \citep{Utkarsh2015}.

\subsection{Uniaxial stretching of a softening bar}

\begin{figure}[h!] \centering
  \includegraphics[width=0.75\textwidth]{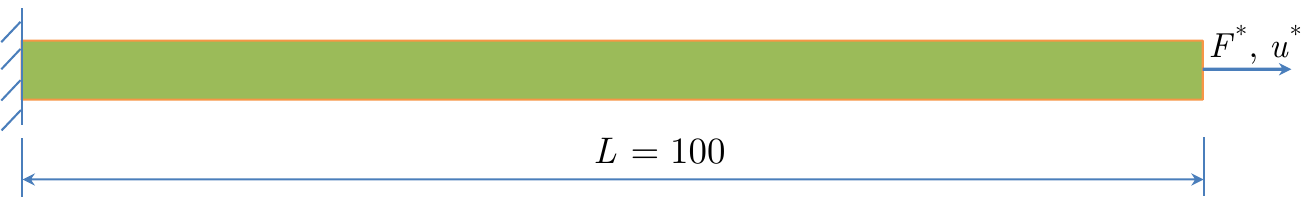}
  \caption{Uniaxial stretching of a softening bar: Geometry (unit of length: mm), loading and boundary conditions.}
	\label{fig:uniaxial_stretching_geometry}
\end{figure}

Let us first consider a bar of length $L = 100$ mm and of cross section $A$ under uniaxial stretching. As shown in \cref{fig:uniaxial_stretching_geometry}, the left edge of the bar is fixed, while the right end is stretched by a monotonically increasing displacement $u^{\ast}$ (with the conjugated force denoted by $F^{\ast}$). The material properties were assumed as: Young's modulus $E_{0} = 3.0 \times 10^{4}$ MPa, Poisson's ratio $\nu_{0} = 0.0$, the failure strength $f_{\text{t}} = 3.0$ MPa and the fracture energy $G_{\text{f}} = 0.12$ N/mm, resulting in an Irwin internal length $l_{\text{ch}} = 400$ mm. Linear, exponential (convex) and \cite{PPR2009} (concave) softening curves were considered.

2-node finite elements with a uniform mesh size $h = 0.05$ mm were used in the spatial discretization of the bar. The Dirichlet condition $d = 0$ was imposed on both ends of the bar to avoid crack initiation at the boundaries. As a consequence, though no imperfection was introduced, crack nucleation can occur at any interior position of the bar due to numerical errors. Moreover, for this 1D problem we did not enforce the crack irreversibility condition $\dot{d} (x) \ge 0$. %since it is automatically fulfilled due to the non-shrinking crack band $\partial D / \partial d_{\ast}$ and the increasing monotonicity $\dot{d}_{\ast} > 0$.

\begin{figure}[h!] \centering
  \subfigure[Linear softening]{
  \includegraphics[width=0.485\textwidth]{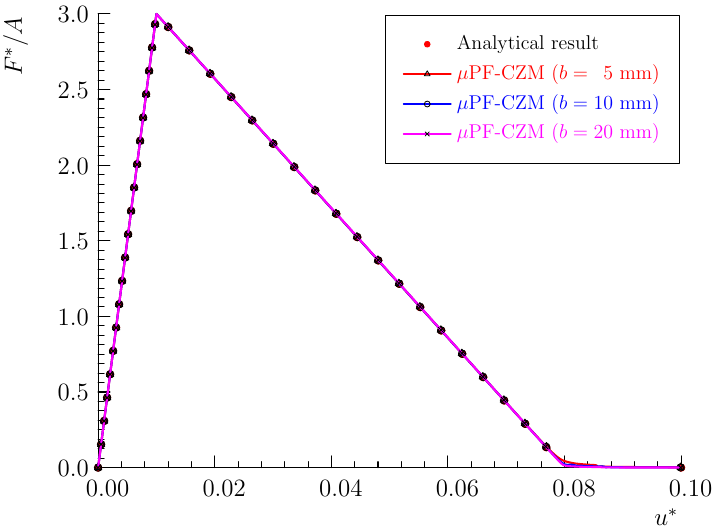}
  \label{fig:uniaxial_stretching_linear_lengths}} \hfill
  \subfigure[Exponential softening]{
  \includegraphics[width=0.485\textwidth]{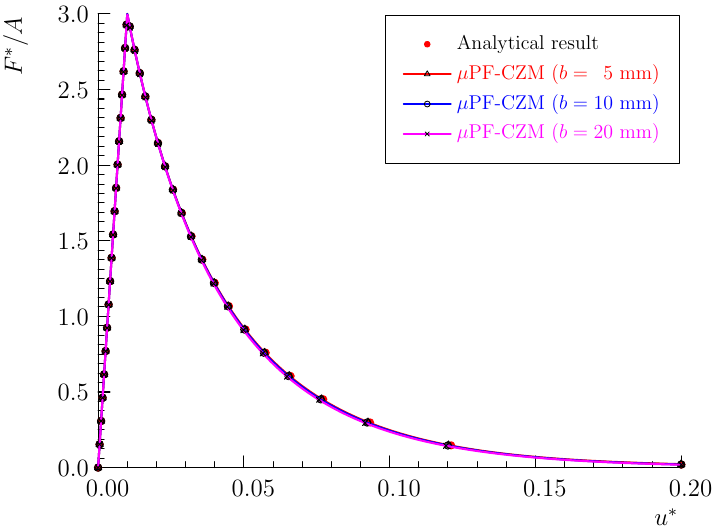}
  \label{fig:uniaxial_stretching_exponential_lengths}}
  \subfigure[\cite{PPR2009} softening with $m = 1.5$]{
  \includegraphics[width=0.485\textwidth]{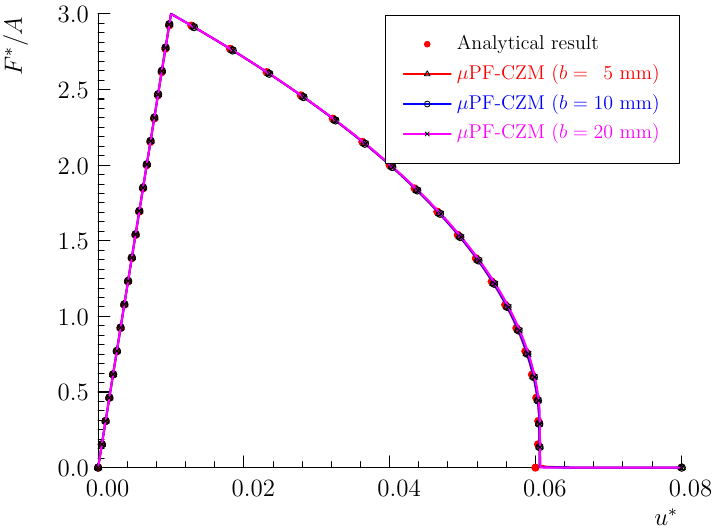}
  \label{fig:uniaxial_stretching_PPR150_lengths}}
  \caption{Uniaxial stretching of a softening bar: Numerical force--displacement curves predicted by the \texttt{$\mu$PF-CZM} ($p = 1.5$) with various values of the length scale parameter $b$.}
	\label{fig:uniaxial_stretching_lengths}
\end{figure}

The length scale insensitivity was first studied. Three different values for the phase-field length scale parameter, i.e., $b = 5$ mm, 10mm and 20 mm, respectively, were considered by the \texttt{$\mu$PF-CZM} with the traction order parameter $p = 1.5$. The predicted force--displacement curves are shown in \cref{fig:uniaxial_stretching_lengths}. As can be seen, the numerical predictions coincide with the analytical results with only negligible discrepancies upon complete fracture. Though the length scale parameter $b$ affects the crack bandwidth $D$, it has no effect on the global fracture behavior of the bar. That is, the length scale insensitivity of the original \texttt{PF-CZM} is preserved by the proposed \texttt{$\mu$PF-CZM}.

\begin{figure}[h!] \centering
  \subfigure[Linear softening]{
  \includegraphics[width=0.485\textwidth]{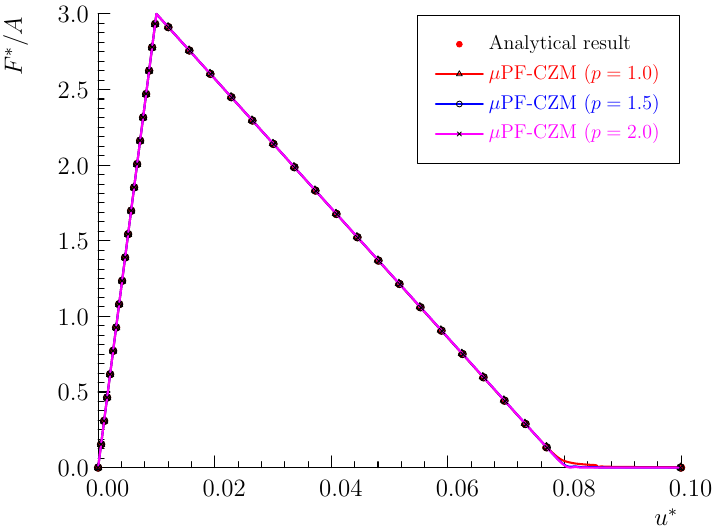}
  \label{fig:uniaxial_stretching_linear_orders}} \hfill
  \subfigure[Exponential softening]{
  \includegraphics[width=0.485\textwidth]{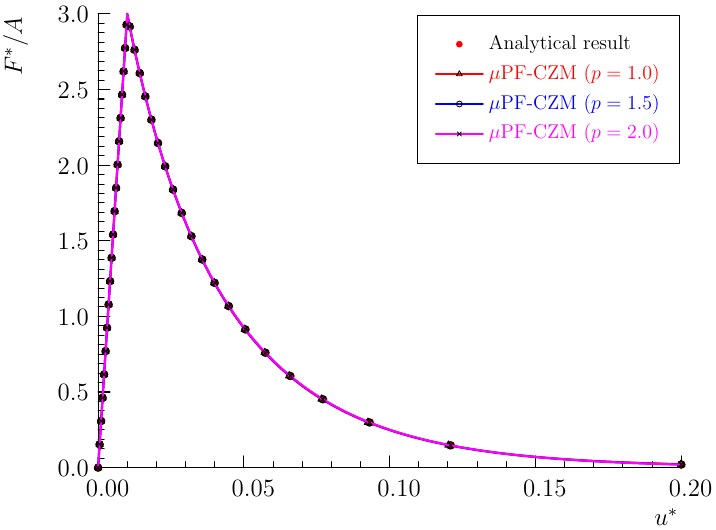}
  \label{fig:uniaxial_stretching_exponential_orders}}
  \subfigure[\cite{PPR2009} softening with $m = 1.5$]{
  \includegraphics[width=0.485\textwidth]{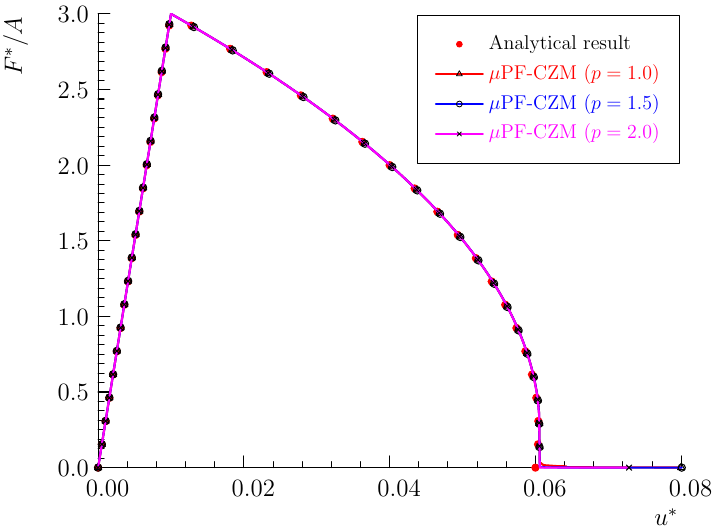}
  \label{fig:uniaxial_stretching_PPR150_orders}}
  \caption{Uniaxial stretching of a softening bar: Numerical force--displacement curves predicted by the \texttt{$\mu$PF-CZM} with various values of the traction order parameter $p \ge 1$ and the identical phase-field length scale $b = 10$ mm.}
	\label{fig:uniaxial_stretching_orders}
\end{figure}

Regarding the insensitivity to the traction order parameter, three different values, i.e., $p = 1.0$, 1.5 and 2.0, respectively, were assumed with the length scale parameter $b = 10$ mm. The numerical force--displacement curves are depicted in \cref{fig:uniaxial_stretching_orders}. Again, the numerical predictions are independent of the exponent $p \ge 1$ and the analytical results are reproduced. In other words, the proposed \texttt{$\mu$PF-CZM} is also insensitive to the traction order parameter $p \ge 1$ though the distribution of the crack phase-field is affected as stated in \cref{sec:optimal-geometric-function}.

As mentioned before, for this 1D problem the condition \eqref{eq:non-shrinking-crack-band} or \eqref{eq:irreversibility-condition} 
for a non-shrinking crack band $\partial D / \partial d_{\ast} \ge 0$ is automatically guaranteed by the optimal geometric function \eqref{eq:optimal-geometric-function} for any value of the traction order parameter $p \ge 1$. Accordingly, the crack irreversibility $\dot{d} \ge 0$ is also fulfilled since the maximum value $d_{\ast}$ is always increasing (i.e., $\dot{d}_{\ast} > 0$). Therefore, it is unnecessary to impose any constraint on the crack phase-field to deal with the crack irreversibility. However, as will be shown in the next example, this is not the case for general 2D/3D problems in which the increasing monotonicity $\dot{d}_{\ast} > 0$ cannot be guaranteed in a unique way.

%In Figures \ref{fig:uniaxial-traction-linear} and \ref{fig:uniaxial-traction-nonlinear}, the numerical results of load $F^{\ast}$ \textit{versus} imposed displacement $u^{\ast}$ are compared against the analytical ones \citep{Wu2011}. It can be seen that, the numerical results are independent from the internal length $b$. As expected, sufficiently fine mesh (i.e., $h \le b/5$) should be used in the discretization. Otherwise, within the localization band the crack phase field of large gradient cannot be represented with sufficient resolution when the specimen is close to breakage, but the overall agreement is still good particularly for the nonlinear softening law. As the mesh is refined, the numerical results converge to the analytical one asymptotically. The above properties are rather important for the proposed phase-field theory be applicable for the numerical modeling of localized failure in solids. 

\subsection{Three-point bending of a notched concrete beam}

\begin{figure}[h] \centering
  \includegraphics[width=0.85\textwidth]{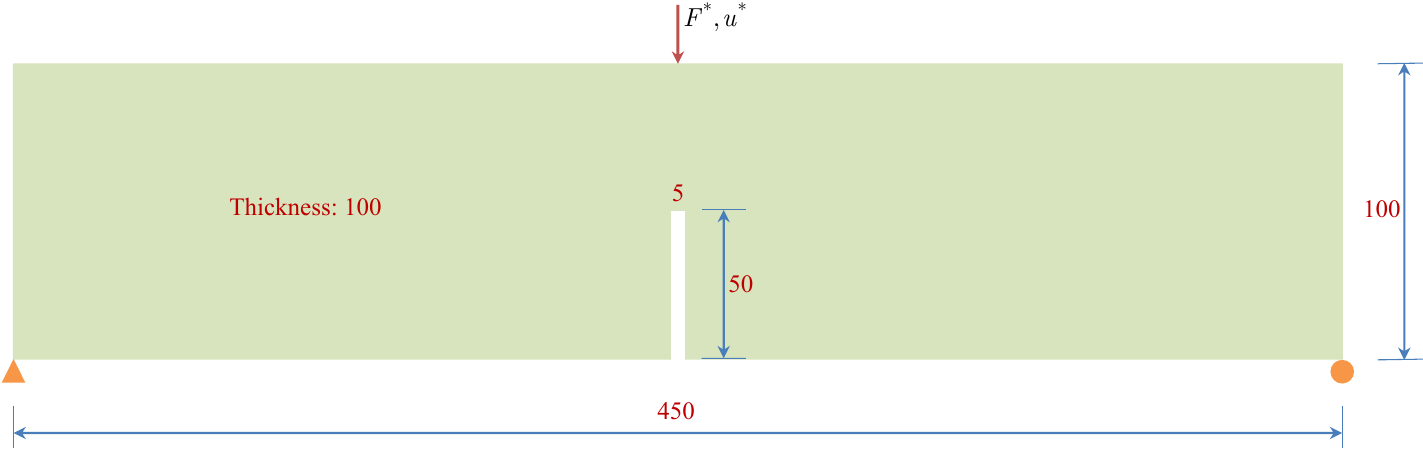}
  \caption{Three-point bending of a notched concrete beam: Geometry (unit of length: mm), loading and boundary conditions.}
	\label{fig:3pt-bending-beam-geometry}
\end{figure}

The next example is the three-point bending test on a notched concrete beam in \citep{Rots1988}. The geometry, loading and boundary conditions of the specimen are shown \cref{fig:3pt-bending-beam-geometry}. The material properties were taken from \citet{Rots1988}: Young's modulus $E_{0} = 2.0 \times 10^{4}$ MPa, Poisson's ratio $\nu_{0} = 0.2$, the failure strength $f_{\text{t}} = 2.4$ MPa and the fracture energy $G_{\text{f}} = 0.113$ N/mm, resulting in an Irwin internal length $l_{\text{ch}} = 392.4$ mm. %Different softening curves were considered. %Mode-I failure of the beam was caused by a crack emerging from the notch tip and propagating vertically along the line of symmetry. 

\begin{figure}[h] \centering
  \includegraphics[width=0.85\textwidth]{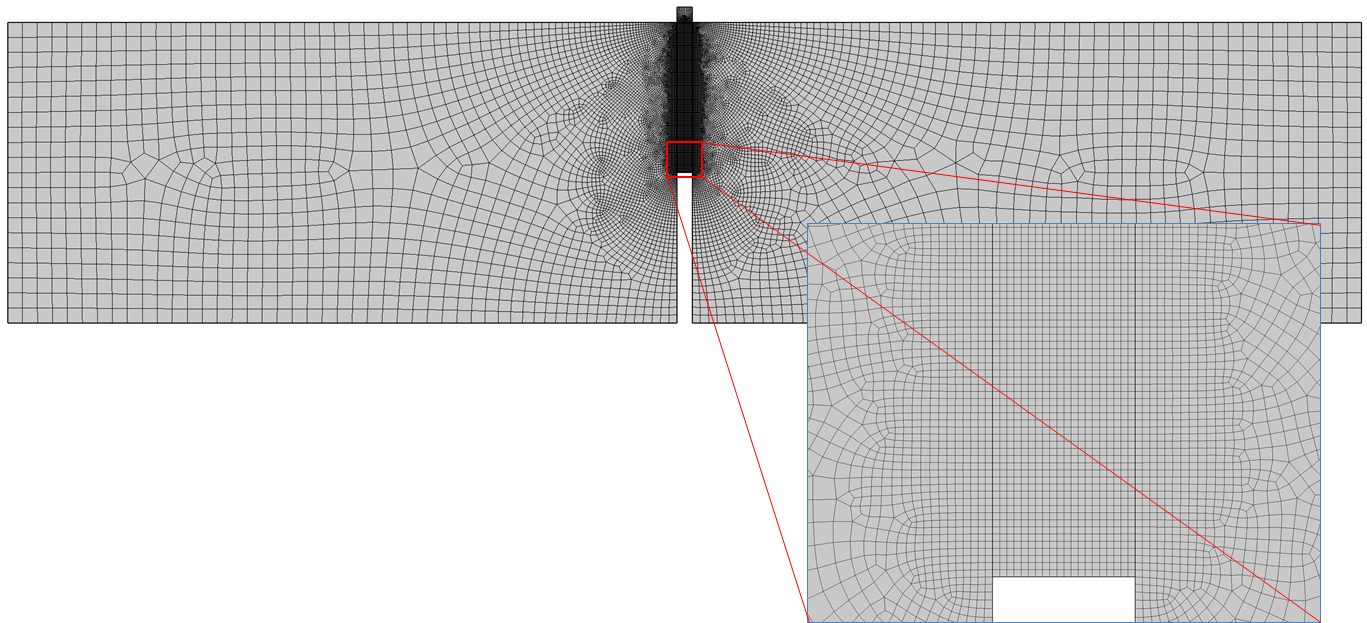}
  \caption{Three-point bending of a notched concrete beam: Finite element mesh.}
	\label{fig:3pt-bending-beam-mesh}
\end{figure}

As shown in Figure \ref{fig:3pt-bending-beam-mesh}, quadrilateral 4-node (Q4) finite elements with the mesh size $h = 0.25 \; \text{mm} \le b / 5$ were used around the center of the beam such that the phase-field length scale can be sufficiently resolved. The CMOD-based indirect displacement control \citep{Wu2018b} was adopted to track the post-peak softening behavior.

%The numerically predicted crack paths are shown in Figure \ref{fig:bending_3pt_crack_paths}. Without any extra crack tracking strategy, the crack propagates vertically from the notch to the top of beam, though the localization bandwidth dependent on the internal length scale $b$ as expected. 

\begin{figure}[h!] \centering
  \includegraphics[width=0.55\textwidth]{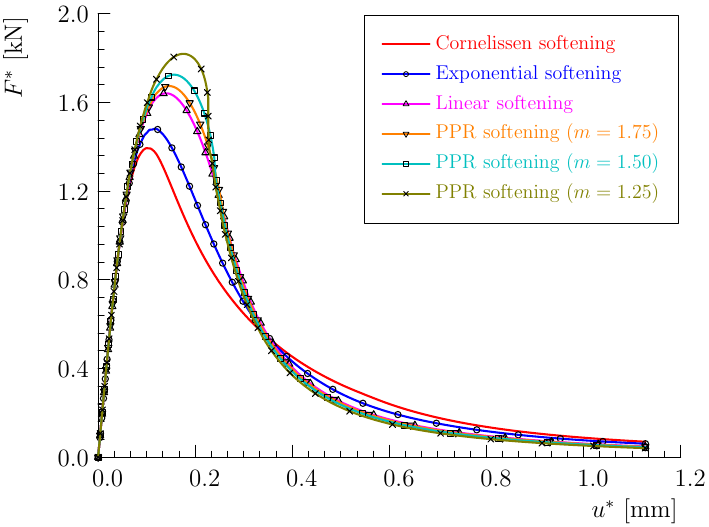}
  \caption{Three-point bending of a notched concrete beam: Applied force--displacement curves predicted by the lowest-order \texttt{$\mu$PF-CZM} (i.e., $p = 1$) with various softening laws}
  \label{fig:gpfczm-3pt-comparison}
\end{figure}

\Cref{fig:gpfczm-3pt-comparison} depicts the applied force--displacement curves predicted by the \texttt{$\mu$PF-CZM} with various softening laws. The phase-field length scale parameter $b = 2.5$ mm and the lowest-order formulation with the exponent $p = 1$ were used in the numerical simulations. As can be seen, the softening curve has significant effects on the global responses and in particular, the peak load. This result is justified from \cref{fig:gpfczm-softening-curves-comparison} --- though various softening curves have identical failure strength $f_{\text{t}}$ and fracture energy $G_{\text{f}}$, the initial slopes $k_{0}$ are totally different. Among all the six softening curves considered, the \cite{CHR1986} softening and \cite{PPR2009} softening with $m = 1.25$ have the smallest and the largest initial slopes, respectively, thus predicting the minimum and maximum load capacities. The smaller the initial slope is, the less peak load the beam sustains. This trend is consistent with that drawn from the results given by the CZM \citep{BP1997}. As the \cite{CHR1986} softening gives the closest peak load to the test data, it was adopted subsequently in this example. 

\begin{figure}[h!] \centering
  \includegraphics[width=0.55\textwidth]{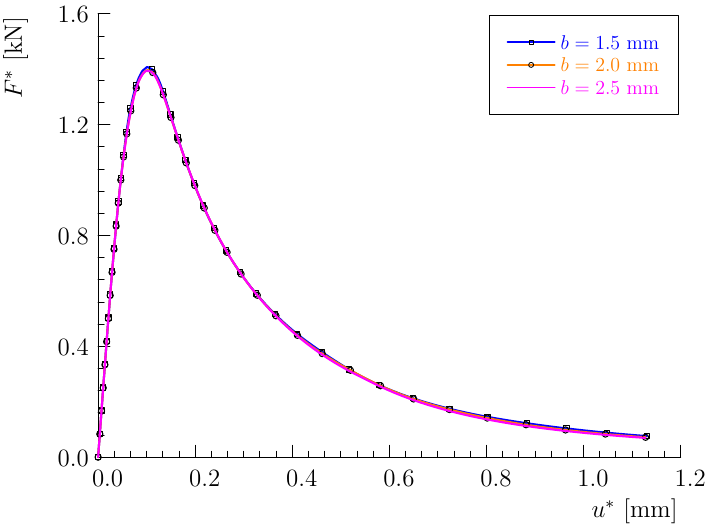}
  \caption{Three-point bending of a notched concrete beam: Applied force--displacement curves predicted by the lowest-order \texttt{$\mu$PF-CZM} (i.e., $p = 1$) with the \cite{CHR1986} softening law and various values of the phase-field length scale parameter.}
  \label{fig:gpfczm-3pt-cornelissen-2nd-lengths}
\end{figure}

\begin{figure}[h!] \centering
  \subfigure[$b = 1.5$ mm]{
  \includegraphics[width=0.3\textwidth]{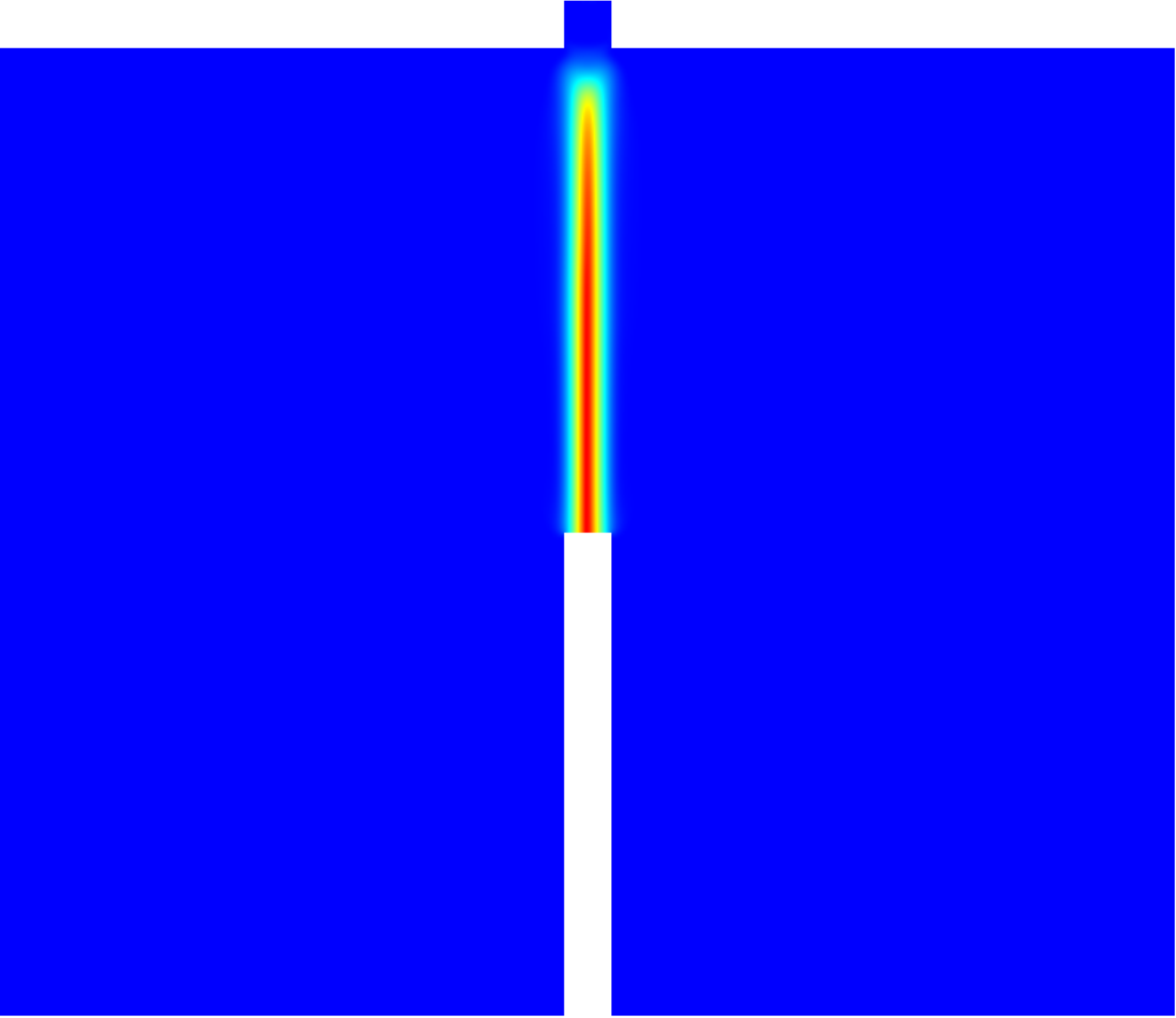}
  \label{fig:gpfczm-3pt-cornelissen-2nd-crack-b15}} \hfill
  \subfigure[$b = 2.0$ mm]{
  \includegraphics[width=0.3\textwidth]{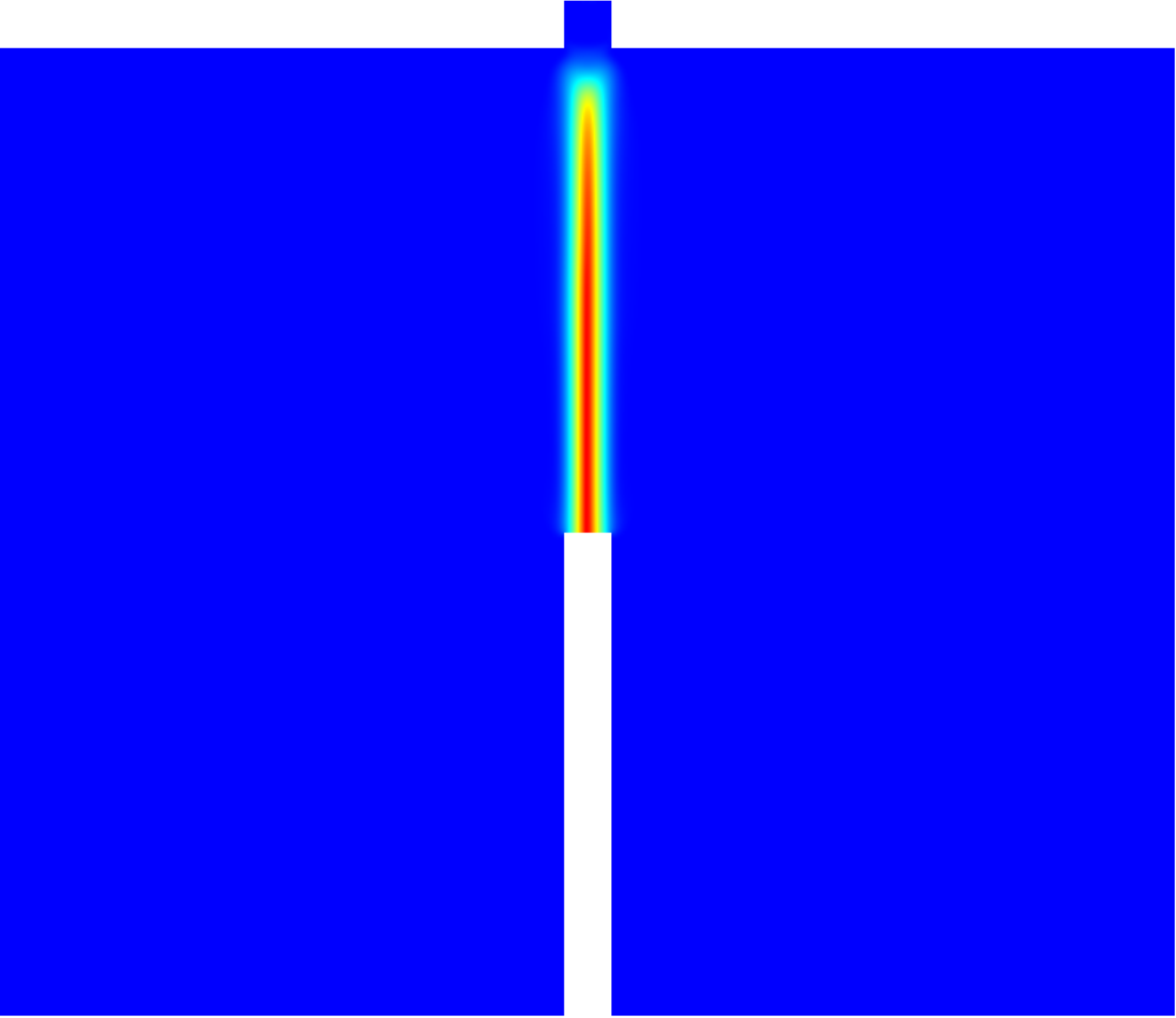}
  \label{fig:gpfczm-3pt-cornelissen-2nd-crack-b20}} \hfill
  \subfigure[$b = 2.5$ mm]{
  \includegraphics[width=0.3\textwidth]{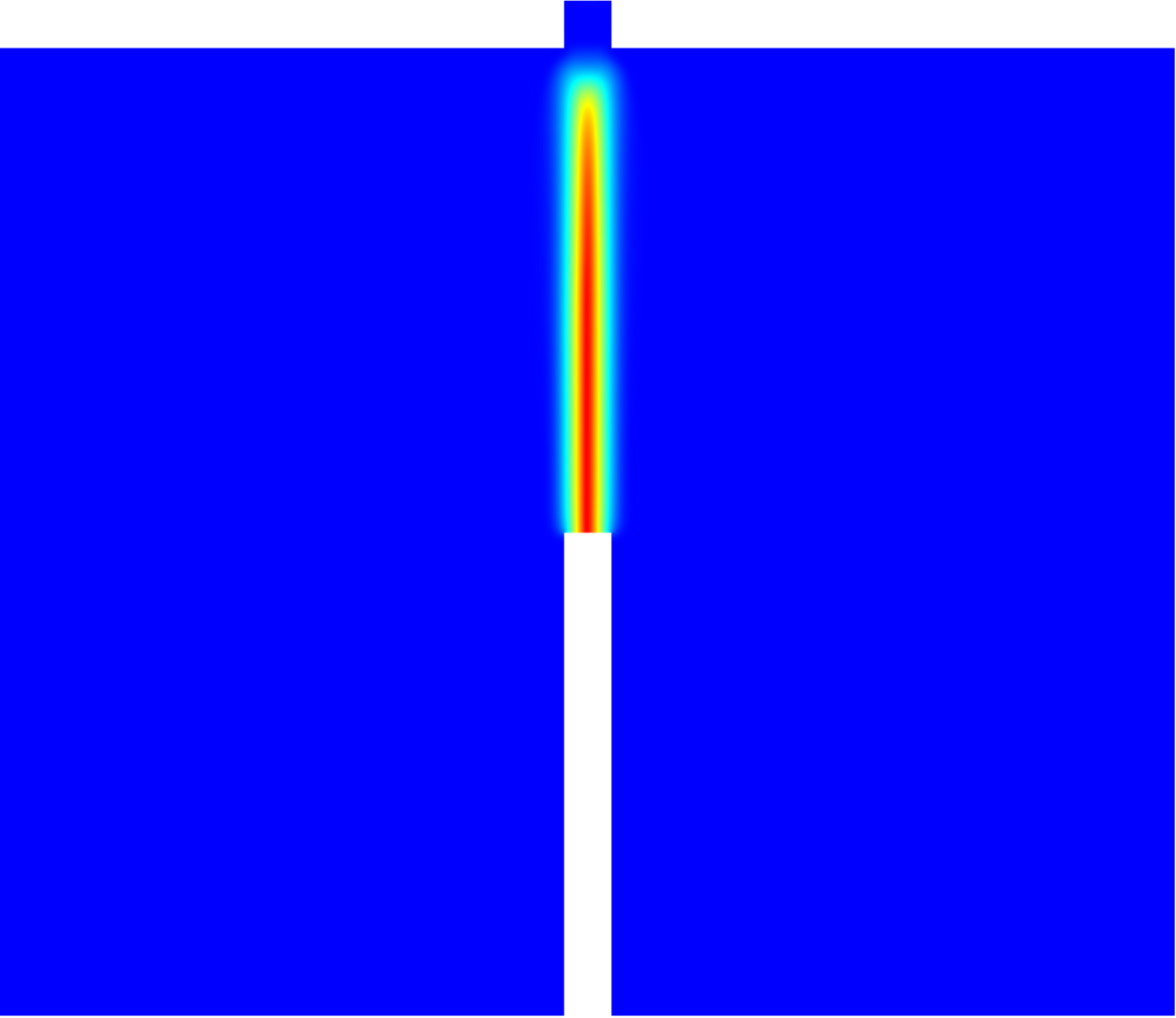}
  \label{fig:gpfczm-3pt-cornelissen-2nd-crack-b25}}  
  \caption{Three-point bending of a notched concrete beam: Ultimate crack profiles for various length scale parameters}
  \label{fig:gpfczm-3pt-cornelissen-2nd-cracks-lengths}  
\end{figure}

Regarding the length scale insensitivity of the proposed \texttt{$\mu$PF-CZM}, three values of the length scale parameter, i.e., $b = 1.5$ mm, 2.0 mm and 2.5 mm, respectively, were considered in the numerical simulations. The predicted force--displacement curves are shown in \Cref{fig:gpfczm-3pt-cornelissen-2nd-lengths}. As can be seen, so long as the length scale parameter $b$ is small enough compared to the characteristic size $L$ (e.g., the span in this example) of the structure, it has negligible effects on the global response. Moreover, the length scale parameter does not affect the crack pattern though it indeed determines the crack bandwidth during the failure process; see \cref{fig:gpfczm-3pt-cornelissen-2nd-cracks-lengths}. In the sense of global responses, the proposed \texttt{$\mu$PF-CZM} is insensitive to the phase-field length scale parameter $b \ll L$. %since it converges to the \cite{Barenblatt1959} CZM for a vanishing value $b \to 0$. 

\begin{figure}[htbp] \centering
  \includegraphics[width=0.55\textwidth]{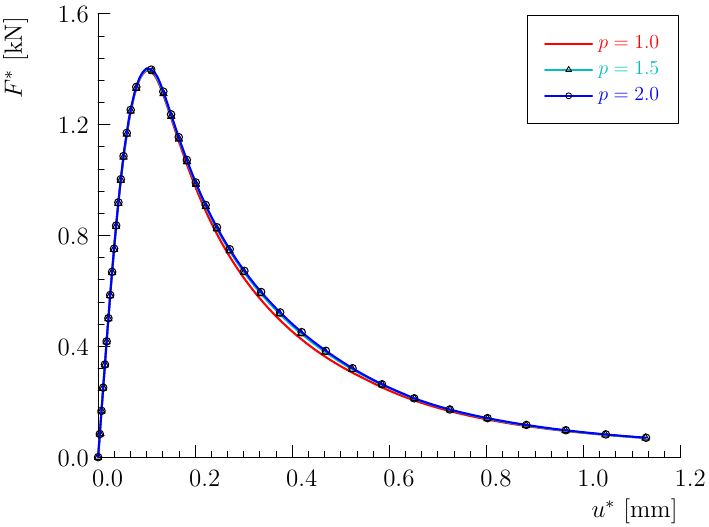}
  \caption{Three-point bending of a notched concrete beam: Applied force--displacement curves predicted by the \texttt{$\mu$PF-CZM} with various traction order parameters $p \ge 1$. }
  \label{fig:gpfczm-3pt-cornelissen-orders}
\end{figure}

\begin{figure}[h!] \centering
  \subfigure[$p = 1.0$]{
  \includegraphics[width=0.3\textwidth]{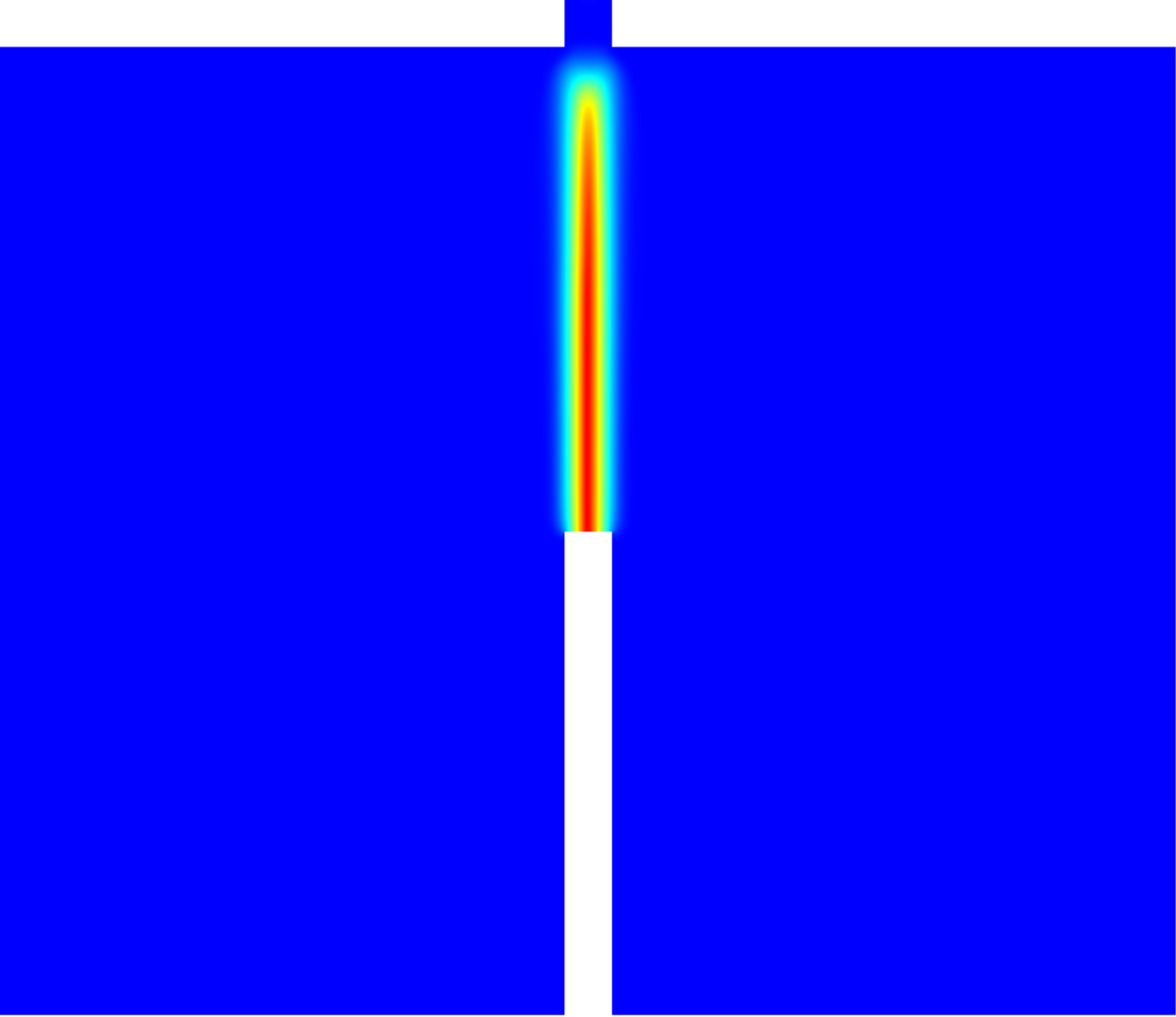}
  \label{fig:gpfczm-3pt-cornelissen-2nd-crack-b25-p1}} \hfill
  \subfigure[$p = 1.5$]{
  \includegraphics[width=0.3\textwidth]{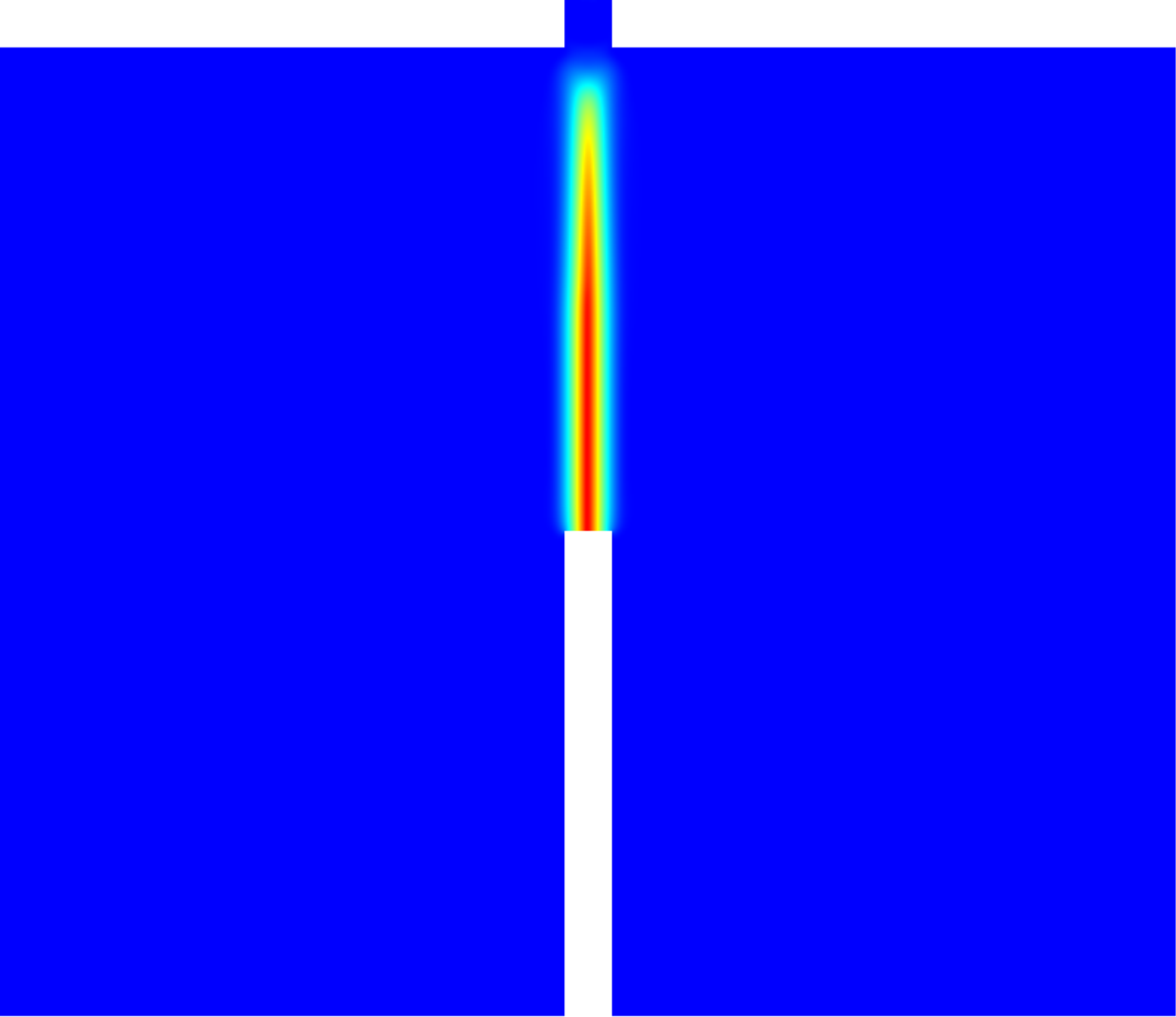}
  \label{fig:gpfczm-3pt-cornelissen-2nd-crack-b25-p15}} \hfill
  \subfigure[$p = 2.0$]{
  \includegraphics[width=0.3\textwidth]{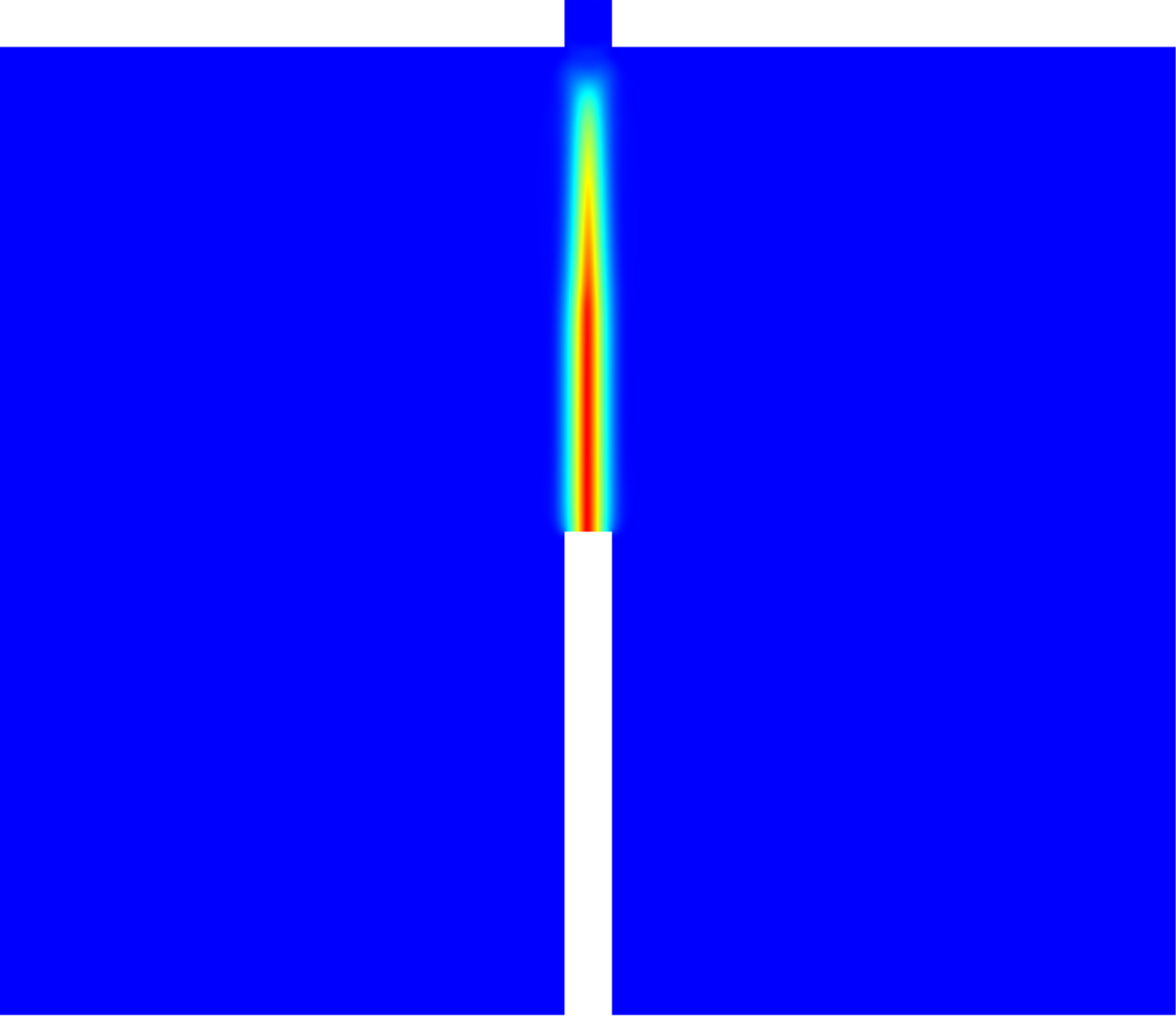}
  \label{fig:gpfczm-3pt-cornelissen-2nd-crack-b25-p2}}
  \caption{Three-point bending of a notched concrete beam: Crack profiles predicted by the \texttt{$\mu$PF-CZM} for various traction order parameters $p \ge 1$}
  \label{fig:gpfczm-3pt-cornelissen-2nd-crack-b25-orders}  
\end{figure}

\cref{fig:gpfczm-3pt-cornelissen-orders} compares the applied force---displacement curve predicted by the \texttt{$\mu$PF-CZM} with various traction order parameters $p = 1.0$, 1.5 and 2.0, respectively, all using the same length scale parameter $b = 2.5$ mm. As can be seen, the global responses almost coincide, showing that the proposed \texttt{$\mu$PF-CZM} is insensitive to the traction order parameter $p \ge 1$. The negligible discrepancy is due to the increasing crack bandwidth of the higher-order \texttt{$\mu$PF-CZM} with the exponent $p > 1$ compared to the constant one of the lowest-order counterpart with $p = 1$. However, the traction order parameter $p \ge 1$ affects a bit the crack profile shown in \cref{fig:gpfczm-3pt-cornelissen-2nd-crack-b25-orders}. At the crack tail, the solid is completely fractured (i.e., $d_{\ast} = 1$) such that the ultimate crack bandwidth is identical as $D_{\text{u}} = \frac{1}{2} \pi b$. Comparatively, around the crack tip the solid is only partially fractured with $d_{\ast} < 1$ and the crack bandwidth $D (d_{\ast})$ is smaller than $D_{\text{u}}$ --- the larger the exponent $p > 1$ is, the smaller the crack bandwidth is. Therefore, a larger exponent $p \ge 1$ results in a more sharp crack tip. This property is favored when determining the position of the crack tip is needed (e.g., in calculation of the crack tip velocity). An exception is the lowest-rder \texttt{$\mu$PF-CZM} with the exponent $p = 1$ in which the crack bandwidth $D (d_{\ast})$ is constant and independent of the maximum value $d_{\ast}$ of the crack phase-field. 

\begin{figure}[h!] \centering
  \subfigure[CMOD = 0.50 mm]{
  \includegraphics[width=0.3\textwidth]{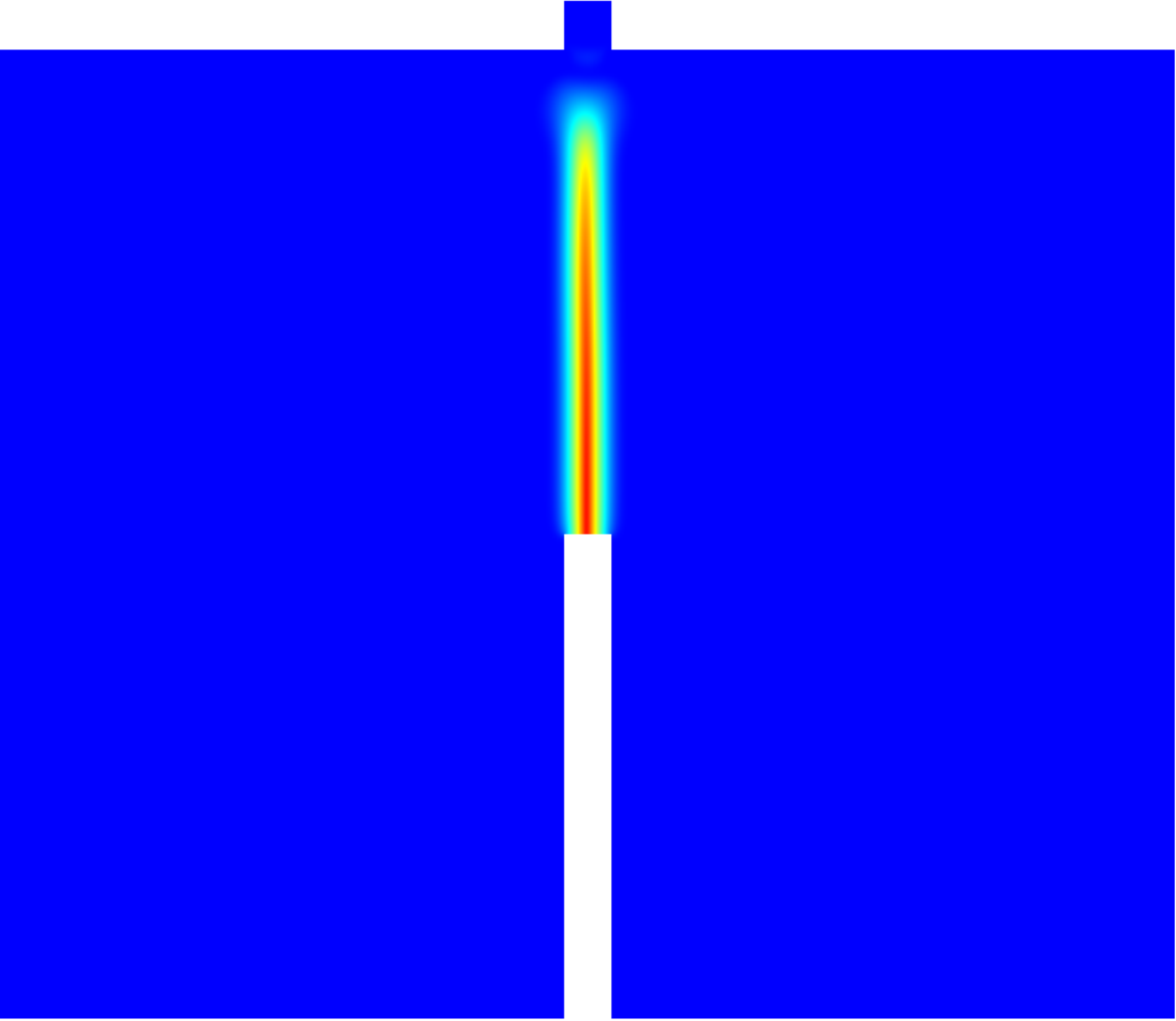}
  \label{fig:gpfczm-3pt-crack-noconstraint-cmod05}} \hfill
  \subfigure[CMOD = 0.85 mm]{
  \includegraphics[width=0.3\textwidth]{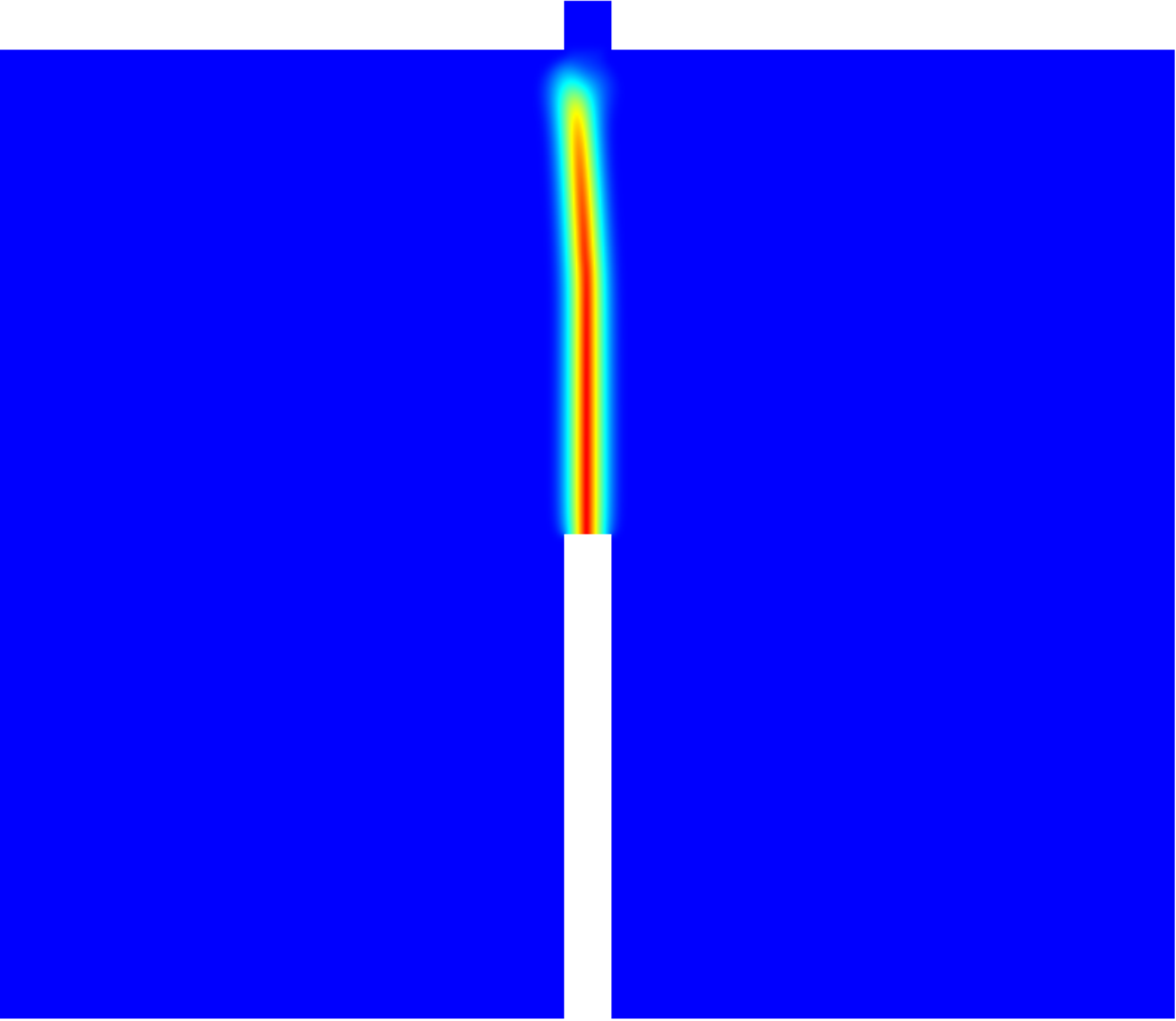}
  \label{fig:gpfczm-3pt-crack-noconstraint-cmod08}} \hfill
  \subfigure[CMOD = 1.00 mm]{
  \includegraphics[width=0.3\textwidth]{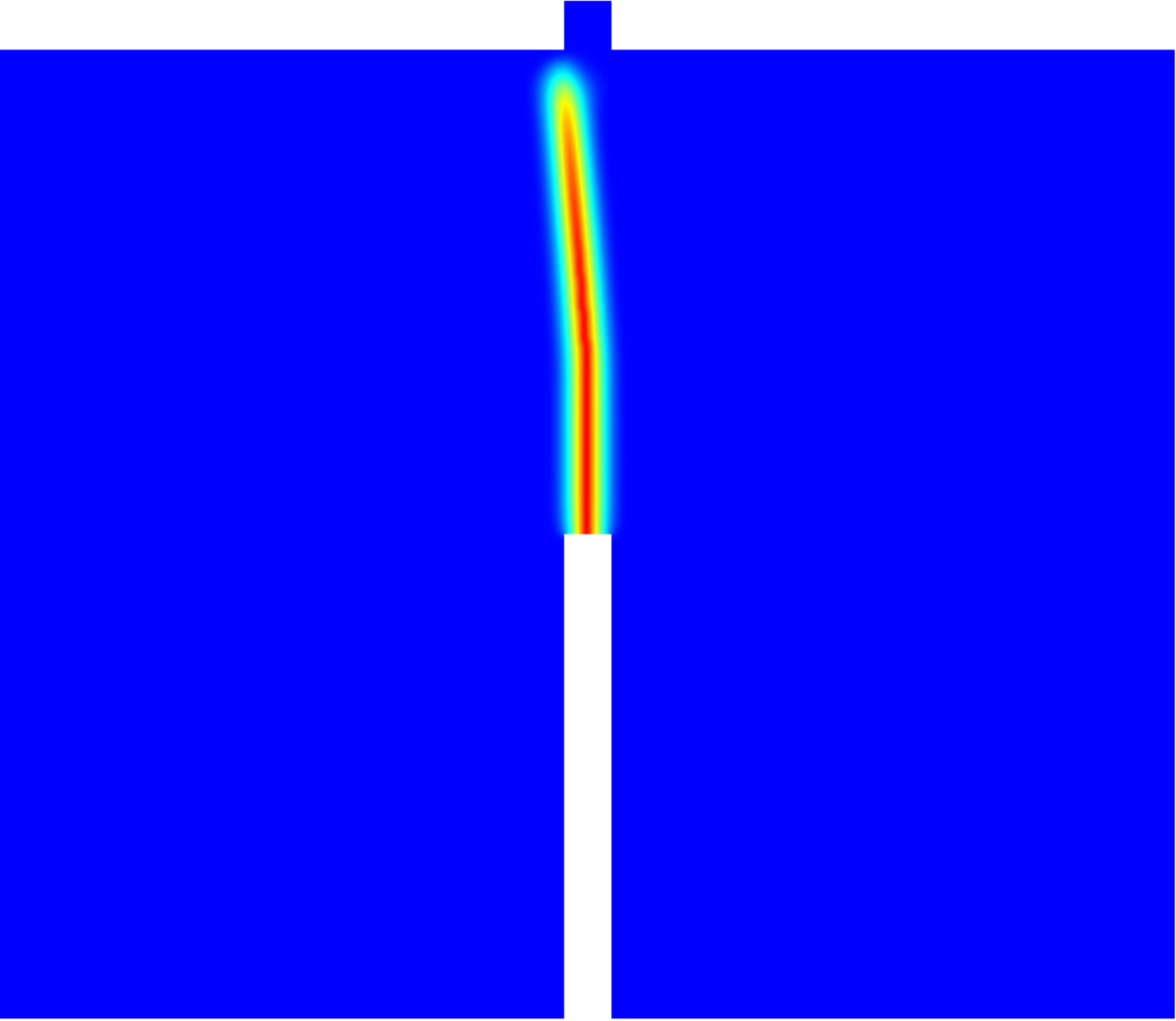}
  \label{fig:gpfczm-3pt-crack-noconstraint-cmod10}}
  \caption{Three-point bending of a notched concrete beam: Evolution of the crack phase-field at various time instants. If the crack irreversibility is not enforced, spurious crack healing occurs.}
  \label{fig:gpfczm-3pt-crack-noconstraint}  
\end{figure}

The above numerical results were obtained with the crack irreversibility condition $\dot{d} \ge 0$ enforced by the bound constraints \eqref{eq:irreversibility-condition-numerical}. If this strategy is removed the predicted crack path does not propagate vertically upward such that the increasing monotonicity $\dot{d}_{\ast} \ge 0$ cannot be guaranteed along the center line of the crack band. As shown in \cref{fig:gpfczm-3pt-crack-noconstraint}, in this 2D problem even if the non-shrinking crack bandwidth is guaranteed, the crack irreversibility is still violated and some previously cracked region is healed spuriously.

\begin{figure}[h!] \centering
  \includegraphics[width=0.55\textwidth]{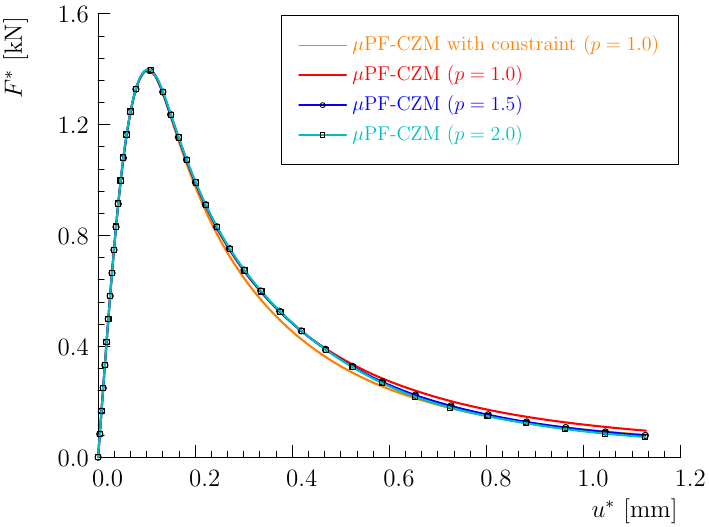}
  \caption{Three-point bending of a notched concrete beam: Applied force--displacement curves predicted by the \texttt{$\mu$PF-CZM} with various traction order parameters $p \ge 1$. No strategy was introduced to deal with the crack irreversibility condition $\dot{d} \ge 0$.}
  \label{fig:gpfczm-3pt-cornelissen-2nd-irreversibility}
\end{figure}

In order to address this issue, only one half of the structure is modeled in the numerical simulations, with symmetric boundary conditions enforced \textit{a priori} such that the crack propagates vertically upward. Without any special strategy introduced to deal with the crack irreversibility, the applied force--displacement curves predicted by the proposed \texttt{$\mu$PF-CZM} with different values of the traction order parameter $p \ge 1$ are depicted in \cref{fig:gpfczm-3pt-cornelissen-2nd-irreversibility}. For the sake of comparison, the numerical result given by the \texttt{$\mu$PF-CZM} with the bound constraints \eqref{eq:irreversibility-condition-numerical} was also presented. As can be seen, now the spurious violation of crack irreversibility is almost removed so long as the increasing monotonicity $\dot{d}_{\ast} > 0$ is guaranteed along the symmetric line. However, for other problems in which the crack path cannot be determined \textit{a priori}, it is necessary to enforce the crack irreversibility condition $\dot{d} \ge 0$ without affecting the expected traction--separation softening law. Therefore, the condition \eqref{eq:non-shrinking-crack-band} or \eqref{eq:irreversibility-condition} is indispensable in such cases.

\subsection{Single-edge notched plate}

Let us now consider the benchmark problem of a single-edge notched plate \citep{MHW2010c}. 

\begin{figure}[h!] \centering
  \subfigure[Tension test]{
  \includegraphics[width=0.4\textwidth]{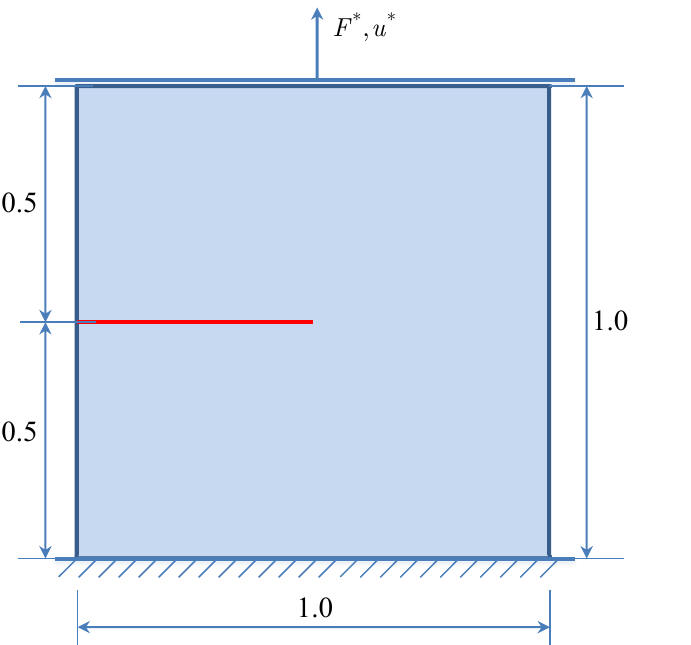}
  \label{fig:SENP-tension-problem}} \hspace{5mm}
  \subfigure[Shear test]{
  \includegraphics[width=0.4\textwidth]{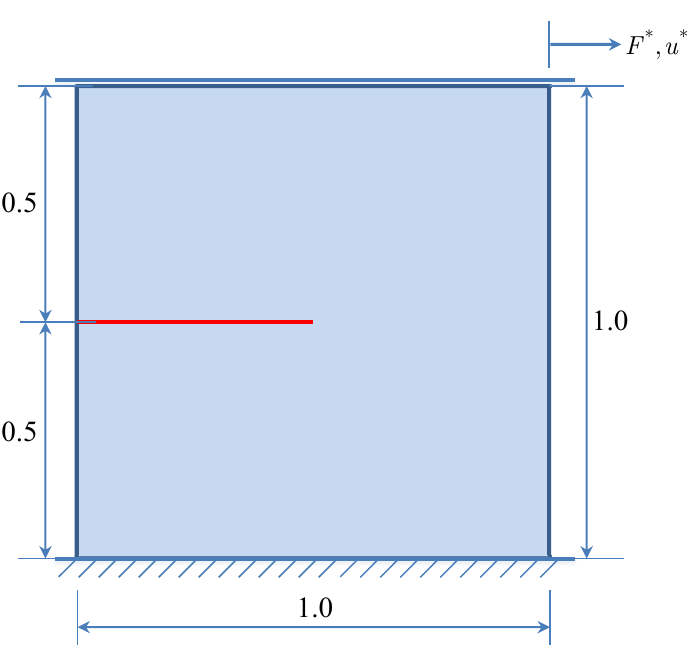}
  \label{fig:SENP-shear-problem}}
  \caption{Single-edge notched plate: Geometry (unit of length: mm), boundary and loading conditions.}
  \label{fig:SENP-problems}
\end{figure}
  
As shown in Figure \ref{fig:SENP-problems}, it is a square plate of length 1 mm, with a unit out-of-plane thickness. A straight horizontal notch of length 0.5 mm is introduced at the mid-height of the left edge. The bottom edge is fixed, while a vertical displacement for the tension test and a horizontal one for the shear test are applied to the top edge, respectively. 

The following material parameters were adopted: Young's modulus $E_{0} = 2.1 \times 10^{5}$ MPa, Poisson's ratio $\nu_{0} = 0.3$, the failure strength $f_{\text{t}} = 2000$ MPa and the fracture energy $G_{\text{f}} = 2.7$ N/mm, resulting in an Irwin internal length $l_{\text{ch}} = 0.095$ mm. As in the original \texttt{PF-CZM} for brittle fracture \citep{WN2018}, linear softening was adopted in the proposed \texttt{$\mu$PF-CZM}. 

In the numerical simulations, two values of the phase-field length scale parameter, i.e., $b = 0.005$ mm and $b = 0.010$ mm, were considered. The mesh size was fixed as $h = 0.001$ mm such that both length scale parameters can be sufficiently resolved. %The initial notch is modeled by the discretized mesh, though it can also be considered by the Dirichlet boundary condition $d (\boldsymbol{x}) = 1$ applied to those nodes on the initial notch.

\subsubsection{Tension test}

\begin{figure}[h!] \centering
  \subfigure[$b = 0.005$ mm]{
  \includegraphics[width=0.3\textwidth]{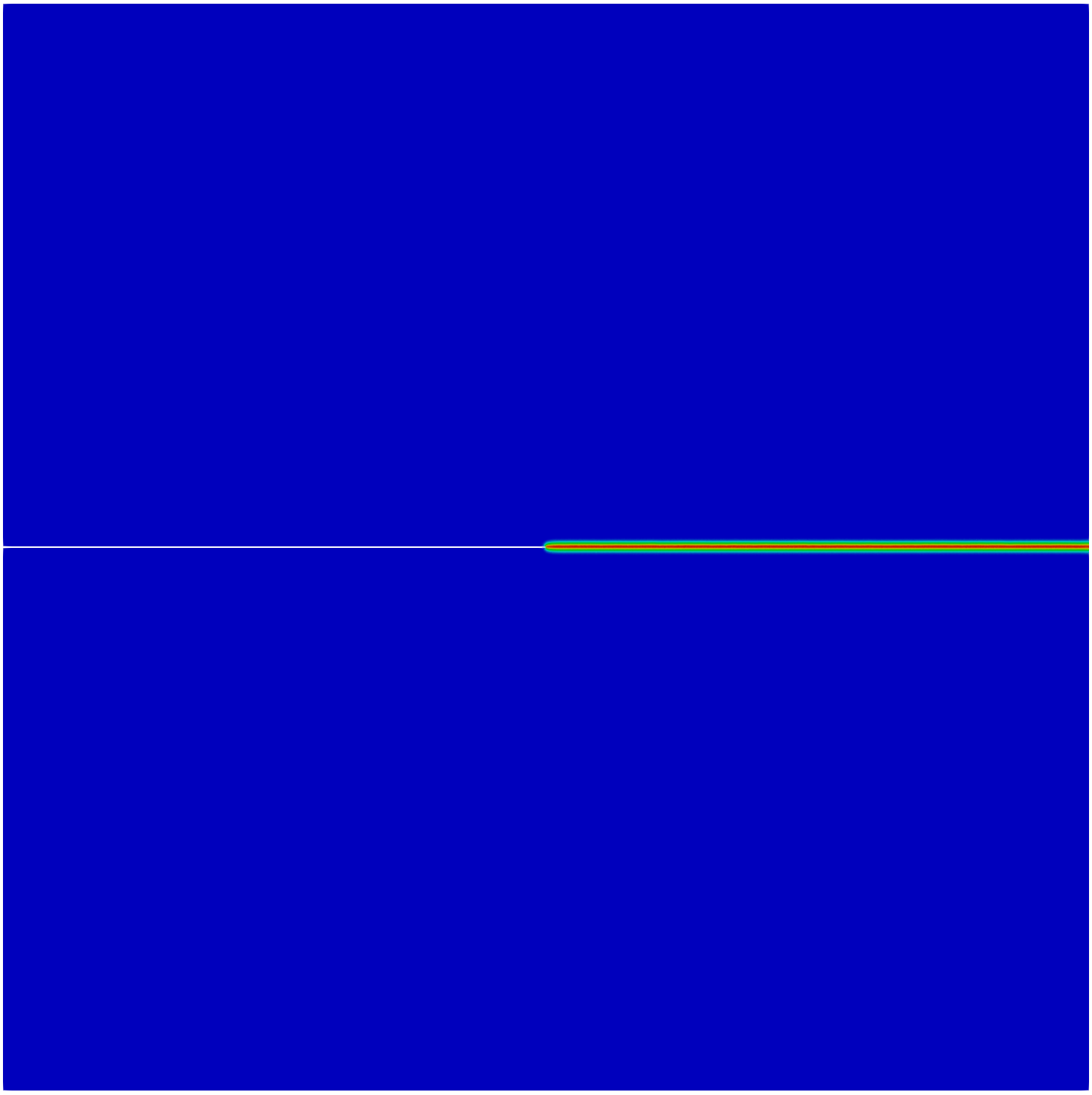}
  \label{fig:SENP-tension-crack-final-b005}} \hspace{5mm}
  \subfigure[$b = 0.010$ mm]{
  \includegraphics[width=0.3\textwidth]{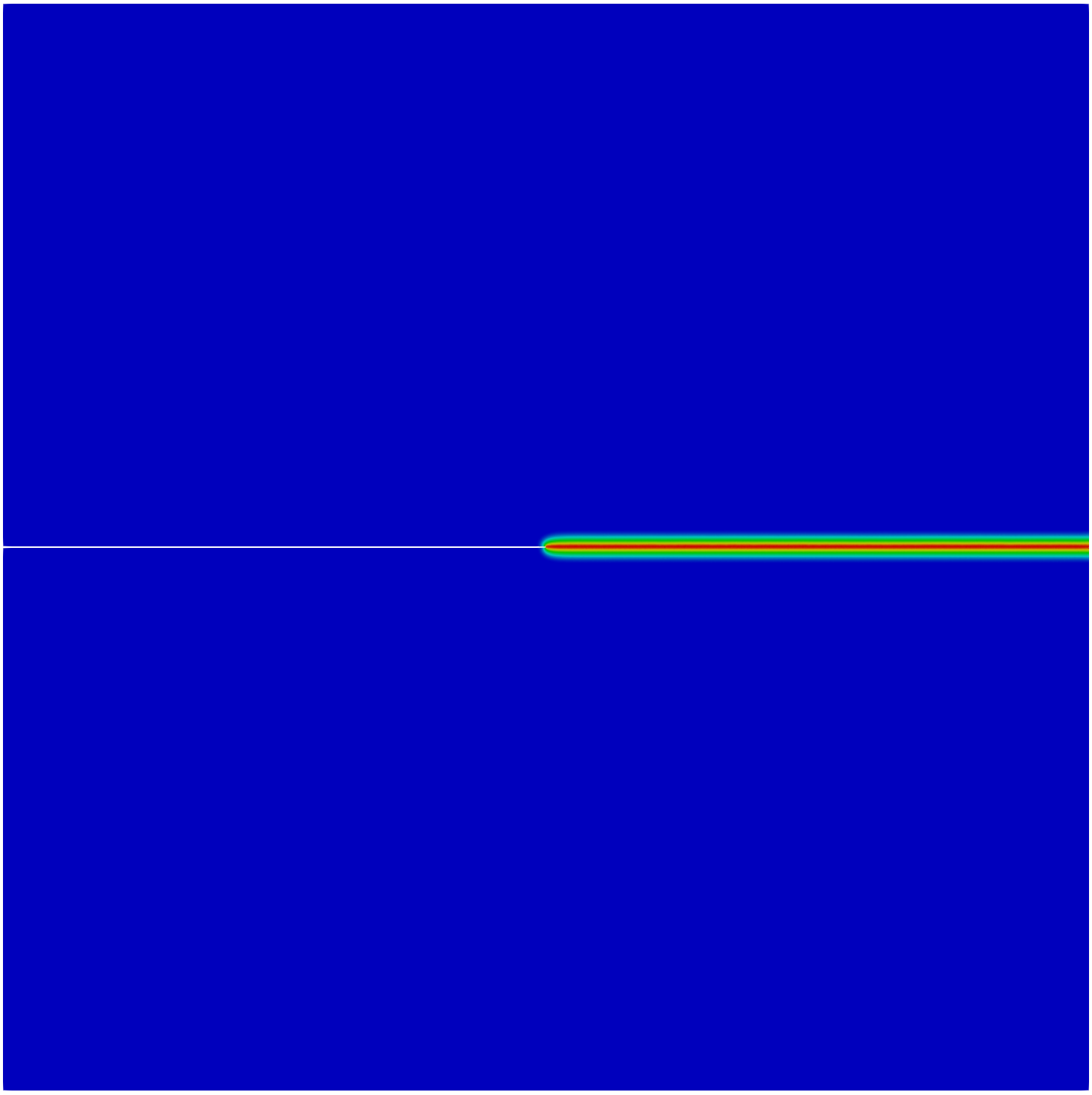}
  \label{fig:SENP-tension-crack-final-b010}}   
  \caption{Single-edge notched plate under tension: Ultimate crack profiles at displacement $u^{\ast} = 0.006$ mm.}
  \label{fig:SENP-tension-cracks-final-lengths}
\end{figure}

The tension test is first discussed. \Cref{fig:SENP-tension-cracks-final-lengths} depicts the crack patterns at displacement $u^{\ast} = 0.006$ mm (complete fracture) given by the \texttt{$\mu$PF-CZM}. Note that the ultimate crack profiles for various traction order parameters $p \ge 1$ coincide. For both values of the length scale parameter, the crack propagates horizontally to the right edge. As expected, the crack bandwidth is affected by (proportional to) the length scale parameter $b$. 

\begin{figure}[h!] \centering
  \subfigure[$p = 1.0$]{
  \includegraphics[width=0.475\textwidth]{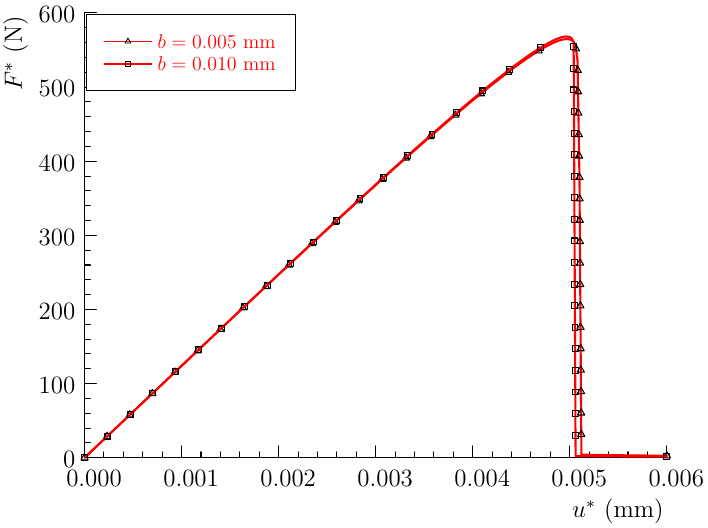}
  \label{fig:SENP-tension-load-disp-lengths-p10}} \hfill
  \subfigure[$p = 1.5$]{
  \includegraphics[width=0.475\textwidth]{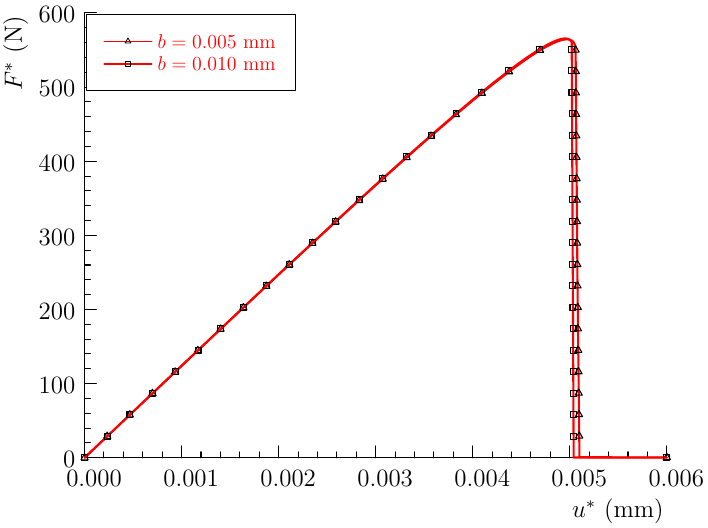}
  \label{fig:SENP-tension-load-disp-lengths-p15}}\\
  \subfigure[$p = 2.0$]{
  \includegraphics[width=0.475\textwidth]{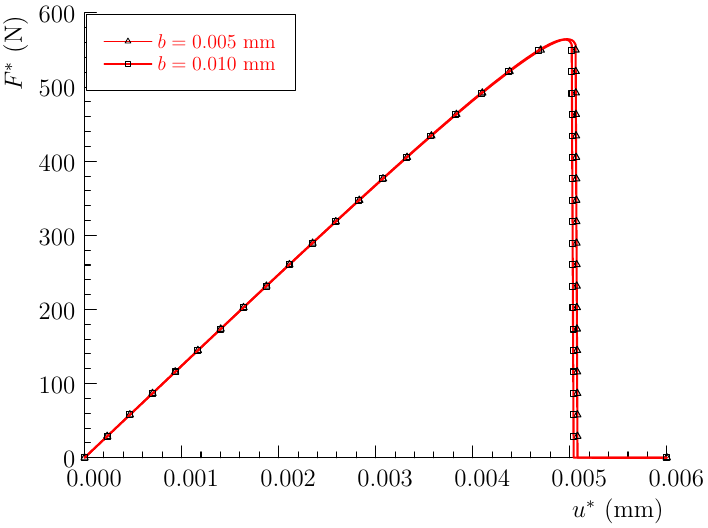}
  \label{fig:SENP-tension-load-disp-lengths-p20}}
  \caption{Single-edge notched plate under tension: Numerical force--displacement curves for various traction order parameters.}
  \label{fig:SENP-tension-load-disp-lengths}  
%\end{figure}
%\begin{figure}[t!] \centering

  \subfigure[$b = 0.005$ mm]{
  \includegraphics[width=0.475\textwidth]{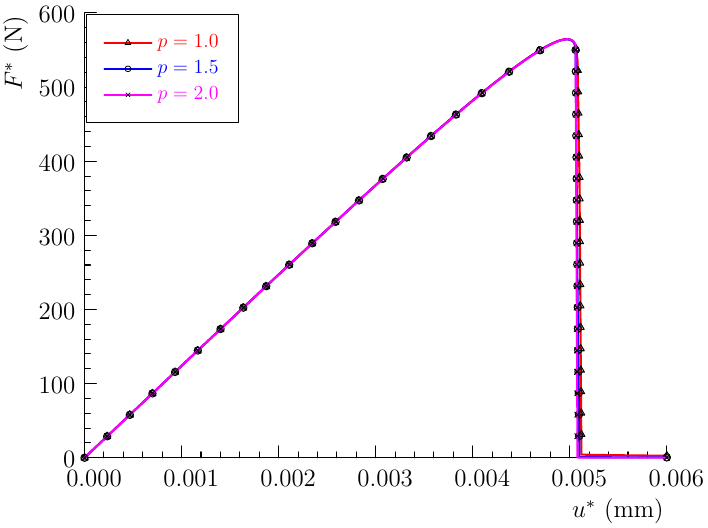}
  \label{fig:SENP-tension-load-disp-orders-b005}} \hfill
  \subfigure[$b = 0.010$ mm]{
  \includegraphics[width=0.475\textwidth]{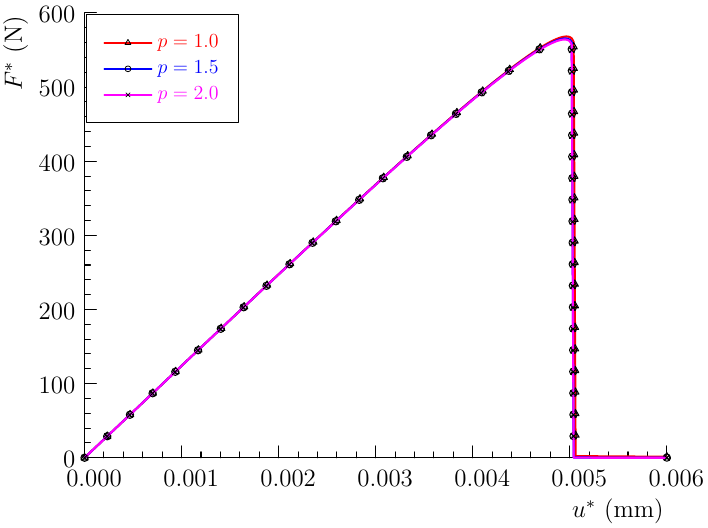}
  \label{fig:SENP-tension-load-disp-orders-b010}}
  \caption{Single-edge notched plate under tension: Numerical force--displacement curves for various length scale parameters.}
  \label{fig:SENP-tension-load-disp-orders}
\end{figure}

For various phase-field length scale parameters, the predicted force--displacement curves are shown in \cref{fig:SENP-tension-load-disp-lengths}. As can be seen, the numerical global responses are independent of the length scale parameter for various traction order parameters $p \ge 1$. The length scale insensitivity of the original \texttt{PF-CZM} is preserved even though the latter is recovered only for the lowest-order \texttt{$\mu$PF-CZM} with the exponent $p = 1$.

\Cref{fig:SENP-tension-load-disp-orders} depicts the force--displacement curves given by the \texttt{$\mu$PF-CZM} with various traction order parameters using the length scale $b = 0.005$ mm and $b = 0.010$ mm, respectively. As expected, the numerical force--displacement curves are independent of the traction order parameter $p \ge 1$ upon the optimal geometric function \eqref{eq:optimal-geometric-function}.

\clearpage
\begin{figure}[h!] \centering
  \subfigure[$p = 1.0$]{
  \includegraphics[width=0.3\textwidth]{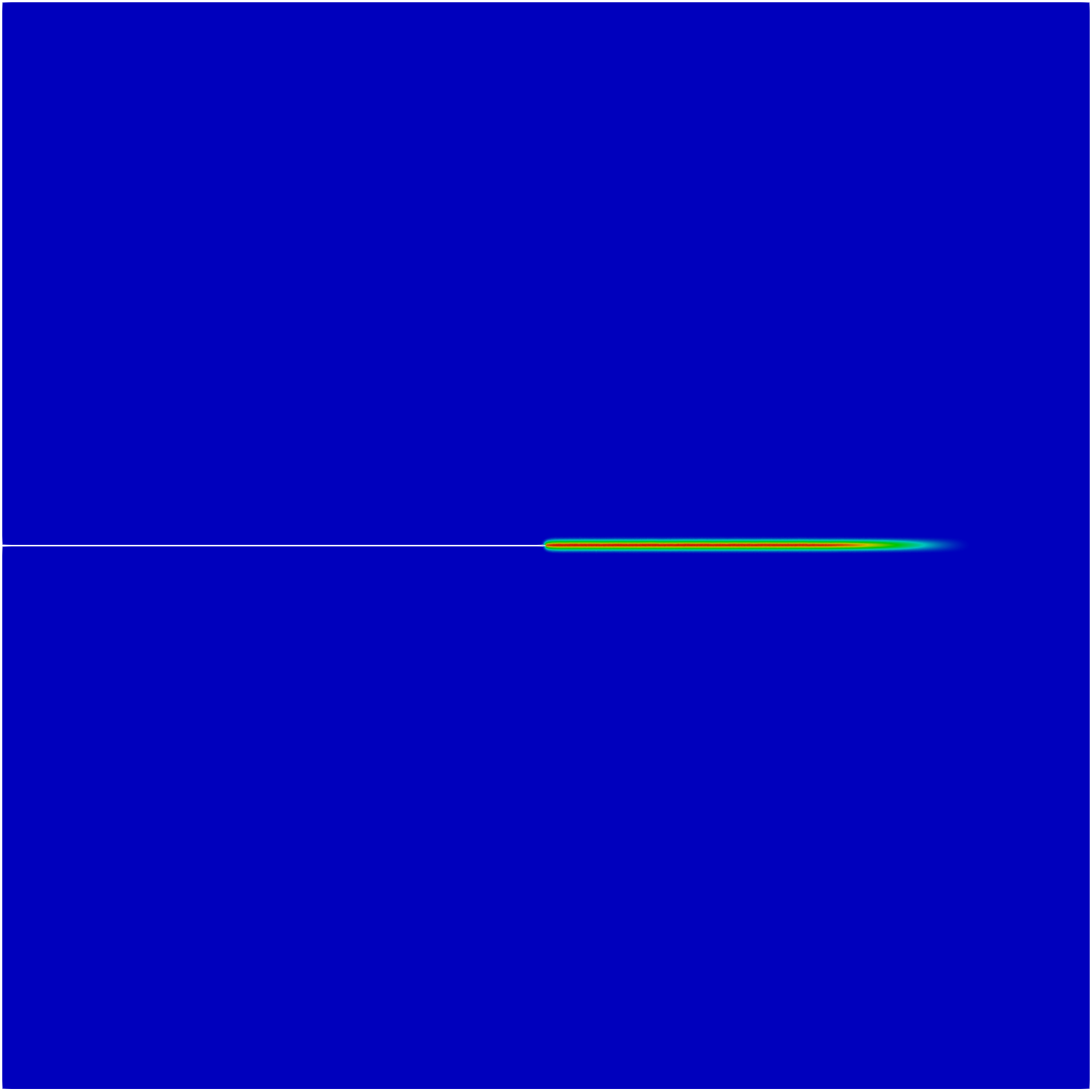}
  \label{fig:SENP-tension-crack-p10-b005}} \hfill
  \subfigure[$p = 1.5$]{
  \includegraphics[width=0.3\textwidth]{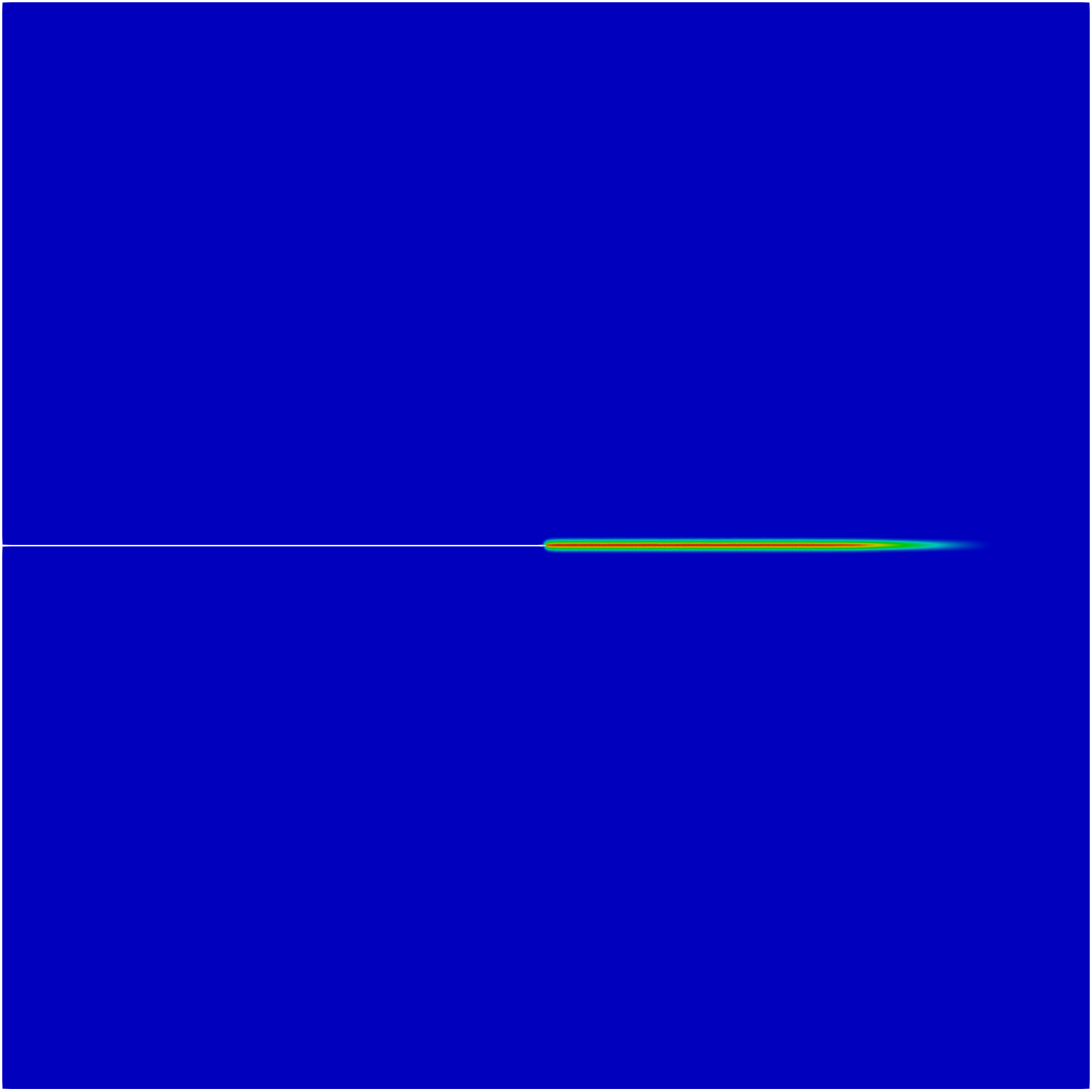}
  \label{fig:SENP-tension-crack-p15-b005}} \hfill 
  \subfigure[$p = 2.0$]{
  \includegraphics[width=0.3\textwidth]{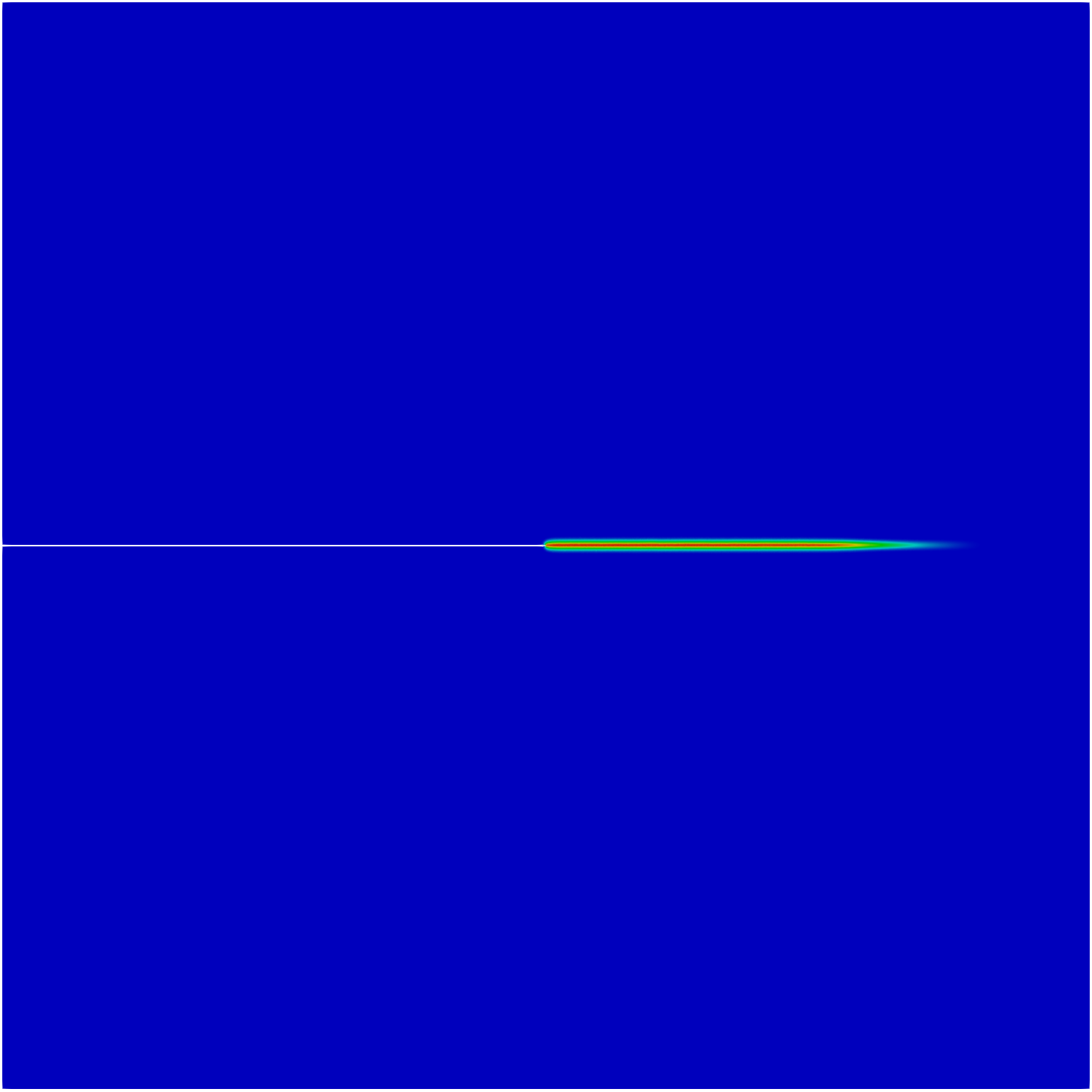}
  \label{fig:SENP-tension-crack-p20-b005}}
  \caption{Single-edge notched plate under tension: Crack profiles at the intermediate instant for various traction order parameters $p \ge 1$.}
  \label{fig:SENP-tension-crack-profiles-orders}
\end{figure}

For this problem of brittle fracture, the traction order parameter $p \ge 1$ also affects the crack profiles at the intermediate stage. As shown in \cref{fig:SENP-tension-crack-profiles-orders}, the same trend still applies --- the larger the exponent $p \ge 1$ is, the more sharp the crack tip is. However, as the crack propagates abruptly and the maximum value of the crack phase-field approaches $d_{\ast} = 1$ rather quickly once crack initiations occurs, the difference between the crack profiles given by the \texttt{$\mu$PF-CZM} with various traction order parameters $p \ge 1$ is not so significant as that for cohesive fracture.

\subsubsection{Shear test}

\begin{figure}[h!] \centering
  \subfigure[$b = 0.005$ mm]{
  \includegraphics[width=0.3\textwidth]{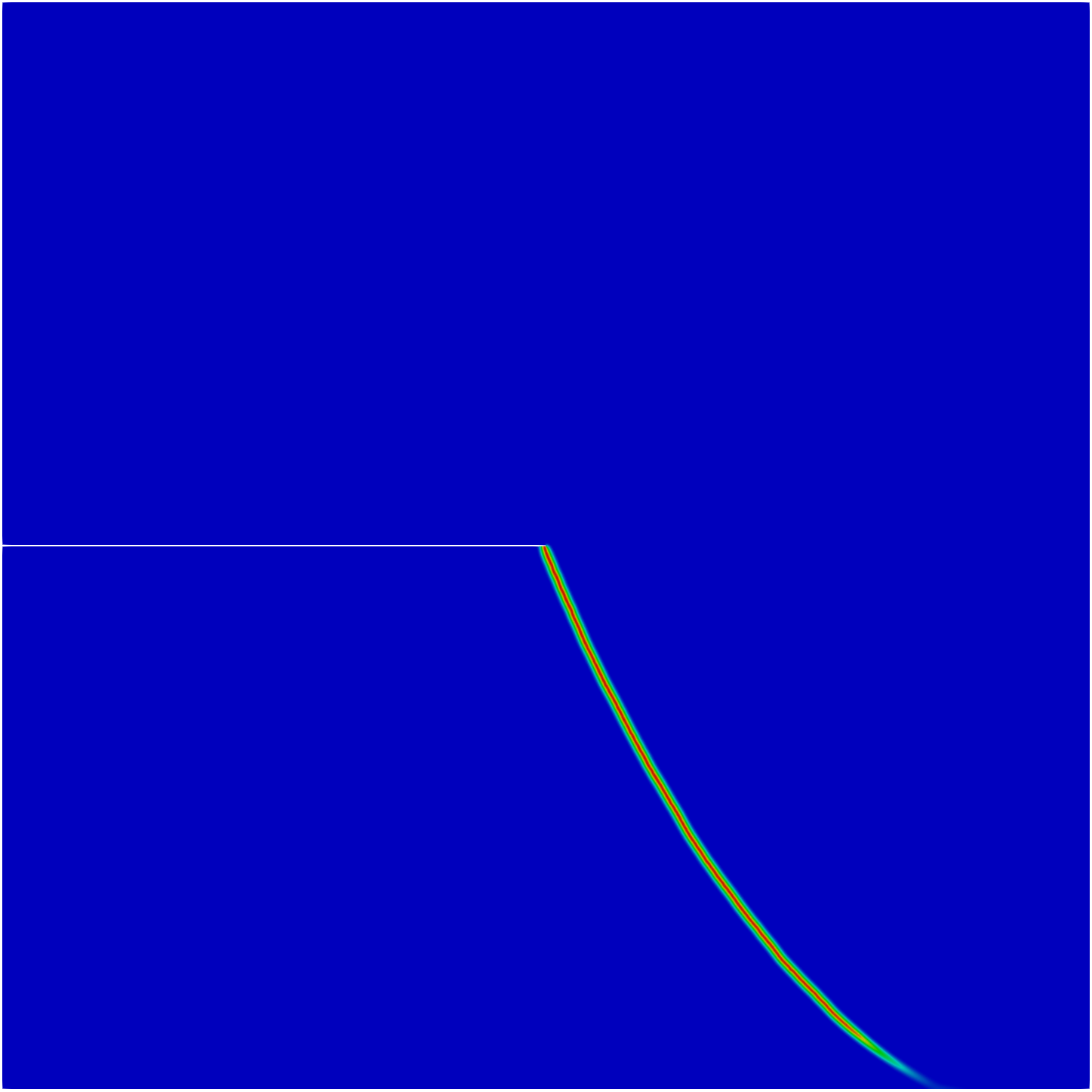}
  \label{fig:SENP-shear-crack-final-p15-b005}} \hspace{5mm}
  \subfigure[$b = 0.010$ mm]{
  \includegraphics[width=0.3\textwidth]{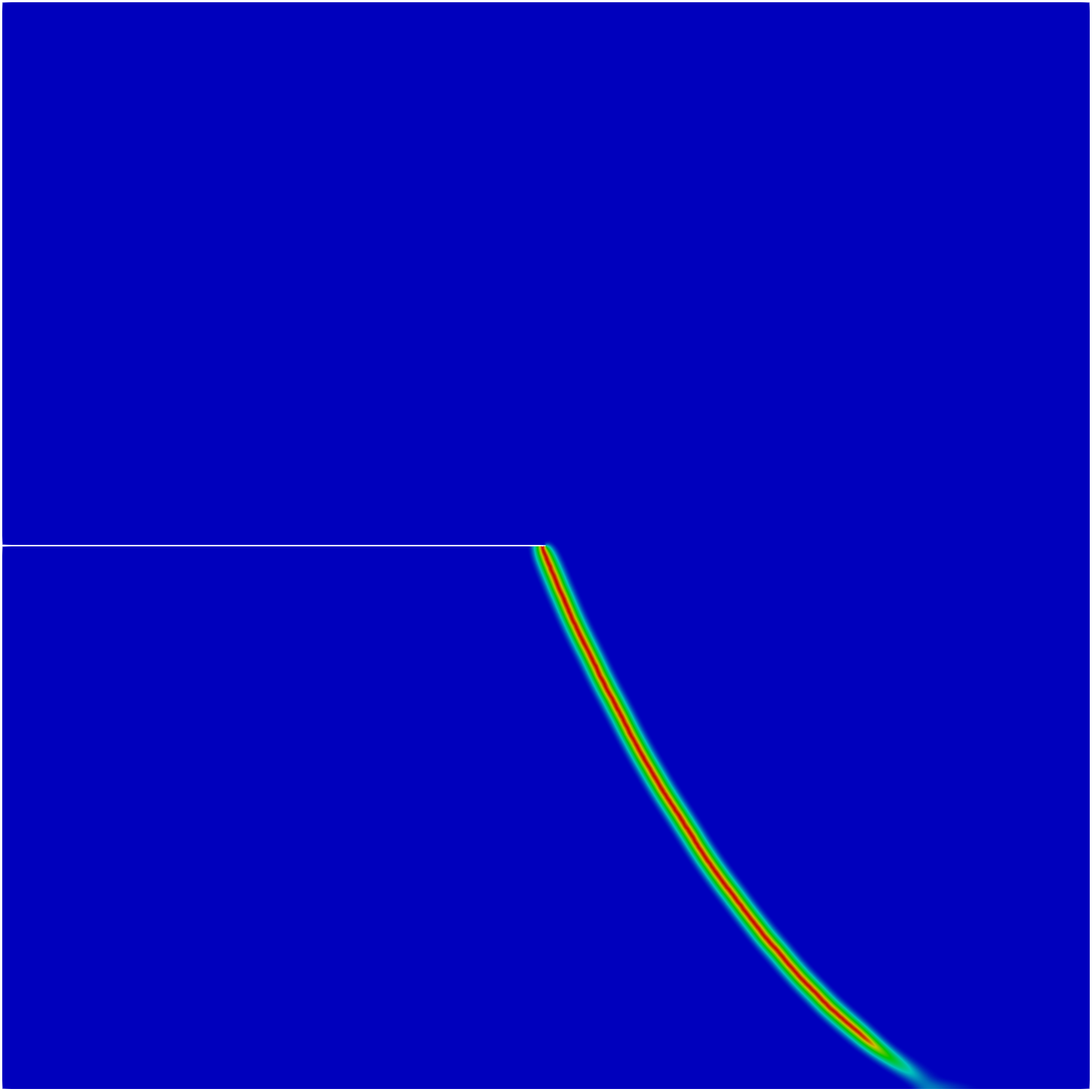}
  \label{fig:SENP-shear-crack-final-p15-b010}}   
  \caption{Single-edge notched plate under shear: Crack profiles at $u^{\ast} = 0.02$ mm predicted by the \texttt{$\mu$PF-CZM} with the exponent $p = 1.5$.}
  \label{fig:SENP-shear-cracks-final-lengths}    
\end{figure}

Regarding the shear test, the predicted crack patterns at displacement $u^{\ast} = 0.02$ mm are shown in \cref{fig:SENP-shear-cracks-final-lengths}. Similarly to those by the \texttt{AT2} model \citep{MWH2010a,AGL2015} and by the original \texttt{PF-CZM} \citep{WN2018}, the crack nucleates at the pre-notch tip and propagates downward to the right bottom corner. The length scale parameter affects only the crack bandwidth, but the crack pattern is not influenced. 

\begin{figure}[h!] \centering

  \subfigure[$p = 1.0$]{
  \includegraphics[width=0.475\textwidth]{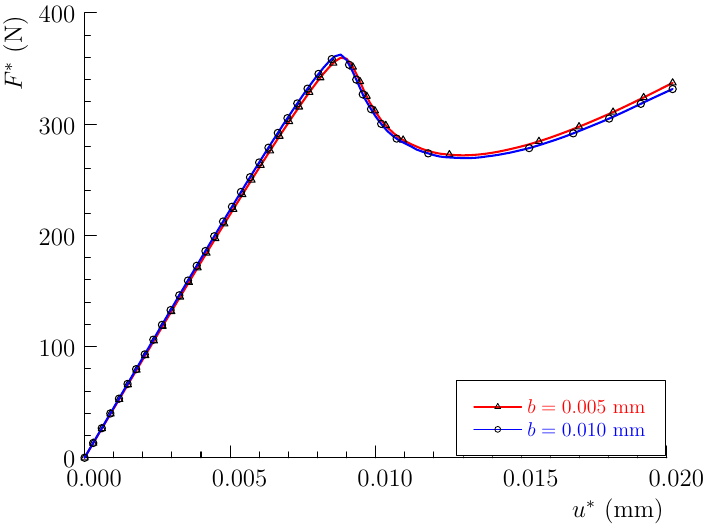}
  \label{fig:SENP-shear-load-disp-p10}} \hfill
  \subfigure[$p = 1.5$]{
  \includegraphics[width=0.475\textwidth]{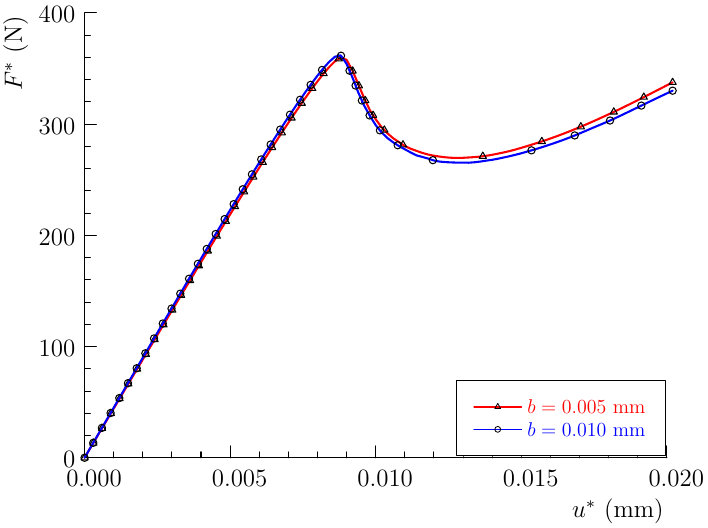}
  \label{fig:SENP-shear-load-disp-p15}}
  \subfigure[$p = 2.0$]{
  \includegraphics[width=0.475\textwidth]{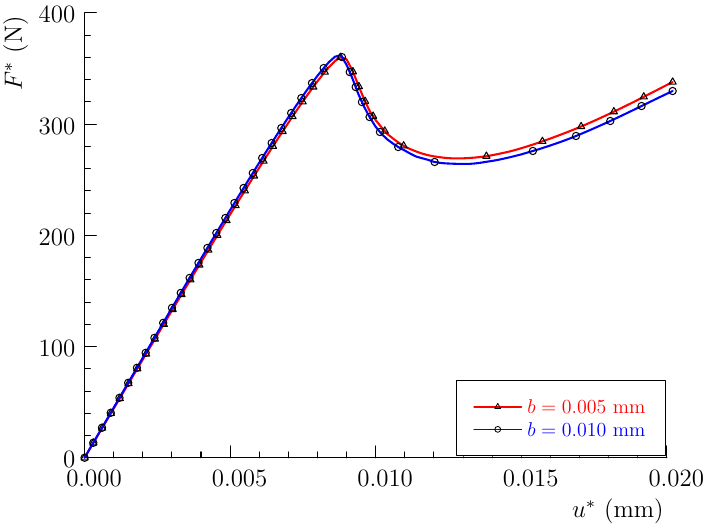}
  \label{fig:SENP-shear-load-disp-p20}}
  \caption{Single-edge notched plate under tension: Force--displacement curves for various traction order parameters.}
  \label{fig:SENP-shear-load-disp-lengths}
%\end{figure}
%
%\begin{figure}[h!] \centering

  \subfigure[$b = 0.005$ mm]{
  \includegraphics[width=0.475\textwidth]{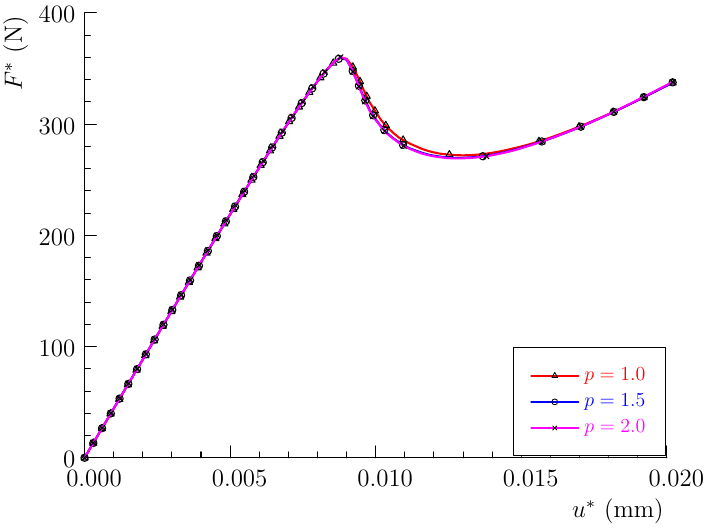}
  \label{fig:SENP-shear-load-disp-b005}} \hfill
  \subfigure[$b = 0.010$ mm]{
  \includegraphics[width=0.475\textwidth]{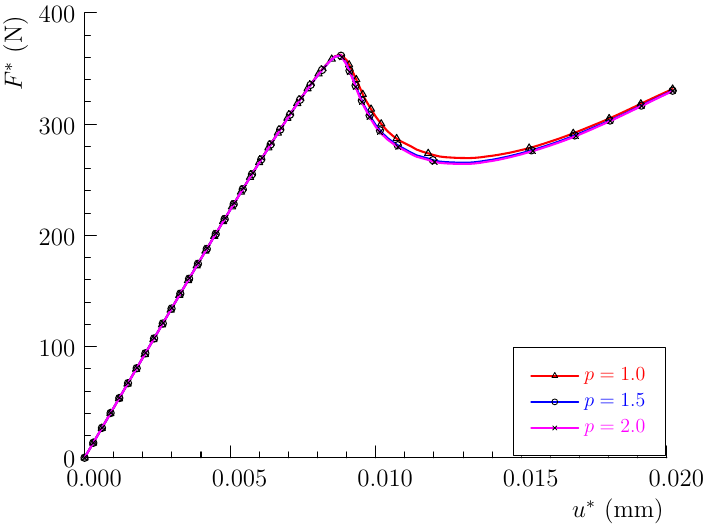}
  \label{fig:SENP-shear-load-disp-b010}}   
  \caption{Single-edge notched plate under shear: Force--displacement curves for various length scale parameters.}
  \label{fig:SENP-shear-load-disp-orders}
\end{figure}

\clearpage
For various values of the phase-field length scale parameter, \cref{fig:SENP-shear-load-disp-lengths} presents the force--displacement curves predicted by the \texttt{$\mu$PF-CZM} with the exponent $p = 1.5$. Consistent with those reported in the literature \citep{AGL2015,WN2018}, the post-peak behavior decays initially and the force increases again. As expected, the length scale parameter has negligible influences on the global numerical responses. In order words, the proposed \texttt{$\mu$PF-CZM} with various traction order parameters $p \ge 1$ is length scale insensitive as the original \texttt{PF-CZM}.

For each length scale parameter $b = 0.005$ mm and $b = 0.010$ mm, the predicted force--displacement curves with various traction order parameters are shown in \cref{fig:SENP-shear-load-disp-orders}. As expected, for brittle fracture the numerical global responses are also independent of the traction order parameter $p \ge 1$ upon the optimal geometric function \eqref{eq:optimal-geometric-function}.

\begin{figure}[htbp] \centering
  \subfigure[$p = 1.0$]{
  \includegraphics[width=0.3\textwidth]{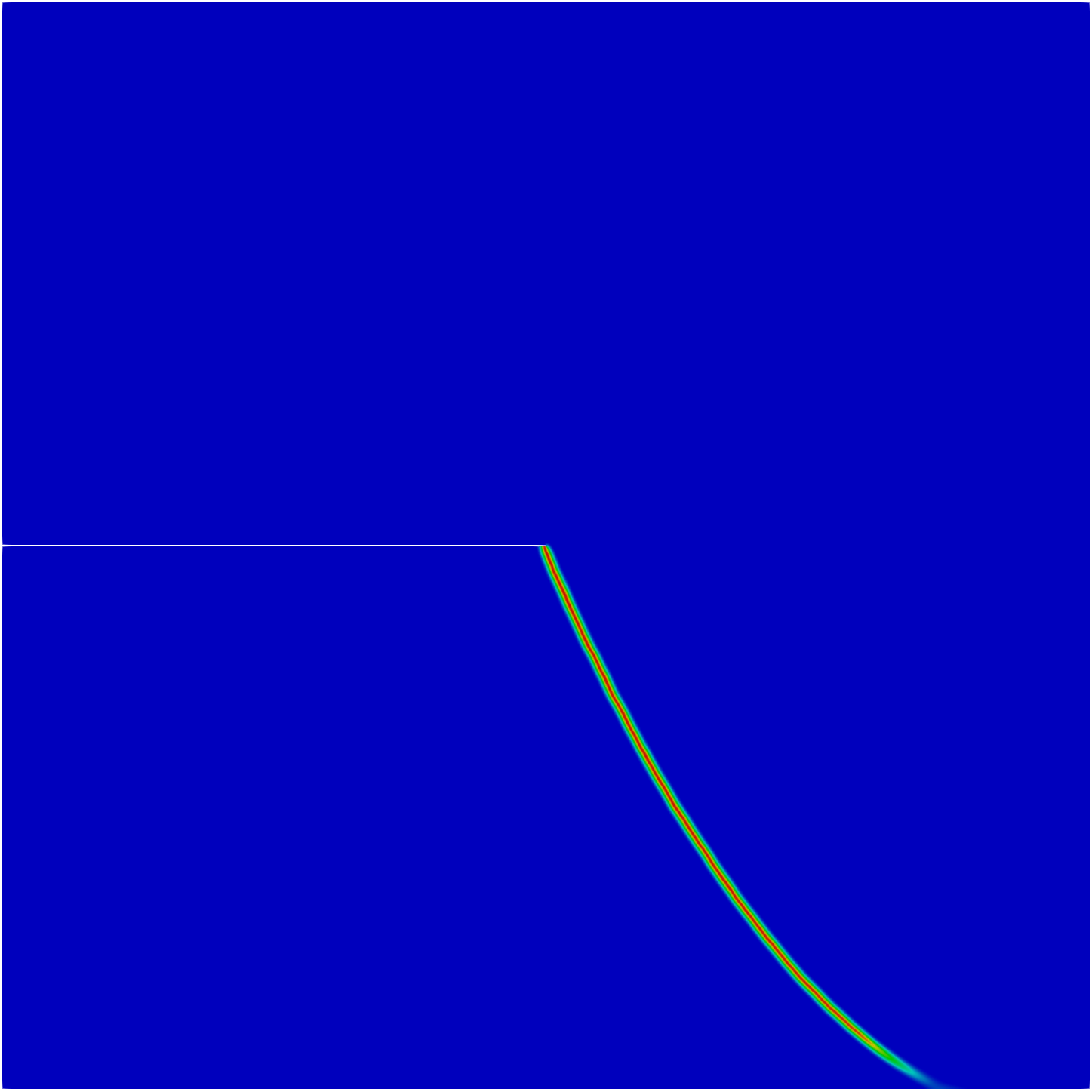}
  \label{fig:SENP-shear-crack-p10-b005}} \hfill
  \subfigure[$p = 1.5$]{
  \includegraphics[width=0.3\textwidth]{SENP-shear-crack-p15-b005.png}
  \label{fig:SENP-shear-crack-p15-b005}} \hfill 
  \subfigure[$p = 2.0$]{
  \includegraphics[width=0.3\textwidth]{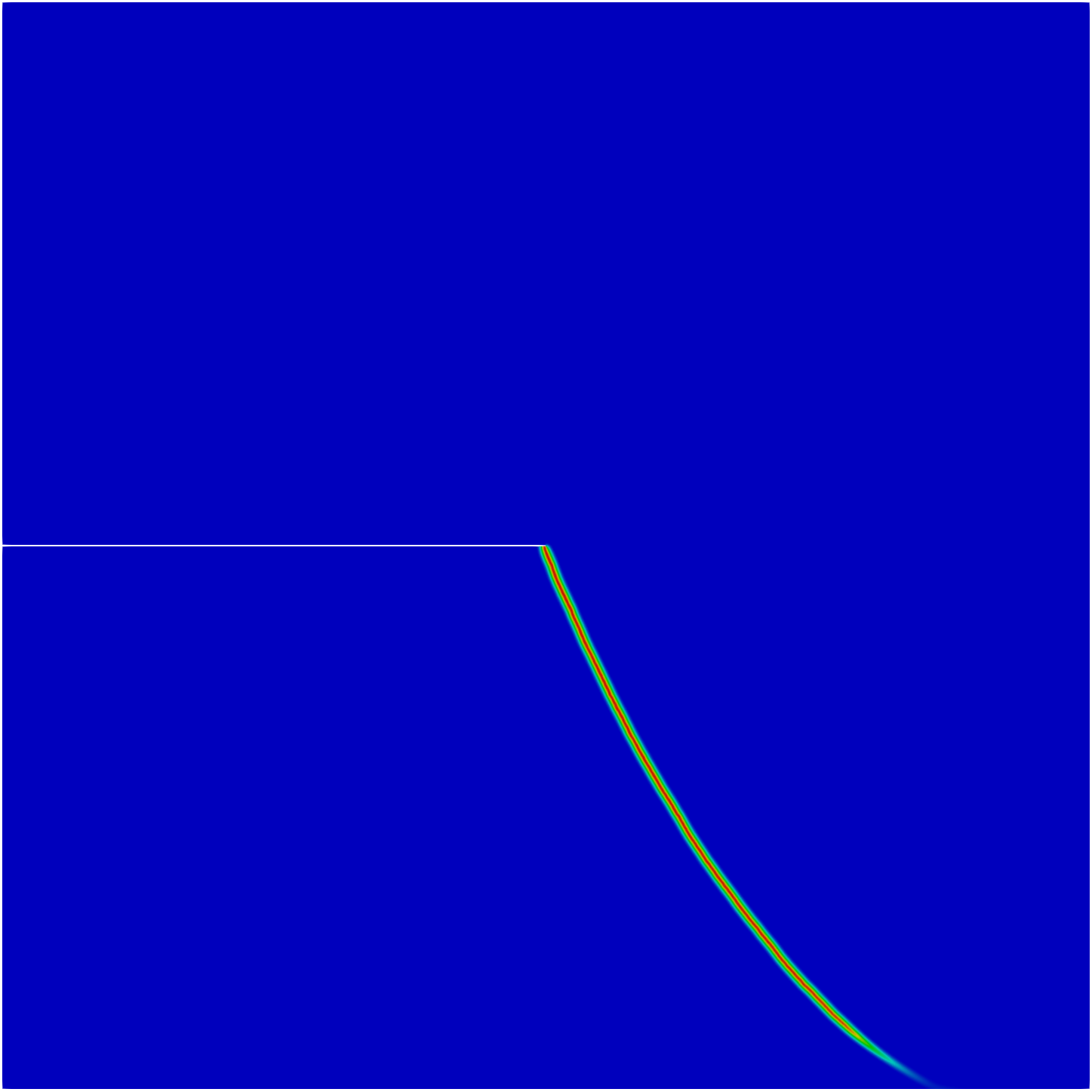}
  \label{fig:SENP-shear-crack-p20-b005}}
  \caption{Single-edge notched plate under tension: crack profiles at $u^{\ast} = 0.02$ mm for various traction order parameters $p \ge 1$.}
  \label{fig:SENP-shear-crack-profiles-orders}
\end{figure}

\Cref{fig:SENP-shear-crack-profiles-orders} presents the crack profiles predicted by the \texttt{$\mu$PF-CZM} with various traction order parameters $p \ge 1$ using the same length scale parameter $b = 0.005$ mm. As can be seen, the crack tails are completely fractured (i.e., $d_{\ast} = 1$) such that the crack profiles therein are almost identical. Comparatively, the crack tip is more sharp for a larger value of the exponent $p \ge 1$. Again, for this brittle fracture involving rapid crack propagation once crack nucleates, the crack tip is limited to a small region and the differences between various traction order parameters are almost negligible.

\subsection{Double cantilever beam test}

\begin{figure}[h!] \centering
  \includegraphics[width=0.75\textwidth]{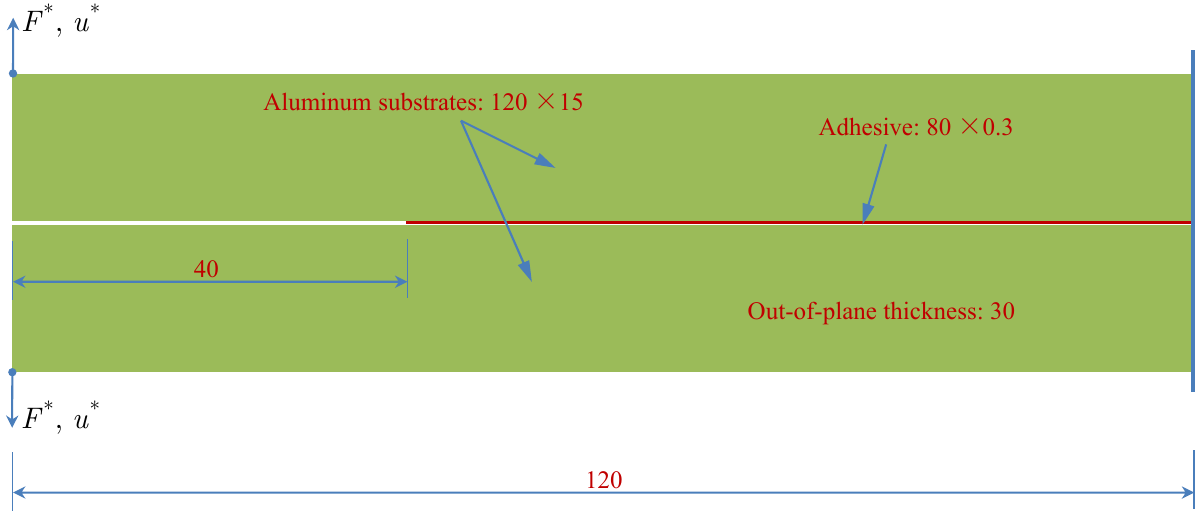}
  \caption{Double cantilever beam (DCB) test: Geometry (unit of length: mm), loading and boundary conditions.}
	\label{fig:dcb-geometry}
\end{figure}

This final example demonstrates the capability of the proposed \texttt{$\mu$PF-CZM} in simulation of cohesive fracture with concave softening responses. The double cantilever beam (DCB) test reported in \cite{PN2000} and studied in \cite{AFLMP2009,GLML2023} was considered. The geometry, boundary and loading condition of the DCB specimen is shown in \cref{fig:dcb-geometry}.

The specimen consists of two aluminum substrates partially bonded with a methacrylate adhesive interface. The dimensions of each aluminum substrate are 120 mm $\times$ 15 mm, and those of the adhesive interface are 80 mm $\times$ 0.3 mm, respectively. The notch is 40 mm long measured from the left edge of the beams. Plane strain of an out-of-plane thickness 30 mm was assumed to match the initial response of the test reported in \cite{AFLMP2009}. 

The aluminum substrates were modeled using a linear elastic material with Young's modulus 70 GPa and Poisson's ratio 0.3. The adhesive layer was modeled by the proposed \texttt{$\mu$PF-CZM} with the following material parameters: Young's modulus $E_{0} = 880$ MPa, Poisson's ratio $\nu_{0} = 0.15$, the failure strength $f_{\text{t}} = 3.3$ MPa and the fracture energy $G_{\text{f}} = 0.523$ N/mm, with an Irwin internal length 42.26 mm. The initial stiffness of the adhesive layer is relatively high, but it softens approaching failure. This softening behavior is caused by development and propagation of microcracks within the adhesive, leading to reduction of its stiffness and load-carrying capacity. This type of behavior is characteristic of materials that exhibit concave softening curve. The \cite{PPR2009} softening curve \eqref{eq:PPR-softening} with various values of the exponent $m \in [1, 2]$ was adopted. For the sake of comparison, other softening curves were also considered.

Two values of the phase-field length scale parameter, i.e., $b = 0.05$ mm and $b = 0.08$ mm, were adopted in the numerical simulations. In order to resolve the length scale properly, the adhesive layer was discretized using piece-wise linear triangular elements with the mesh size $h = \frac{1}{5} b$ in this critical area.

\begin{figure}[h!] \centering
  \subfigure[$p = 1.0$]{
  \includegraphics[width=0.475\textwidth]{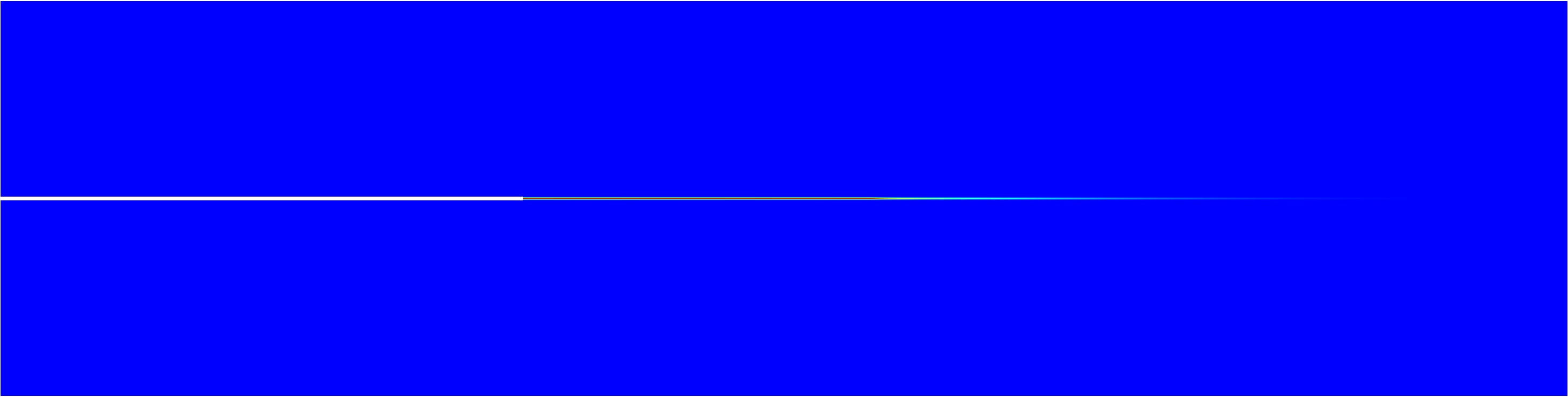}
  \label{fig:dcb-crack-profile-p10}} \hfill
  \subfigure[$p = 1.5$]{
  \includegraphics[width=0.475\textwidth]{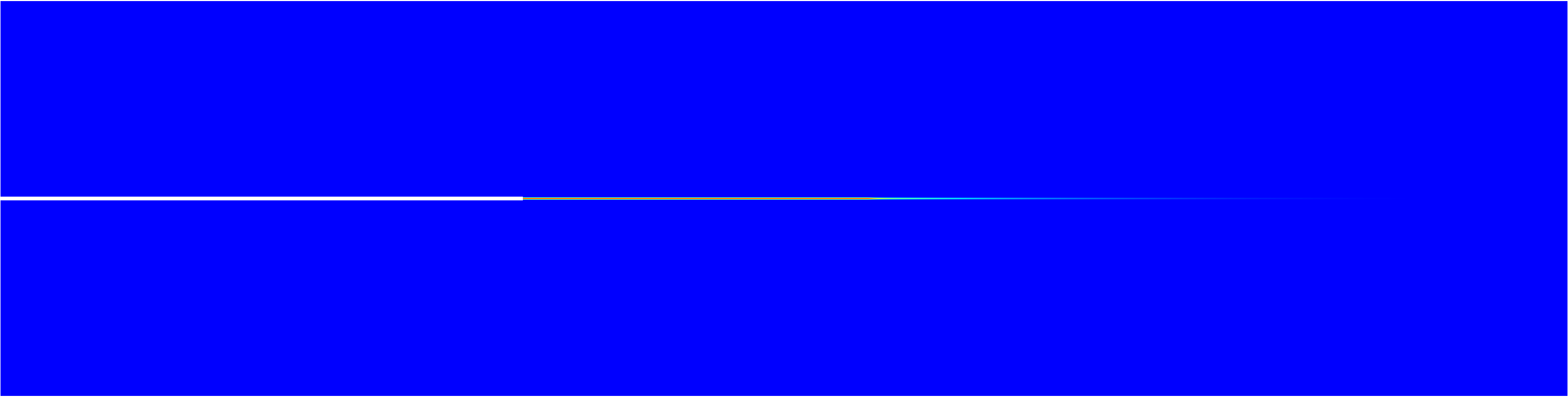}
  \label{fig:dcb-crack-profile-p15}}
  \subfigure[$p = 2.0$]{
  \includegraphics[width=0.475\textwidth]{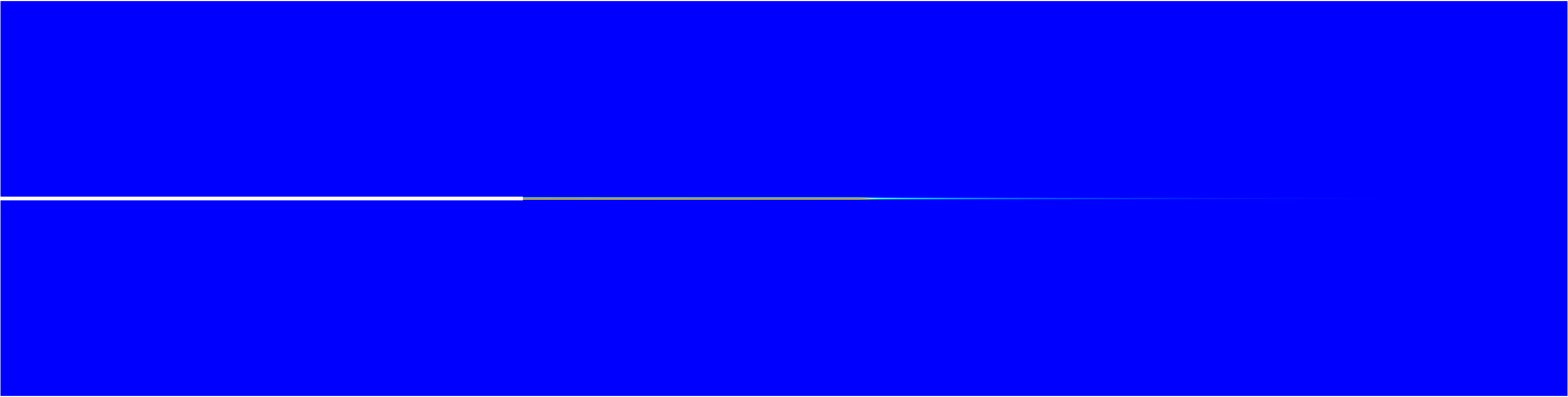}
  \label{fig:dcb-crack-profile-p20}} \hfill
  \subfigure[Zoomed crack profile around the notch]{
  \includegraphics[width=0.475\textwidth]{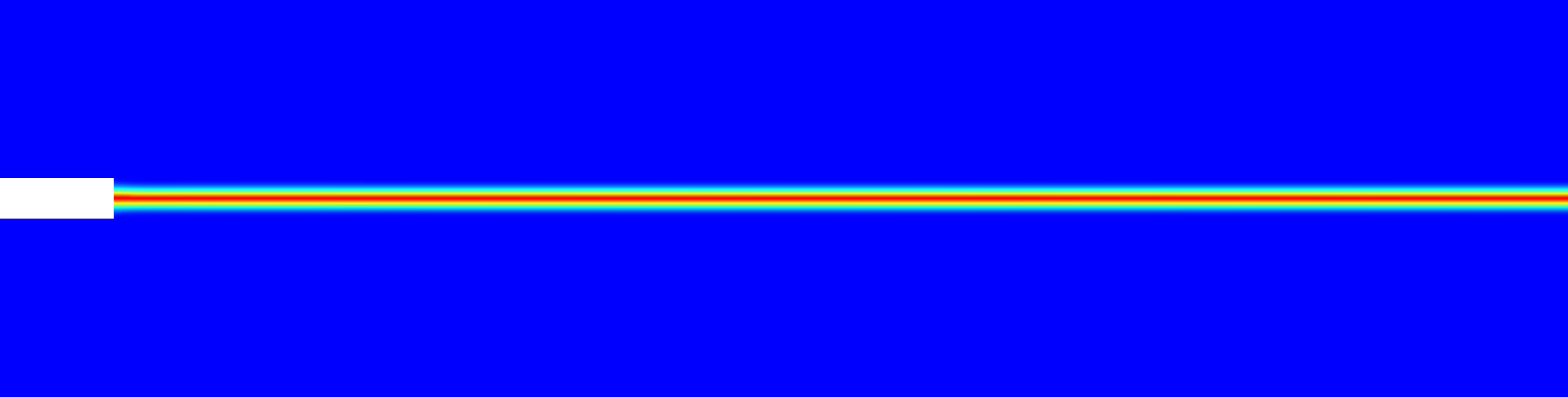}
  \label{fig:dcb-crack-profile-zoomed}}
  \caption{Double cantilever beam (DCB) test: Crack profiles at $u^{\ast} = 0.6$ mm predicted by the \texttt{$\mu$PF-CZM} with various traction order parameters and the same length scale parameter $b = 0.08$ mm.}
  \label{fig:dcb-crack-profiles-b008}
\end{figure}

The crack patterns predicted by the \texttt{$\mu$PF-CZM} with the length scale $b = 0.08$ mm are presented in \cref{fig:dcb-crack-profiles-b008}. As expected, a crack propagates horizontally within the adhesive layer between two substrates. The traction order parameter $p \ge 1$ has negligible effects on the crack profile. 

\begin{figure}[h!] \centering
  \subfigure[Force--deflection curves for various softening curves]{
  \includegraphics[width=0.485\textwidth]{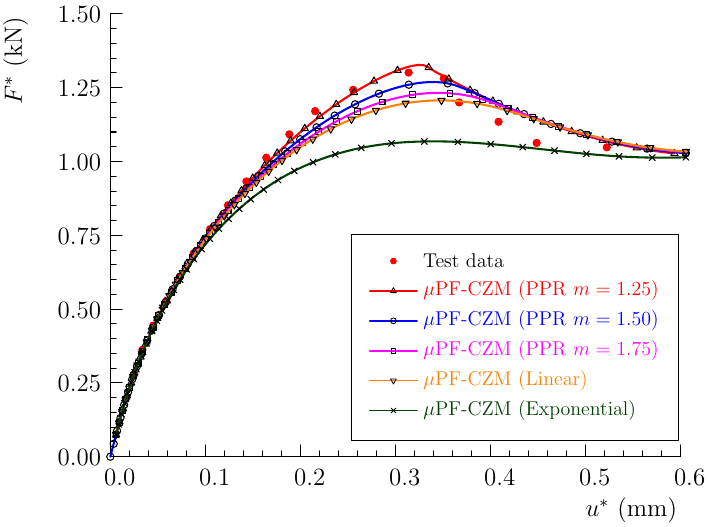}
  \label{fig:dcb-load-disp}}
  \subfigure[Force--deflection curves for the \cite{PPR2009} softening]{
  \includegraphics[width=0.485\textwidth]{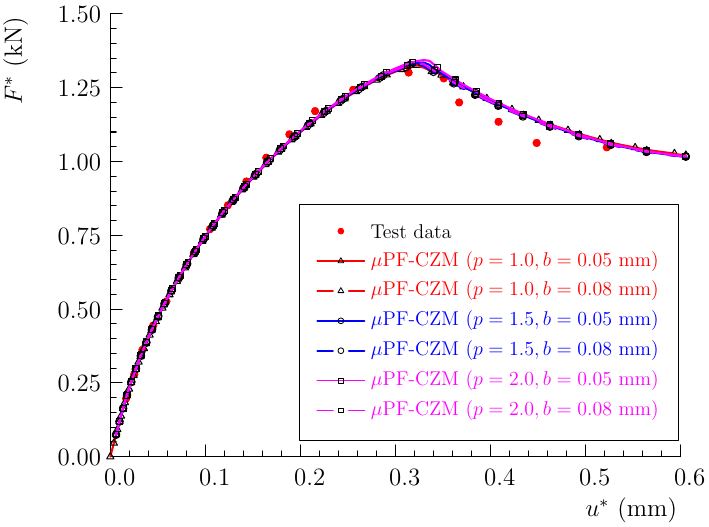}
  \label{fig:dcb-load-disp-ppr-m125}}
  \caption{Double cantilever beam (DCB) test: Force--deflection curves.}
  \label{fig:dcb-load-disp-all}
\end{figure}

\Cref{fig:dcb-load-disp} compares the force--deflection curve predicted by the \texttt{$\mu$PF-CZM} with various softening curves and an identical length scale $b = 0.05$ mm against the test data. As can be seen, the numerical results given by the \texttt{$\mu$PF-CZM} with various parameters almost coincide before the deflection $u^{\ast} \approx 1.0$ mm, after which the curves start to deviate dependent on the initial slope of each softening curve. For the softening curve with the largest initial slope $k_{0} < 0$ (i.e., the \cite{PPR2009} softening with $m = 1.25$), the predicted load capacity is largest and the force--displacement curve around the peak tends to be sharpest. Comparatively, as the initial slope $k_{0}$ decreases, the response tends to be more rounded and the peak load becomes smaller. Except for that of the exponential softening which is convex, all the other numerical predictions converge again in the post-peak regime when the deflection reaches $u^{\ast} \approx 0.4$ mm. 

As the predicted force--displacement curve agrees best with the test result, the \cite{PPR2009} softening curve with the exponent $m = 1.25$ was used in the numerical simulations to study the effect of the traction order parameter $p \ge 1$ and the length scale $b$. All the numerical predictions presented in \cref{fig:dcb-load-disp-ppr-m125} agree well with the test data. Once again, the insensitivity to both the length scale and traction order parameters is validated even in the modeling of concave softening behavior.

\section{Conclusions}
\label{sec:conclusions}

This work extends the unified phase-field theory for brittle and cohesive fracture to more general scenarios. With the introduction of an extra dissipation function for the crack driving force, in addition to the geometric function for the phase-field regularization and the degradation function for the constitutive relation, the theoretical and application scopes are broadened greatly. Despite the loss of variational consistency, the resulting formulation is still thermodynamically consistent. 

Moreover, a generalized phase-field cohesive zone model (\texttt{$\mu$PF-CZM}) is proposed. Not only concave softening behavior, e.g., the \cite{PPR2009} softening for adhesives, but also high-order cohesive traction, both being restrictions of the previous works, can be properly dealt with. The involved characteristic functions are no longer postulated \textit{a priori} as in the original \texttt{PF-CZM}, but are analytically determined in a universal, optimal and rationalized manner, for almost any specific traction--separation law. Specifically, upon assuming an arguably simplest relationship between the dissipation and geometric functions, all the involved functions are determined in closed-form solution relying only on the least conditions for a non-shrinking crack band and length scale insensitivity. In particular, with an optimal geometric function, the crack irreversibility can be enforced without affecting the target traction--separation law. The global responses are insensitive not only to the incorporated phase-field length scale as in the original \texttt{PF-CZM}, but also to the traction order exponent, though the crack bandwidth and profile might be affected by both. Last but not the least, some previous works can be recovered as its particular cases. For instance, for linear softening the lowest-order \texttt{$\mu$PF-CZM} coincides with the original \texttt{PF-CZM} \citep{WN2018}; the \texttt{PF-CZM} with integral transform \citep{FFL2021} and the so-called extended gradient damage model \citep{XR2024} are both the particular cases of the lowest-order \texttt{$\mu$PF-CZM} with a linear parameterized traction.

The proposed \texttt{$\mu$PF-CZM} still falls into the unified phase-field theory for fracture. Accordingly, the previous numerical implementation based on the multi-field finite element method and alternate minimization algorithm are adopted with minor modifications. The \texttt{$\mu$PF-CZM} is validated by several representative numerical examples of brittle and cohesive fracture. The first 1D example shows that, as the non-shrinking crack band is automatically guaranteed, the crack irreversibility can be enforced without affecting the target traction--separation law whatever the softening curve is convex, linear or concave. In the second example, the effects of the softening curve on the global responses of cohesive fracture in quasi-brittle solids are studied. It is found that the smaller the initial slope of the softening curve is, the minimal peak load the structure sustains. This conclusion is consistent with that drawn from the numerical results given by the CZM \citep{BP1997}. Moreover, for 2D problems the crack irreversibility has to be considered since guaranteeing a non-shrinking crack band alone is generally insufficient. The third example is concerned with application of the proposed \texttt{$\mu$PF-CZM} with linear softening to the modeling of brittle fracture. Both the global responses and the crack patterns are well captured as in the original \texttt{PF-CZM}. The final example demonstrates the capability of the proposed \texttt{$\mu$PF-CZM} in simulation of cohesive fracture with concave softening behavior, removing the limitation of the original \texttt{PF-CZM} \citep{GLML2023}. Remarkably, the insensitivity of the proposed \texttt{$\mu$PF-CZM} to the incorporated length scale and the traction order exponent is verified in all the examples.

Extension of the proposed \texttt{$\mu$PF-CZM} to more complex scenarios is straightforward. For instance, the modeling of fracture in finite deformation setting and in composites \citep{LYZL2020,LZL2020}, mixed-mode fracture, dynamic fracture, multiphysical fracture \citep{MGN2018,KNMP2020,CMMP2021}, etc., among many others, can be considered in the forthcoming work.

\section*{Acknowledgments}
	
This work is supported by National Natural Science Foundation of China (52125801), State Key Laboratory of Disaster Reduction in Civil Engineering (SLDRCE20-01), and Guangdong Provincial Key Laboratory of Modern Civil Engineering Technology (2021B1212040003) to the author (J.Y. Wu).

%\section*{References}

%\begin{spacing}{1.0}
%\bibliographystyle{elsarticle-num}
\bibliographystyle{elsarticle-harv}
%\setlength{\bibsep}{0.5ex}  % vertical spacing between references
%\footnotesize % font size
\bibliography{references}

\clearpage
\begin{appendix}

\section{Solution to Abel's equation}
\label{sec:solution-Abel-equation}

Let us consider the following Abel's equation
\begin{align}
	y (d)
		= \int_{0}^{d}	\dfrac{\varrho (\vartheta)}{\sqrt{r (d) 
		- r (\vartheta)}} \; \td \vartheta
\end{align}
The kernel function $\varrho (d)$ is solved as \citep{WW1996,PM2008,FFL2021}
\begin{align}
	\varrho (d) 
		= \dfrac{1}{\pi} \dfrac{\partial}{\partial d} \Bigg( 
			\int_{0}^{d} \dfrac{y (\vartheta) r' (\vartheta)}{
			\sqrt{r (d) - r (\vartheta)}} \; \td \vartheta \Bigg)
\end{align}
for the first derivative $r' (\vartheta) = \partial r / \partial \vartheta$. 

As far as \cref{eq:crack-opening-gpfczm-general,eq:crack-opening-gpfczm-final} are concerned, it follows that 
\begin{align}
	r (\vartheta)
		=-\big(1 - \vartheta \big)^{2p}, \qquad
	r' (\vartheta)
		= 2p \big(1 - \vartheta \big)^{2p - 1}
\end{align}
After some straightforward manipulations, the solutions given in \cref{eq:solution-Abel-equation} and \cref{eq:cracking-function-Abel}$_{2}$ are obtained.

\section{Softening curves}
\label{sec:softening-curves}

In this work, the following softening curves are considered; see \cref{fig:gpfczm-softening-curves-comparison}.
\begin{itemize}
\item Linear softening: The linear softening law is expressed as
\begin{align}\label{eq:linear-softening-cohesive}
%	\mathcal{G} (\itw)
%		= \begin{cases}
%				G_{\text{f}} \dfrac{\itw}{\itw_{\text{cL}}} \Big( 2 
%					- \dfrac{\itw}{\itw_{\text{cL}}} \Big) & \qquad \text{if} \;
%					\itw \le \itw_{\text{cL}} \\
%				G_{\text{f}} & \qquad \text{if} \; \itw \ge \itw_{\text{cL}}
%			\end{cases} \qquad
  \sigma (\itw)
		= f_{\text{t}} \max \Big( 1 - \dfrac{f_{\text{t}}}{2 G_{\text{f}}} 
			\itw, 0 \Big)
\end{align}
with the following initial slope $k_{0}$ and ultimate crack opening $\itw_{\text{c}}$
\begin{align}
	k_{0}
	  = k_{0\text{L}}
	  =-\dfrac{f_{\text{t}}^{2}}{2 G_{\text{f}}}, \qquad
	\itw_{\text{c}}
	  = \itw_{\text{cL}}
		= \dfrac{2 G_{\text{f}}}{f_{\text{t}}}	  	
\end{align}

\item Exponential softening: The exponential softening curve is given by
\begin{align}\label{eq:softening-curve-exponential}
%	\mathcal{G} (\itw)
%		= G_{\text{f}} \bigg[ 1 - \exp \Big(-\dfrac{f_{\text{t}}}{G_{\text{f}}} 
%			\itw \Big) \bigg], \qquad
	\sigma (\itw)
		= f_{\text{t}} \exp \Big(-\dfrac{f_{\text{t}}}{G_{\text{f}}} 
			\itw \Big)
\end{align}
with the following initial slope $k_{0}$ and ultimate crack opening $\itw_{\text{c}}$
\begin{align}
	k_{0}
	  =-\dfrac{f_{\text{t}}^{2}}{G_{\text{f}}}, \qquad
	\itw_{\text{c}}
		=+\infty
\end{align}

%\item Hyperbolic softening: Let us consider the following hyperbolic softening curve
%\begin{align}\label{eq:softening-curve-hyperbolic}
%%	\mathcal{G} (\itw)
%%		= G_{\text{f}} \bigg[ 1 - \Big(1 + \dfrac{f_{\text{t}}}{G_{\text{f}}} 
%%			\itw \Big)^{-1} \bigg], \qquad
%	\sigma (\itw)
%		= f_{\text{t}} \Big(1 + \dfrac{f_{\text{t}}}{G_{\text{f}}} 
%			\itw \Big)^{-2}
%\end{align}
%with the following initial slope and ultimate crack opening
%\begin{align}
%	k_{0}
%	  =-\dfrac{2 f_{\text{t}}^{2}}{G_{\text{f}}}, \qquad
%	\itw_{\text{c}}
%		=+\infty
%\end{align}

\item \cite{CHR1986} softening for concrete. The softening curve is expressed in terms of a normalized crack opening $\tilde{\itw} := \itw / \itw_{\text{c}}$ 
\begin{align}\label{eq:softening-curve-Cornelissen}
  \sigma (\itw)
    = f_{\text{t}} \Big[ \big( 1.0 + \eta_{1}^{3} \tilde{\itw}^{3} \big) 
	    \exp \big(-\eta_{2} \tilde{\itw} \big) 
    - \tilde{\itw} \big(1.0 + \eta_{1}^{3} \big) 
	    \exp \big(-\eta_{2} \big) \Big]
\end{align}
with the initial slope $k_{0}$ and ultimate crack opening $\itw_{\text{c}}$ given by
\begin{align}
  k_{0}
%    =-6.9574 \dfrac{f_{\text{t}}}{\itw_{\text{c}}}
    =-1.3546 \dfrac{f_{\text{t}}^{2}}{G_{\text{f}}}, \qquad
  \itw_{\text{c}}
    = 5.1361 \dfrac{G_{\text{f}}}{f_{\text{t}}}
\end{align}
where the typical values $\eta_{1} = 3.0$ and $\eta_{2} = 6.93$ have been considered for normal concrete. 

\item \cite{PPR2009} softening: The traction--separation law is expressed as
\begin{align}\label{eq:PPR-softening}
	\sigma (\itw)
		= f_{\text{t}} \Big(1 - \dfrac{f_{\text{t}}}{m G_{\text{f}}} 
			\itw \Big)^{m - 1}
\end{align}
with the initial slope $k_{0}$ and ultimate crack opening $\itw_{c}$ given by
\begin{align}
	k_{0}
		=-\dfrac{m-1}{m} \dfrac{f_{\text{t}}^{2}}{G_{\text{f}}}, \qquad
	\itw_{\text{c}}
		= m \dfrac{G_{\text{f}}}{f_{\text{t}}}
\end{align}
where the exponent $m > 1$ controls the shape of the softening curve, i.e., convex for $m > 2$, concave for $1 < m < 2$ and linear for $m = 2$.

\end{itemize}

\section{Coefficients and characteristic functions in the polynomial softening curves}
\label{sec:coefficients-polynomial-softening-curves}

The least square fitting is used to determine the coefficients $c_{n} \; (n = 0, 1, \cdots, N)$ in the polynomial function \eqref{eq:polynomial-softening}. In order to reproduce the crucial characteristics (e.g., the fracture energy $G_{\text{f}}$, the ultimate crack opening $\itw_{\text{c}}$, the initial slope $k_{0}$, etc.) of a specific softening curve $\sigma (\itw)$, the fitting coefficients $c_{n} \; (n = 0, 1, \cdots, N)$ need to satisfy the following conditions
\begin{align}\label{eq:polynomial-fitting-conditions}
\begin{cases}
	\itw_{\text{c}}
		= \itw (d_{\ast} = 1)
		=-\itw_{\text{cL}} c_{0} 
		& \qquad \Longrightarrow \qquad
	c_{0}
		=-\dfrac{\itw_{\text{c}}}{\itw_{\text{cL}}} \vspace{2mm} \\
	0
		= \itw (d_{\ast} = 0)
		= \displaystyle \sum_{n = 0}^{N} c_{n} 
		& \qquad \Longrightarrow \qquad
	\displaystyle \sum_{i = 0}^{N} c_{n}
		= 0 \vspace{2mm} \\
	G_{\text{f}}
		= \displaystyle \int_{0}^{\itw_{\text{c}}} \sigma \; \td \itw 
		= \int_{0}^{1} \sigma (\vartheta) \dfrac{\partial \itw (\vartheta)}{
			\partial \vartheta} \; \td \vartheta 
		& \qquad \Longrightarrow \qquad
	\displaystyle \sum_{n = 0}^{N} \dfrac{n}{n + 1} c_{n}
		= \dfrac{1}{2} \vspace{2mm} \\ 
	k_{0}
		= \dfrac{\partial \sigma}{\partial \itw} \bigg|_{d_{\ast} = 0}
		= \bigg( \displaystyle \sum_{n = 0}^{N} n \cdot c_{n} \bigg)^{-1}
			k_{0\text{L}}
		&	\qquad \Longrightarrow \qquad
	\displaystyle \sum_{n = 0}^{N} n \cdot c_{n}
		= \dfrac{k_{0\text{L}}}{k_{0}}
\end{cases}
\end{align}
Note that the last condition \eqref{eq:polynomial-fitting-conditions}$_{4}$ was frequently ignored in the literature, resulting in inaccurate evaluation of the peak load \citep{BP1997}. %for the initial slope $k_{0\text{L}} =-\frac{1}{2} f_{\text{t}}^{2} / G_{\text{f}}$ of the linear softening curve. 

\begin{itemize}
\item \cite{CHR1986} softening. The parameterized traction \eqref{eq:polynomial-softening}$_{1}$ and the softening curve \eqref{eq:softening-curve-Cornelissen} gives the following normalized crack opening $\bar{\itw} (d_{\ast})$
\begin{align}
  \big( 1 - d_{\ast} \big)^{p}
    = \Big[ 1.0 + \big( 0.3894 \eta_{1} \big)^{3} 
    	\bar{\itw}^{3} \Big] \exp \big(-0.3894 \eta_{2} \bar{\itw} \big) 
    - 0.3894 \bar{\itw} \big(1.0 + \eta_{1}^{3} \big) 
	    \exp \big(-\eta_{2} \big)
\end{align}

The normalized crack opening $\bar{\itw} (d_{\ast})$ is fitted by the sixth-order polynomial \eqref{eq:polynomial-softening} with the following coefficients $c_{n}$
\begin{subequations}
\begin{align}
	c_{0}
	 &=-2.5681, \quad
	c_{1}
		= \phantom{-} 14.8193, \quad
	c_{2}
		=-40.4105  \\
	c_{3}
	 &= \; 57.3515, \quad
	c_{4}
	  = -40.4200, \quad
	c_{5}
		= \phantom{-}11.3700, \quad
	c_{6}
		=-0.1423
\end{align}
\end{subequations}
such that
\begin{subequations}
\begin{align}
	\bar{c}_{1}
	 &= \; 101.6763, \quad
	\bar{c}_{2}
		=-40.4105, \quad
	\bar{c}_{3}
	  =-129.1615 \\
	\bar{c}_{4}
	 &=-60.6300, \quad
	\bar{c}_{5}
		= \phantom{-}30.0532, \quad
	c_{6}
		= \; -0.2668
\end{align}
\end{subequations}
for the coefficients $\bar{c}_{n}$ in the function $\varXi (d)$.

\item \cite{PPR2009} softening. The normalized crack opening $\bar{\itw} (d_{\ast})$ for the softening curve \eqref{eq:PPR-softening} is expressed as
\begin{align}
%\begin{cases}
%	\sigma (d_{\ast})
%		= \big(1 - d_{\ast} \big) f_{\text{t}} \\
	\itw (d_{\ast})
		= \itw_{\text{cL}} \dfrac{m}{2} \Big[ 1 - \big(1 - d_{\ast} 
			\big)^{\frac{p}{m - 1}} \Big]
	\qquad \Longrightarrow \qquad
	\bar{\itw} (d_{\ast}) 
		= \dfrac{m}{2} \Big[ 1 - \big(1 - d_{\ast} 
			\big)^{\frac{p}{m - 1}} \Big]
%\end{cases}
\end{align}
For various values of the exponent $m \in (0, 2)$, the coefficients $c_{n}$ of the sixth-order polynomial fitting \eqref{eq:polynomial-softening} are given by
\begin{itemize}
\item $m = 1.25$: 
\begin{align}
	c_{0}
	 =-0.625, \quad 
	c_{1}
		= 0, \quad 
	c_{2}
		= 0, \quad 
	c_{3}
	  = 0, \quad 
	c_{4}
		= 0.625, \quad 
	c_{5}
		= 0, \quad 
	c_{6}
		= 0
\end{align}
which lead to 
\begin{align}
	\bar{c}_{1}
	  = 1.5625, \quad
	\bar{c}_{2}
		= 0, \quad
	\bar{c}_{3}
	  =-0.9375, \quad 
	\bar{c}_{4}
	  = 0.9375, \quad 
	\bar{c}_{5}
		= 0, \quad 
	\bar{c}_{6}
		= 0
\end{align}
for the coefficients $\bar{c}_{n}$ in the function $\varXi (d)$.

\item $m = 1.50$:
\begin{align}
	c_{0}
	 =-0.750, \quad 
	c_{1}
		= 0, \quad 
	c_{2}
		= 0.75, \quad 
	c_{3}
	  = 0, \quad 
	c_{4}
		= 0, \quad 
	c_{5}
		= 0, \quad 
	c_{6}
		= 0
\end{align}
such that
\begin{align}
	\bar{c}_{1}
	  = 0.75, \quad
	\bar{c}_{2}
		= 0.75, \quad
	\bar{c}_{3}
	  = 0, \quad 
	\bar{c}_{4}
	  = 0, \quad 
	\bar{c}_{5}
		= 0, \quad 
	\bar{c}_{6}
		= 0
\end{align}
for the coefficients $\bar{c}_{n}$ in the function $\varXi (d)$.

\item $m = 1.75$
\begin{align}
	c_{0}
	 &=-0.8750, \qquad
	c_{1}
		= 0.2561, \qquad
	c_{2}
		= \phantom{-}1.7740 \notag \\
	c_{3}
	 &=-3.2461, \qquad
	c_{4}
		= 4.0717, \qquad
	c_{5}
		=-2.6982, \qquad
	c_{6}
		= 0.7176
\end{align}
It then follows that
\begin{align}
	\bar{c}_{1}
	 &=-8.0599, \qquad
	\bar{c}_{2}
		= \phantom{-}1.7740, \qquad
	\bar{c}_{3}
	  = 14.7847 \notag \\
	\bar{c}_{4}
	 &=\phantom{-}6.1076, \qquad
	\bar{c}_{5}
		=-5.8497, \qquad
	\bar{c}_{6}
		=\phantom{-1}1.3455
\end{align}
for the coefficients $\bar{c}_{n}$ in the function $\varXi (d)$.

\end{itemize}

\end{itemize}

The functions $\beta_{n} (d)$ involved in the sixth-order polynomial fitting \eqref{eq:cracking-function-polynomial} are expressed as %\citep{FFL2021}
\begin{subequations}
\begin{align}
	\beta_{0}
	 &=-\dfrac{1}{s} \\
	\beta_{1}
	 &=-\dfrac{s_{1}^{2}}{s} \\
	\beta_{2}
	 &= s_{1}^{2} \bigg[ -\dfrac{1}{s} + \text{arctanh} (s) \bigg] \\
	\beta_{3}
	 &= \dfrac{s_{1}^{2}}{s} \big( 1 - 2 s_{1}^{2} \big) \\
	\beta_{4}
	 &= \dfrac{s_{1}^{2}}{2} \bigg[ \dfrac{1 - 3 s_{1}^{2}}{s}
	 	+ 3 s_{1}^{2} \cdot \text{arctanh} (s) \bigg] \\
	\beta_{5}
	 &= \dfrac{s_{1}^{2}}{3 s} \big( 1 + 4 s_{1}^{2} - 8 s_{1}^{4} \big) \\
	\beta_{6}
	 &= \dfrac{s_{1}^{2}}{8} \bigg[ \dfrac{1}{s} \big( 2 + 5 s_{1}^{2} 
	 	- 15 s_{1}^{4} \big) + 15 s_{1}^{4} \cdot \text{arctanh} (s) 
	 		\bigg]
\end{align}
\end{subequations}
with the functions
\begin{align}\label{eq:functions-6th-fitting-s-s1}
	s 
		= \sqrt{1 - \big(1 - d \big)^{2 p}}, \qquad
	s_{1} 
		= \big(1 - d \big)^{p}
\end{align}

\section{Summary of the involved characteristic functions}
\label{sec:characteristic-functions}

The characteristic functions $\alpha (d)$, $\omega (d)$ and $\varpi (d)$ and the derivatives involved in the proposed \texttt{$\mu$PF-CZM} are summarized as follows:
\begin{itemize}
\item Optimal geometric function
\begin{align}
	\alpha (d)
		= 2 d - d^{2}, \qquad
	\alpha' (d)
		= 2 - 2 d, \qquad
	\alpha'' (d)
		=-2
\end{align}

\item Degradation function
\begin{align}
	\omega (d)
	  = \dfrac{1}{1 + \phi (d)}, \qquad
	\omega' (d)
		=-\omega^{2} (d) \phi' (d), \qquad
	\phi (d)
	  = \dfrac{ a_{0} p \sqrt{\alpha (d)}}{\big(1 - d \big)^{p + 1}}
			\varXi (d)		
\end{align}
\begin{align}\label{eq:crack-function-derivative}
	\phi' (d)
		= a_{0} p \dfrac{\big(1 - d \big) \alpha (d) \varXi' (d)
		+ \varXi (d) \Big[ \frac{1}{2} \big(1 - d \big) \alpha' (d) 
		+ \big( p + 1 \big) \alpha (d) \Big]}{\big(1 - d \big)^{p + 2} 
			\sqrt{\alpha (d)}}
\end{align}
where the function $\varXi (d)$ and its derivative $\varXi' (d)$ depend on the specific softening curve
\begin{align}
	\varXi (d)
		= \begin{cases}
				s (d) & \quad
					\text{Linear softening} \\
				\frac{1}{2} \text{arctanh} \big( s(d) \big) & \quad
					\text{Exponential softening} \\	
				\bar{c}_{1} s + \bar{c}_{3} s^{3}
						+ \bar{c}_{5} s^{5} + \big( \bar{c}_{2} s_{1}^{2}	
						+	\bar{c}_{4} s_{1}^{4} + \bar{c}_{6} s_{1}^{6} \big)
							\; \text{arctanh} \big( s(d) \big) & \quad
							\text{Polynomial softening}
			\end{cases}
\end{align}
\begin{align}
	\varXi' (d)
		= \begin{cases}
				\dfrac{p}{s (d)} \big(1 - d \big)^{2p - 1} & \quad
					\text{Linear softening} \vspace{2mm} \\
				\dfrac{p}{s (d)} \dfrac{1}{2} \big( 1 - d \big)^{-1} & \quad
					\text{Exponential softening} \vspace{2mm} \\	
				\big( \bar{c}_{1} + 3 \bar{c}_{3} s^{2}
						+ 5 \bar{c}_{5} s^{4} \big) \dfrac{p}{s (d)} 
						  \big(1 - d \big)^{2p - 1} \\
					  + \big( \bar{c}_{2} s_{1}^{2}	+	\bar{c}_{4} s_{1}^{4} 
						+ \bar{c}_{6} s_{1}^{6} \big) 
							\dfrac{p}{s (d)} \big( 1 - d \big)^{-1}  \\ 
						- 2 \big( \bar{c}_{2} s_{1}	
						+	2 \bar{c}_{4} s_{1}^{3} + 3 \bar{c}_{6} s_{1}^{5} \big)
							\; \text{arctanh} \big( s(d) \big) p 
							\big(1 - d \big)^{p - 1} & \quad
							\text{Polynomial softening}
			\end{cases}
\end{align}
for the functions $s (d)$ and $s_{1} (d)$ introduced in \cref{eq:functions-6th-fitting-s-s1}.

%\begin{itemize}
%\item Linear softening
%\begin{align}
%	\varXi (d)
%		= s (d), \qquad 
%%		= \sqrt{1 - (1 - d)^{2p}}
%	\varXi' (d)
%		= s' (d)
%\end{align}
%with
%\begin{align}
%	s (d) 
%		= \sqrt{1 - (1 - d)^{2 p}}, \qquad
%	s' (d)
%		= p \dfrac{\big(1 - d \big)^{2p - 1}}{s (d)}
%\end{align}
%
%\item Exponential softening
%\begin{align}
%	\varXi (d)
%		= \dfrac{1}{2} \text{arctanh} \big( s(d) \big), \qquad
%	\varXi' (d)
%		= \dfrac{1}{2} \dfrac{p}{(1 - d) s (d)}
%\end{align}
%
%\item Polynomial softening
%\begin{align}
%	\varXi (d)
%		= \bar{c}_{1} s - \bar{c}_{3} s^{3}
%		+ \bar{c}_{5} s^{5} + \big( \bar{c}_{2} s_{1}^{2}	
%		+	\bar{c}_{4} s_{1}^{4} + \bar{c}_{6} s_{1}^{6} \big)
%			\; \text{arctanh} \big( s(d) \big)
%\end{align}
%\begin{align}
%	\varXi' (d)
%	 &= \big( \bar{c}_{1} - 3 \bar{c}_{3} s^{2}
%		+ 5 \bar{c}_{5} s^{4} \big) s' (d)
%		+ 2 \big( \bar{c}_{2} s_{1}	
%		+	2 \bar{c}_{4} s_{1}^{3} + 3 \bar{c}_{6} s_{1}^{5} \big)
%			s'_{1} \; \text{arctanh} \big( s(d) \big) \notag \\
%	 &+ \big( \bar{c}_{2} s_{1}^{2}	+	\bar{c}_{4} s_{1}^{4} 
%		+ \bar{c}_{6} s_{1}^{6} \big) 
%			\dfrac{p}{(1 - d) s (d)}
%\end{align}
%with
%\begin{align}
%	s_{1} (d) 
%		= (1 - d)^{p}, \qquad
%	s'_{1} (d) 
%		=-p \big(1 - d \big)^{p - 1}
%\end{align}
%\end{itemize}
Upon crack initiation, i.e., $d = 0$, the derivative $\phi' (d)$ is evaluated as
\begin{align}
	\phi' (0)
		= \lim_{d \to 0} \phi' (d)
		= a_{0} p \lim_{d \to 0} \Bigg[ \sqrt{\alpha (d)} \varXi' (d) 
		+ \dfrac{\varXi (d)}{\sqrt{\alpha (d)}} \Bigg]
		= \dfrac{2}{\bar{k}_{0}} a_{0} p^{\frac{3}{2}}
\end{align}
or, specifically,
\begin{align}
	\phi' (0)
		= \begin{cases}
				2 a_{0} p^{\frac{3}{2}} & \qquad 
						\text{Linear softening} \\
				\phantom{2} a_{0} p^{\frac{3}{2}} & \qquad 
						\text{Exponential softening} \\
				2 a_{0} p^{\frac{3}{2}} \big( \bar{c}_{1} + \bar{c}_{2}
					+ \bar{c}_{4} + \bar{c}_{6} \big) & \qquad 
						\text{Polynomial softening} \\
			\end{cases}
\end{align}

\item Derivatives of the dissipation function
\begin{align}
	\varpi' (d)
		=-\omega^{2} (d) \mu' (d) \le 0, \qquad
	\varpi'' (d)
%	  =-\omega (d) \Big[ 2 \mu' (d) \omega' (d) 
%		+ \omega (d) \mu'' (d) \Big]
	  = \omega^{2} (d) \Big[ 2 \omega (d) \mu' (d) \phi' (d)
		- \mu'' (d) \Big]		
			\ge 0
\end{align}
with
\begin{align}
 &\mu (d)
	  = a_{0} \dfrac{\alpha (d)}{\big(1 - d \big)^{2p}}, \qquad
	\mu' (d)
		= a_{0} \dfrac{2p \alpha (d) + \big(1 - d \big) \alpha' (d)}{
			\big(1 - d \big)^{2p + 1}} \\
 &\mu'' (d)
	  = a_{0} \dfrac{2p \big(2p + 1 \big) \alpha (d) 
	  + 4p \big(1 - d \big) \alpha' (d)
	  + \big(1 - d \big)^{2} \alpha'' (d)}{\big(1 - d \big)^{2p + 2}}
\end{align}
Note that the dissipation function $\varpi (d)$ itself is not needed for the proposed \texttt{$\mu$PF-CZM}.

\end{itemize}

\end{appendix}

\end{document}